\documentclass[conference]{IEEEtran}
\ifCLASSINFOpdf
\usepackage[pdftex]{graphicx}
\else
\fi
\usepackage{sansmath}

\usepackage[cmex10]{amsmath}
\usepackage{tabularx}
\usepackage{xcolor}
\usepackage{url}

\usepackage{booktabs}

\usepackage{multirow}
\usepackage{amsthm} 
\usepackage{amssymb} 

\usepackage{scrextend} 

\usepackage[english]{datetime2} 

\usepackage{placeins}

\newtheorem{mydef}{Definition}
\newtheorem{quest}{Question}

\usepackage{array}

\usepackage[tight,footnotesize]{subfigure}
\begin{document}
	%
	\title{Decrypting Distributed Ledger Design - \\Taxonomy, Classification and Blockchain Community Evaluation}

	\author{\IEEEauthorblockN{Mark C. Ballandies}
		\IEEEauthorblockA{Computational Social Science \\
			ETH Zurich\\
			Zurich, Switzerland\\ 
			bmark@ethz.ch}
		\and
		\IEEEauthorblockN{Marcus M. Dapp}
		\IEEEauthorblockA{Computational Social Science \\
			ETH Zurich\\
			Zurich, Switzerland\\ 
			mdapp@ethz.ch}
		\and
		\IEEEauthorblockN{Evangelos Pournaras}
		\IEEEauthorblockA{School of Computing\\
			University of Leeds\\
			Leeds LS2 9JT UK\\ 
			e.pournaras@leeds.ac.uk}}

	
	%


	\maketitle

	\begin{abstract}
		More than 1000 distributed ledger technology (DLT) systems raising \$600 billion in investment in 2016 feature the unprecedented and disruptive potential of blockchain technology. A systematic and data-driven analysis, comparison and rigorous evaluation of the different design choices of distributed ledgers and their implications is a challenge. The rapidly evolving nature of the blockchain landscape hinders reaching a common understanding of the techno-socio-economic design space of distributed ledgers and the crypto\-economies they support.
	To fill this gap, this paper makes the following contributions: (i) A conceptual architecture of DLT systems with which (ii) a taxonomy is designed and (iii) a rigorous classification of DLT systems is made using real-world data and wisdom of the crowd. (iv) A DLT design guideline is the end result of applying machine learning methodologies on the classification data. Compared to related work and as defined in earlier taxonomy theory, the proposed taxonomy is highly comprehensive, robust, explanatory and extensible. The findings of this paper can provide new insights and better understanding of the key design choices evolving the modeling complexity of DLT systems, while identifying opportunities for new research contributions and business innovation. 
	\end{abstract}
	

	%
	\IEEEpeerreviewmaketitle

	\section{Introduction}

Over 1000 systems have emerged in recent years from distributed ledger technology (DLT), raising \$600 billion in investment in 2016 \cite{tapscott2017blockchain}. They power a large spectrum of novel distributed applications making
use of data immutability, integrity, fair access, transparency, non-repudiation
of transactions \cite{xu2017taxonomy} and crypto\-currencies. These
applications include improving supply-chains~\cite{Korpela2017},
creating self-sovereign identities\footnote{Decentralized identities, owned and controlled by the individual represented
	through the identity.}~\cite{Manohar2018}, establishing peer-to-peer energy markets~\cite{Kang2017},
securing digital voting~\cite{Kshetri2018,Pournaras2019} and enabling international
financial transactions \cite{xu2017taxonomy}. The most well-known DLT system is Bitcoin, featuring a novel
consensus mechanism\footnote{Bitcoin uses a Nakamoto consensus, see Section \ref{sec:consensus}.}
and a crypto\-economic design\footnote{In particular, paying a block reward (Section \ref{sec:tax_token})
	and transaction fees (Section \ref{sec:consensus}) to its consensus
	participants.} (CED), which enables untrusted parties to reach consensus~\cite{bonneau2015sok}.
Bitcoin is the first public DLT system, which prevents
double-spending\footnote{Faulty transactions of the same token to two different receivers.}
and Sybil attacks\footnote{Setup of fake identities to insert faulty information into the distributed
	ledger.} \cite{tschorsch2016bitcoin}.

A \emph{distributed ledger} (DL) is a distributed data structure,
whose entries are written by the participants of a DLT system after
reaching consensus on the validity of the entries. A \emph{consensus
	mechanism} is usually an integral part of a distributed ledger system
and guarantees system reliability: all written entries are validated
without a trusted third party. Distributed ledgers are usually designed
to support secure \emph{cryptoeconomies}, which are capable of operating
cross-border, without depending on a particular political structure
or legal system. These cryptoeconomies rely on digital currencies
referred to as \emph{tokens} and cryptographic techniques to regulate
how value is exchanged between the participating actors~\cite{babbitt2014crypto,davidson2016economics}.
The options and choices of a cryptoeconomy are referred to as \textit{cryptoeconomic
	design} (CED) and this plays a key role in the stability of a DLT
system in terms of convergence, liveness, and fairness~\cite{bonneau2015sok}. 

Nevertheless, making system design choices in this rapidly evolving technological landscape to meet the requirements of a broad spectrum of distributed applications is complex and challenging. The lack of a common and insightful conceptual framework for DLT has
been cited as a significant barrier in this regard~\cite{notheisen2017breaking}.
Moreover, the system configuration space of distributed ledgers and
the cryptoeconomies they support is large, which has implications on the applicability as well as cost-effectiveness of DLT systems in real-world applications~\cite{xu2017taxonomy}. To date, these configurations have not been rigorously formalized to guide researchers and practitioners on how to design DLT systems~\cite{bonneau2015sok,samuel2016layered}. In particular, the broad spectrum and complexity of key design choices have not been determined.  
It has been argued that this lack of a clear positioning of DLT systems leads to a fragmentation in the blockchain
community and a duplication of effort~\cite{tasca2017ontology}. The
significance of this challenge is reflected in the recent taxonomies
of distributed ledgers \cite{xu2017taxonomy,tasca2017ontology,yeow2018decentralized,xu2016blockchain,wieninger2019development}.

This paper derives a \textit{useful}\footnote{Usefulness is defined in Nickerson et. al \cite{nickerson2013method} and formerly introduced in Section \ref{sec:feedback_taxonomy}.} taxonomy of DLT systems from a novel conceptual
architecture. This taxonomy is then utilized to classify 50 viable and actively maintained DLT systems. In contrast to earlier work, a novel evaluation
methodology is employed that solicits feedback from the blockchain
community and constructively uses it to validate and further improve
the proposed taxonomy and classification. 
Moreover, the classification data are utilized to reason about key design choices in the observed DLT systems, which then, in turn, determine a design guideline for these systems.

\begin{figure*}[!htb]
	 \label{fig:concept}
	\caption{An overview of the conceptual architecture containing the four key
		concepts of DLT systems and their relationship: action, consensus,
		distributed ledger and token.}
	\includegraphics[width=1.0\textwidth]{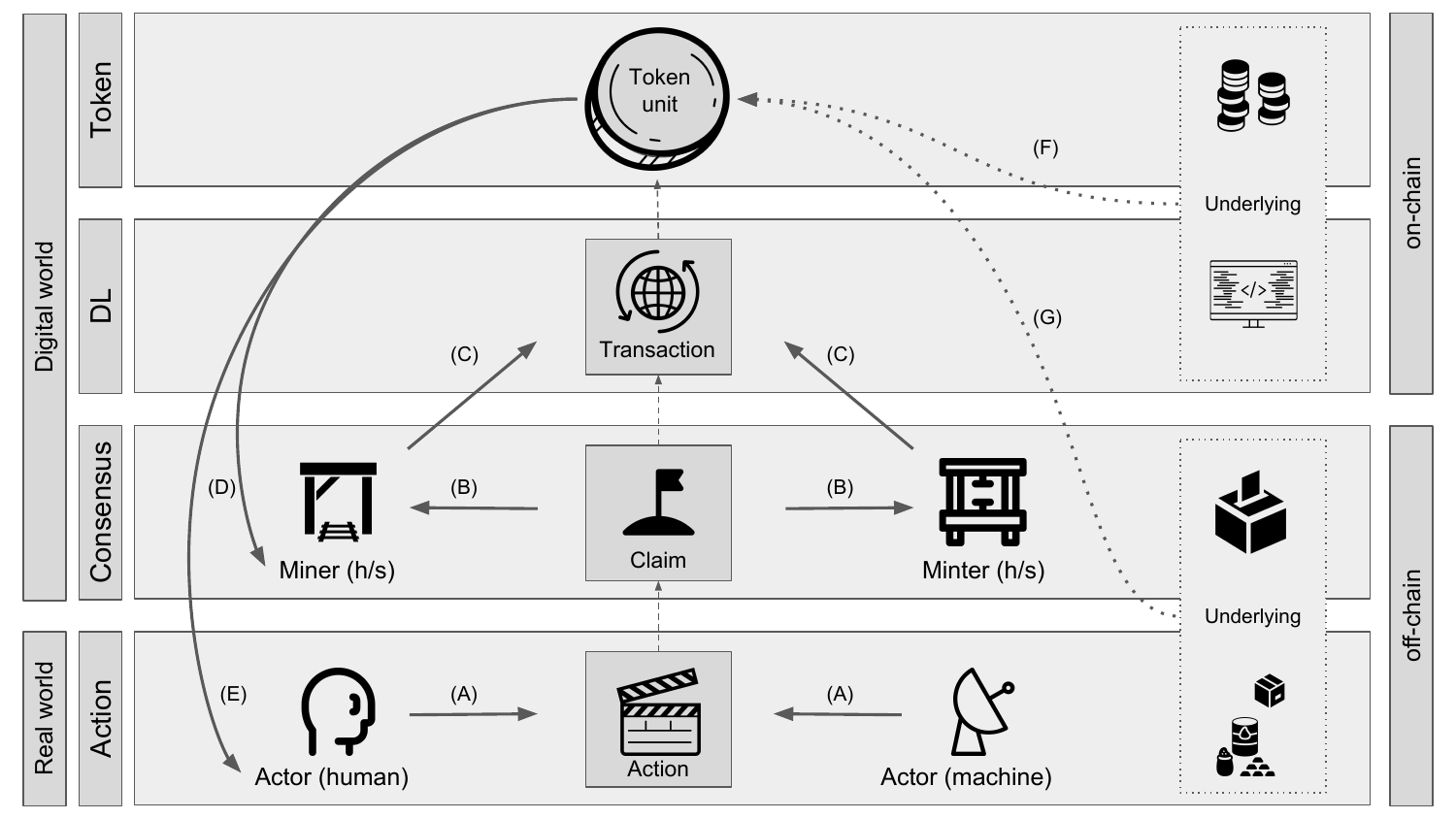}	        
\end{figure*}

The contributions of this paper are outlined as follows:
\begin{enumerate}
	\item \textit{A conceptual architecture} that models DLT systems with four components. The architecture (Figure \ref{fig:concept}) defines minimal and insightful design elements to illustrate the inner mechanics of distributed ledgers and the interrelationships of their components. 
	\item \textit{A taxonomy} of distributed ledgers that formalizes a set of
	19 descriptive and qualitative attributes, including a set of possible
	values for each attribute. They illustrate the four DLT components
	in more detail (Figure \ref{fig:tax_ontology}) and provide deeper
	insight into cryptoeconomic concepts such as utility token, public blockchain,
	etc. 
	\item \textit{A classification} of 50 DLT systems, including Bitcoin and
	Ethereum, backed by an extensive literature review. 
	\item \textit{A taxonomy evaluation criterion} referred to as `expressiveness'
	derived from earlier theory on taxonomies. 
	\item \textit{Crowdsourced feedback}  from the blockchain
	community to further assess and improve the taxonomy and classification.
	\item \textit{Design guideline} for DLT systems (Figure \ref{fig:design_guideline}), which is constructed by reasoning based on empirical data of viable, actively maintained and academically referenced DLT systems. This guideline structures the modeling complexity of DLT systems by grouping similar attribute values of the taxonomy into a characteristic design choice.
\end{enumerate}

This paper is organized as follows: In Section \ref{sec:backgorund},
terminology and recent taxonomies for DLT systems are discussed. A
conceptual architecture for DLT systems is introduced in Section \ref{sec:conceptmodel},
while a taxonomy is outlined in Section \ref{sec:taxonomy}. Thereafter,
Section \ref{sec:experimental_methodology} illustrates the methodology of the conducted experiments and Section \ref{sec:experimental_evaluation} presents the evaluation.
Section \ref{sec:summary_findings_design_guideline} summarizes the findings and derives based on these a design guideline for DLT systems.
Finally, in Section \ref{sec:conclusion}
a conclusion is drawn and an outlook on future work is given.

\section{Background and Literature Review}
\label{sec:backgorund}


\begin{table*}[!t]
	\caption{Comparative overview of earlier work outlining the landscape of distributed ledgers.}
	\resizebox{\textwidth}{!}{
		\label{tab:lit_review}
		\begin{tabular}{llcccccccccc}
			\toprule
			1. &     2.    & \textbf{3.} & \textbf{4.} & \textbf{5.} & \textbf{6.} & \textbf{7.}& \textbf{8.} & \textbf{9.} & \textbf{10.} & \textbf{11.} & \textbf{12.}  \\
			ID &     Paper    & \textbf{Concept} & \textbf{Attributes} & \textbf{\begin{tabular}[c]{@{}c@{}}Consensus \\ Incentivization \end{tabular}} & \textbf{\begin{tabular}[c]{@{}c@{}}Diff.\\ DL\end{tabular}} & \textbf{CED} & \textbf{\begin{tabular}[c]{@{}c@{}}Access rights\\to transactions\end{tabular}} & \textbf{\begin{tabular}[c]{@{}c@{}}Token \\ properties\end{tabular}} & \textbf{Classification} & \textbf{\begin{tabular}[c]{@{}c@{}}Community \\ Evaluation \end{tabular}} & \textbf{\begin{tabular}[c]{@{}c@{}}Quantitative \\ Analysis \end{tabular}} \\ \midrule
			1    & Tasca et. al (2017) \cite{tasca2017ontology}                 & -       & 30         & yes                                                                     & -                                                  & yes & yes                                                                     & yes                                                         & -              & -   & -       \\
			2    & Comuzzi et. al (2018) \cite{comuzzi2018towards}                                 & -       & 8          & yes                                                                     & -                                                  & yes & yes                                                                     & -                                                           & -              & -     &-     \\
			3    & Xu et. al (2017) \cite{xu2017taxonomy}                                        & -       & 13         & -                                                                       & yes                                                & yes & yes                                                                     & -                                                           & -              & -         & - \\
			4    & Xu et. al (2016) \cite{xu2016blockchain}                                                             & yes     & 7          & -                                                                       & -                                                  & yes   & yes                                                                     & -                                                           & -              & -  &-        \\
			5    & Yeow et. al (2018) \cite{yeow2018decentralized} & -       & 4          & -                                                                       & yes                                                & -   & -                                                                       & -                                                           & yes            & -   &-       \\
			6    & Okada et. al (2017) \cite{okada2017proposed}                & -       & 4          & yes                                                                     & -                                                  & yes & -                                                                       & -                                                           & -              & - &-         \\
			
			7    & Wieninger et. al (2019) \cite{wieninger2019development}                                                                     & -       & 11   & yes                                                                       & -                                                  & yes   & yes                                                                       & yes                                                           & -              & -&-    \\
			8    & Dinh et. al (2018) \cite{dinh2018untangling}                                                                     & yes       & 9   & -                                                                       & -                                                  & yes   & yes                                                                       & -                                                           & yes (partial)              & -&-    \\
			9    & De Kruijff et. al (2017) \cite{de2017towards}                                                                     & -       & 6 (many)   & -                                                                       & -                                                  & -   & -                                                                       & -                                                           & -              & -&-    \\
			10    & Sarkintudu et. al (2018) \cite{sarkintudu2018taxonomy}                                                                     & -      & 5   & -                                                                       & -                                                  & -   & -                                                                       & -                                                           & -              & -&-    \\
			\midrule
			& This paper & yes       & 19   & yes                                                                       & yes                                                  & yes   & yes                                                                       & yes                                                           & yes              & yes & yes   \\        
			
			\bottomrule     
		\end{tabular}
	}
\end{table*}

DLT systems use different \textit{types of distributed ledgers} as
data structures. In particular, the literature distinguishes between
distributed ledgers (DL) and blockchains \cite{xu2017taxonomy,samuel2016layered},
the latter representing one way to implement the former. Another type
of distributed ledger is the directed acyclic graph \cite{yeow2018decentralized}.

The entries of a distributed ledger contain \textit{transactions}.
Any type of transaction can be stored, ranging from cryptographically
signed financial transactions, to hashes of digital assets, and Turing-complete
executable programs \cite{xu2017taxonomy}, i.e. smart contracts. DLT systems often provide \textit{access rights to these transactions}, which determine
who can initiate transactions, write them to the distributed ledger,
and read them again from the ledger \cite{xu2017taxonomy}. In addition,
DLT systems utilize so-called \textit{tokens} \cite{xu2016blockchain}, which are defined as a unit of value issued within a DLT system and which can be used as a medium of exchange or unit of account (see Section \ref{sec:tax_token}). These tokens
span a multi-dimensional incentive system via which they can promote self-organization \cite{kleineberg2016social} and thus lead to benefits in society \cite{kewell2017blockchain}, such as contributing solutions for the UN Sustainable Development Goals (SDGs) \cite{dierksmeier2018cryptocurrencies}. 
Hence tokens are identified as another key component of DLT systems in addition
to the distributed ledger \cite{morisse2015cryptocurrencies}.
These components can be modeled independently, resulting in systems
that do not necessarily maintain a native distributed ledger. In
such cases, a token is defined while another system is used to provide
the infrastructure for a distributed ledger. For instance, the Aragon
system does not maintain a natively developed distributed ledger \cite{dhillon_decentralized_2017}.

The ability to define the type of transactions, access rights and
tokens is used to regulate the behavior of users, i.e. by limiting
and granting access rights to system services or by incentivizing
specific actions with tokens. These socio-economic choices not only
influence aspects of the system stability, such as the correctness,
liveness and fairness of the consensus mechanism \cite{bonneau2015sok},
but also determine how complex cryptoeconomies emerge \cite{babbitt2014crypto,davidson2016economics}.
In other words, cryptoeconomic design (CED) plays a key role in enabling
DLT systems to reach stability and underpin how the economies form.

A DLT system has to reach \textit{consensus} before a transaction
can be permanently written to its ledger \cite{xu2016blockchain}.
This consensus mechanism is a functional element of any DLT system
\cite{samuel2016layered}, as it enables a decentralized network to take decisions about the validity of entries in the distributed
ledger \cite{sankar2017survey}. In particular, in the context of
DLT systems, consensus prevents token units from being spent twice
\cite{mingxiao2017review} and Sybill attacks \cite{tschorsch2016bitcoin},
which is where fake identities are used to inject false information
into the distributed ledger.


Recent ontologies and taxonomies have been proposed to structure the design space of DLT systems. A comparative summary of earlier work is shown in
Table \ref{tab:lit_review}. Column 3 of that table depicts if the paper utilizes a conceptual architecture to construct the taxonomy. Nickerson et al. \cite{nickerson2013method} suggest to conceptualize the domain of interest for which a taxonomy is developed. In such a conceptual architecture, the attributes of a taxonomy should be positioned such that these are mutually exclusive and collectively exhaustive \cite{nickerson2013method}.
Nevertheless, only Paper 4 and 8 in Table~\ref{tab:lit_review} provide a conceptual
architecture (Column 3 in Table~\ref{tab:lit_review}) that determines
the choice of some of the attributes. For instance, Paper 4 distinguishes
between on-chain and off-chain components~\cite{xu2016blockchain}:
attributes of the DLT system, which exist on the distributed ledger
(e.g. permission management) vs. attributes, which exist outside (e.g.
control, data). 

Taxonomy theory identifies, that a useful taxonomy should be concise and robust \cite{nickerson2013method}, hence using a limited number of attributes, which differentiates the objects of interest.
The number of attributes listed in the papers varies considerably,
from $4$ to $30$ (Column 4 in Table~\ref{tab:lit_review}). One
explanation is that the papers focus on different aspects of DLT systems
and thus study different (sub)sets of attributes. For instance, Yeow
et. al \cite{yeow2018decentralized} (Paper 5 in Table~\ref{tab:lit_review})
focus on Internet of Things applications of DLT systems and only use
four attributes, whereas Tasca et. al (Paper 1 in Table~\ref{tab:lit_review})
design a taxonomy to model all types of DLT systems and hence use
30 attributes \cite{tasca2017ontology}. Nevertheless, none of the papers justifies the number of selected attributes. In particular, their impact on conciseness and robustness of the taxonomy is not evaluated. Also, several of the attributes
potentially overlap with each other conceptually due to the aforementioned
lack of a conceptual architecture.

Consensus is identified as a core feature of DLT systems~\cite{sankar2017survey}
and as such, it is incorporated in all papers listed in Table~\ref{tab:lit_review}.
For this reason, it is omitted from this table. Nevertheless, just
four papers consider schemes to incentivize participation in the
consensus mechanism (Column 5 in Table \ref{tab:lit_review}).

Moreover, only Paper 3 and 5 distinguish between different types of
distributed ledgers (Column 6 in Table \ref{tab:lit_review}). For
instance, Xu. et al. differentiate between blockchains and directed
acyclic graphs \cite{xu2017taxonomy}. Nevertheless, some of the most
recent contributions solely include blockchain-based DLT systems \cite{xu2017taxonomy,tasca2017ontology,wieninger2019development,comuzzi2018towards}.

Seven papers include cryptoeconomic design in their taxonomy (Column
7 in Table \ref{tab:lit_review}). In particular, six papers consider
access rights to transactions (Column 8 in Table \ref{tab:lit_review}). 
Only Paper 1 and 7 derive a taxonomy, which includes tokens and their properties
(Column 9 in Table \ref{tab:lit_review}).

Paper 5 and 8 illustrate a classification of DLT systems based on their proposed taxonomy (Column 10 in Table
\ref{tab:lit_review}). For instance, 
Paper 5 illustrates the classification
of $28$ DLT systems. The authors rely on three attributes: data
structure, scalable consensus ledger, and transaction model \cite{yeow2018decentralized}.
However, neither of the two papers introduces a formal methodology to select the classified DLT systems, which lowers their objectivity. Also, without a formal selection methodology it is not guaranteed that the taxonomy enables a comprehensive classification of all known DLT systems, which is a quality criterion of taxonomies \cite{nickerson2013method}.

The \textit{usefulness} of a taxonomy depends on
qualitative criteria studied in taxonomy theory~\cite{nickerson2013method}. An approach to assess the usefulness of a taxonomy is to utilize crowdsourced community feedback and thus the wisdom of the crowd. This
is particularly relevant in the case of DLT systems and the blockchain
community. As the community shapes the blockchain landscape, soliciting
their feedback can provide both, invaluable new insight into the design
of DLT systems and increase the usefulness of a taxonomy. Nevertheless, such an endeavor has not been pursued until nowadays,
as shown in Column 11 of Table \ref{tab:lit_review}.

Finally, a quantitative evaluation and analysis of taxonomy and classification elements by means of statistical or machine learning methods have not been performed so far (Column 12 in Table \ref{tab:lit_review}). This is a missed opportunity, as such an approach can provide more objective insights into the usefulness of taxonomies and identify key design choices in DLT systems that structure the modeling complexity of these systems at design phase, as demonstrated in this paper (Section \ref{sec:machine_learning_analysis}).


The most comprehensive taxonomy, as defined by Nickerson et. al \cite{nickerson2013method} and introduced by Tasca et al. \cite{tasca2017ontology},
is worth a brief discussion. This taxonomy for DLT systems consists
of eight components encompassing a total of $30$ attributes. It includes
a CED with three components: native currencies/tokenization, identity
management, and charging, as well as a reward system. In particular,
the access rights to transactions and the properties of tokens are
discussed. Access rights to participate in the consensus mechanism
and the incentivization of this mechanism are also illustrated. Despite
extensively covering many relevant concepts, this taxonomy lacks a
conceptual architecture to connect the various elements, i.e. no information
is given about how the eight components relate to each other. Moreover,
it remains unclear what explicit criteria the authors use to decide upon these
eight components and $30$ attributes. In particular, as the amount of attributes is four times larger than the average in the other papers (on average $7.4$ attributes, Column 3 in Table \ref{tab:lit_review}) it is unclear if the proposed taxonomy is concise as defined in the context of taxonomy theory \cite{nickerson2013method}. 
In addition, no distinction is explicitly made between CED and DL, or between on-chain and off-chain
aspects. In particular, this taxonomy does not differentiate between
different types of distributed ledgers. This limits its ability to
provide a more granular differentiation of distributed ledgers, which
is a quality indicator for taxonomies \cite{nickerson2013method}.


In summary, a few observations can be made about current DLT system
taxonomies. First, they predominantly focus on the DL and consensus
mechanisms, while largely missing the role of cryptoeconomics and token design,
despite their significant influence on system stability \cite{bonneau2015sok}.
Second, the interrelationships between the different components
as well as the choice of attributes are usually not based on an overarching
conceptual architecture. Third, only two of the papers classify
real-world DLT systems. Nevertheless, these papers neither utilize a rigorous scientific methodology nor quantitatively analyze their classification. As a result, classification is usually not formally validated and the identification of design choices is limited to qualitative criteria. Last but not least,
none of the proposed taxonomies are systematically refined based on
feedback from blockchain practitioners. This complementary external
validation process promises to produce more unbiased taxonomies.

This paper addresses all of the aforementioned limitations identified
in the literature and contributes a useful taxonomy as defined in earlier taxonomy theory~\cite{nickerson2013method}, built
on a solid conceptual architecture, assessed via classifications and
validated by both, feedback from the blockchain community and machine learning methods. 
Moreover, the quantitative analysis of the classification is utilized to identify key design choices in observed DLT systems.

\section{Conceptual Architecture}
\label{sec:conceptmodel}

By studying 50 DLT systems (see Table 1 in Supplementary Material for an overview of these systems), a conceptual architecture
is introduced in this section. The architecture is composed of a set of four
key components and shows, how they relate to each other as well as how they are positioned
in the distributed ledger design space. The architecture is depicted
in Figure \ref{fig:concept}. The four components are illustrated in
the rest of this section.

\textit{Action component}. A human or machine performs an action in
the real world (Arrow A in Figure \ref{fig:concept}), for example
planting a tree or carrying out a monetary transaction. Here, at the
border between the real world and digital world, the action is represented
digitally, and is referred to as claim.

\textit{Consensus component}. Claims are broadcast to all nodes in
the network that can participate in the consensus mechanism (Arrow
B). These nodes (referred to as miners in Bitcoin or minters in
Peercoin) collect these claims to write them to the distributed
ledger.

\textit{Distributed ledger component}. 
Participants in the consensus mechanism combine these
claims into entries (referred to as blocks in Bitcoin) and write them
to the distributed ledger (Arrow C). This representation of the
claim on the distributed ledger is called a transaction. Transactions
and their containing objects (e.g. smart contracts) that exist on the distributed ledger are referred to
as on-chain, in contrast to off-chain objects, which exist on the
Consensus or Action component.

\textit{Token component}. The way token units are created depends on whether
an incentive system is part of the DLT system. If it is, there are
two options: token units are given as rewards to nodes for either
participating in the consensus mechanism (Arrow D) or carrying out
an action (Arrow E). While the inherent properties of such tokens
(e.g. whether supply is capped or not) are determined by the design
of the DLT system, the value of the token units is backed by underlyings,
which are cryptoeconomic assets that reside on-chain (Arrow F,
for example other tokens or executable code) or off-chain (Arrow G,
for example goods, services or commodities). In particular, it has been noted, that the value of cryptoeconomic tokens is important for the ecosystem to be robust \cite{kleineberg2016social}.

\textit{Example Ethereum}. In the case of Ethereum, one type of action involves deploying a piece of code (Arrow
A in Figure \ref{fig:concept}), such as a smart contract. These
actions are collected by miners (Arrow B) and written as a block
to the Ethereum distributed ledger (Arrow C). A miner who successfully
writes a block obtains Ether, which refers to newly created units
of a token that serves as an incentive to mine (Arrow D). The Ether
token has inherent properties, e.g. it has uncapped supply. It also
has value because it enables its owner to access the on-chain computational
power of the Ethereum network (Arrow F).


\section{Taxonomy}
\label{sec:taxonomy}

\begin{figure}[!thb]
	\centering
	\includegraphics[width=1.0\columnwidth]{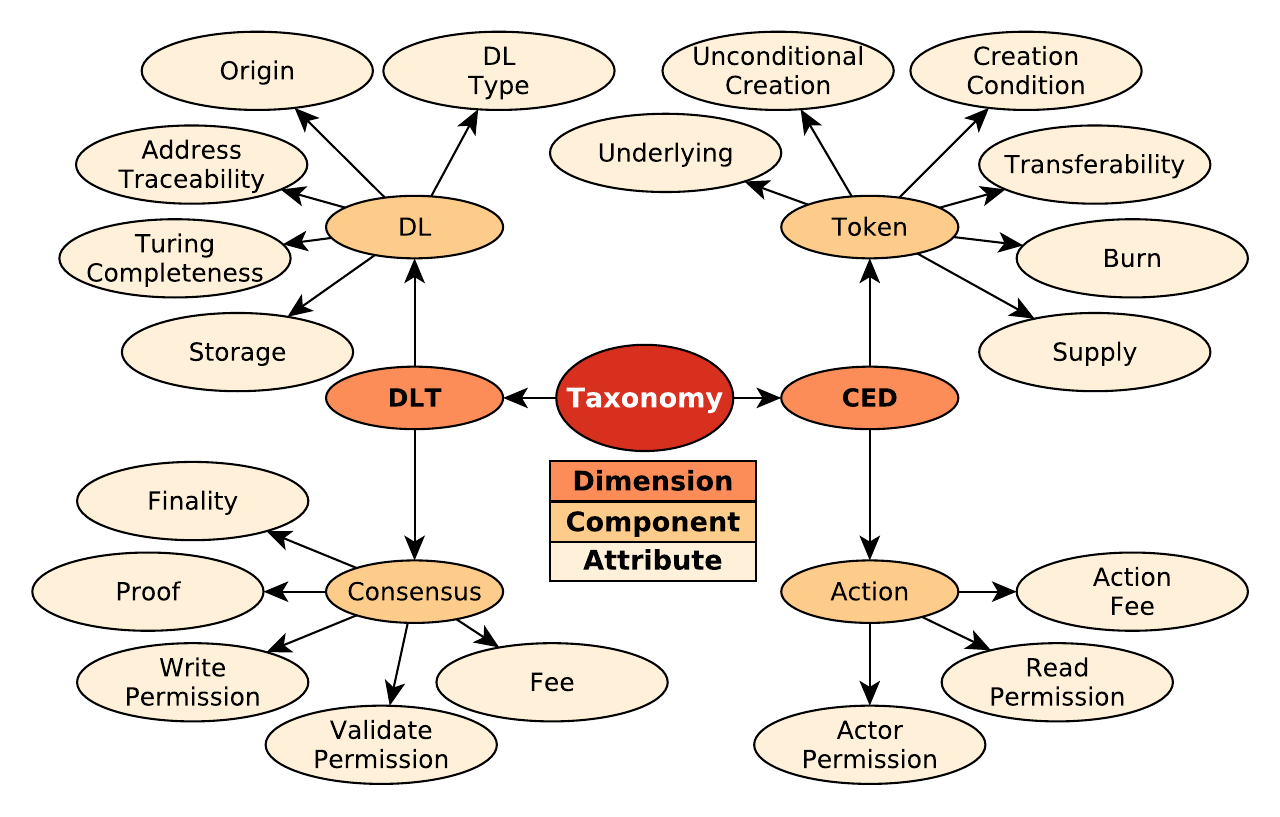}
	\caption{Overview of the taxonomy, depicting the two dimensions of DLT and CED, its four components and 19 attributes.}\label{fig:tax_ontology}
\end{figure}

Based on the conceptual architecture of Section~\ref{sec:conceptmodel},
a taxonomy is designed, using the method proposed by Nickerson et
al. \cite{nickerson2013method}. The goal of the taxonomy is to enable a comprehensive classification of DLT systems that enable the quantitative derivation of key design choices in these systems. For this, the taxonomy illustrates both, the distributed ledger technology (DLT) and the cryptoeconomic design (CED) of academically relevant DLT systems.  
For this, the taxonomy positions the four components from Section \ref{sec:conceptmodel}
across two dimensions (Figure \ref{fig:tax_ontology}).
The first dimension concerns aspects of the system design related
to distributed ledger technology (DLT) -- \textit{Distributed Ledger
	component}, \textit{Consensus component} --, while the second dimension
concerns aspects pertaining to cryptoeconomic design (CED) -- \textit{Action
	component} and \textit{Token component}.


\subsection{Distributed Ledger}

\begin{mydef}
	\textit{A distributed ledger is defined as a distributed data structure, containing entries that serve as digital records of actions.}
\end{mydef}
In the Bitcoin system, an entry in the data structure is called a
block. In the IOTA system, it is called a bundle. An entry contains
a set of transactions (Figure \ref{fig:concept}, DL
component). In Bitcoin, these transactions represent the exchange
of cryptocurrency value.
The attributes of the distributed ledger are \textit{type}, \textit{origin},
\textit{address traceability} and \textit{Turing completeness}.

\subsubsection{Type}
denotes the data structure of the distributed ledger and can be one of the following: \textit{blockchain}, \textit{directed acyclic
	graph} (DAG) or \textit{other}.

The most well-known type is the blockchain; an immutable and append-only
linked list, which has a total order of elements. Several systems
use blockchains, such as Bitcoin \cite{xu2017taxonomy}, Ethereum
\cite{de2017survey} and Litecoin \cite{kalinin_blockchain_2018}.
In contrast to these systems, IOTA uses a directed acyclic graph \cite{yeow2018decentralized}.
This data structure is no longer a linked list, but a directed graph
with no cycles, leading to a partial order of elements. Moreover,
Ripple neither uses a blockchain nor a directed acyclic graph but
instead operates on other consensus-based accounting mechanism \cite{kakushadze2018blockchain}.

\subsubsection{Origin} 
refers to who maintains the distributed ledger. The attribute value can
either be \textit{native}, if the distributed ledger is maintained
by and for the system itself or \textit{external}, if the system uses
a distributed ledger from another DLT system or \textit{hybrid} if the systems maintain their own distributed ledger in combination with a distributed ledger of another DLT system.

The level of maintenance varies between different DLT systems. Bitcoin
develops and maintains its distributed ledger natively, as does NXT
\cite{xu2016blockchain}. In contrast, Aragon \cite{dhillon_decentralized_2017},
Augur \cite{peterson2018augur} \cite{kakushadze2018blockchain} and
Counterparty \cite{yeow2018decentralized} does not maintain a native distributed ledger, opting to use the Ethereum or Bitcoin infrastructure instead. Systems can use a hybrid approach. Factom combines a natively
developed blockchain and its own consensus mechanism with the Bitcoin blockchain \cite{xu2017taxonomy}.

\subsubsection{Address traceability} 
denotes the extent to which different transactions, which originate
from or arrive at the same chain identity, can be linked together.
The value can either be \textit{obfuscatable}, if the distributed
ledger has mechanisms in place to hide such links or \textit{linkable,}
if these links can be inferred with some computational effort.

The level of address traceability varies between the different DLT
systems. Zcash \cite{xu2017taxonomy} and Monero \cite{tasca2017ontology}
are so-called privacy coins, which perform advanced measures to unlink
transactions \cite{bonneau2015sok}. Hence, the on-chain identities
of the actors remain obfuscated. Bitcoin has linkable address traceability
\cite{bonneau2015sok}. In theory, transactions cannot be linked to
a particular chain identity, but it has been shown that this can actually
be achieved with some computational effort \cite{xu2017taxonomy}.
The same applies to Ripple \cite{moreno2017pathshuffle}.

\subsubsection{Turing completeness} 
refers to whether a Turing machine can be simulated by the DL and can  either be \textit{Yes} or \textit{No}.

Some DLs, such as Ethereum, can execute Turing machines. This allows
Turing complete smart contracts to be stored and executed \cite{xu2016blockchain},
in contrast to the Bitcoin blockchain \cite{bonneau2015sok}. 

\subsubsection{Storage} 
denotes whether additional data can be stored on the distributed ledger
beyond the default transaction information.
The attribute value can either be \textit{yes} if data can be stored
or \textit{no}, if additional data cannot be stored.

The distributed ledger of Bitcoin allows arbitrary data to be stored
inside transactions. This allows Bitcoin to be used as a base layer for other DLT systems, such as observed in the Counterparty
system \cite{yeow2018decentralized}. In contrast to Bitcoin, IOTA
does not allow additional data to be stored \cite{iota_transaction}.

\subsection{Consensus}
\label{sec:consensus}
\begin{mydef} \textit{Consensus is the mechanism through which entries are
		written to the distributed ledger, while adhering to a set of rules
		that all participants enforce when an entry containing transactions
		is validated.} \end{mydef}
The attributes of consensus are \textit{finality}, \textit{proof}, \textit{write permission}, \textit{validation permission}
and \textit{fee}. Due to the scope of the taxonomy to enable a comprehensive classification of all components of a DLT system (Figure \ref{fig:concept}), more granular consensus attributes such as verification speed are not considered. Nevertheless, detailed consensus attributes can be found in \cite{mingxiao2017review, cachin2017blockchain}.

\subsubsection{Finality}
refers to the guarantee that past transactions can not be changed or reversed. 
Its value is \textit{deterministic} if consensus is guaranteed to be reached in finite time, or \textit{probabilistic} if there is some
uncertainty over whether consensus can be reached.


Most DLT systems use the Nakamoto consensus \cite{xu2017taxonomy},
which is a Byzantine Fault Tolerance (BFT) algorithm. These types
of algorithms tolerate a class of system failures that belong to the
Byzantine Generals Problem \cite{Lamport1982}. In particular, a consensus
algorithm that has this property prevents under some guarantees\footnote{e.g. that no participant/ cartel of participants controls more than $50 \%$ of the computational capacity of the consensus network or that participants behave rationally} \cite{bonneau2015sok} consensus participants
from writing a false transaction to the distributed ledger.

In contrast to other BFT algorithms, the Nakamoto consensus is probabilistic.
This type of algorithm validates each new entry using the entire history of previous entries:
An entry is accepted if there is a certain number of new
entries referencing it \cite{bonneau2015sok}. For instance, in the case of Bitcoin, a writer validates a transaction
by considering the whole blockchain and then including the transaction
in a new block. As soon as this block is referenced by six other blocks,
it is confirmed, as the probability that a second chain of six blocks
referencing each other, but not referencing this block, is low \cite{xu2017taxonomy}.
Similarly, the directed acyclic graph of IOTA confirms an entry when
it is referenced by a significant number of new entries \cite{yeow2018decentralized}.
On the other hand, Ripple does not use a Nakamoto consensus algorithm
and it is guaranteed that consensus can be reached in a finite period
of time \cite{sankar2017survey}.

\subsubsection{Proof} 

is the evidence used to achieve consensus. The value can
either be \textit{proof-of-work} (PoW), if consensus is achieved using
the processing power of computers; \textit{proof-of-stake} (PoS),
if it is achieved through voting processes linked to (economic) power
in the system; \textit{hybrid}, if it is a combination of the previous
two or \textit{other}, if another form of proof is required.

Participants in the consensus mechanism require proof before accepting the validity of an entry. Bitcoin uses a proof-of-work \cite{xu2016blockchain}, which is the solution to a mathematical puzzle that requires computational
processing power. A proof-of-stake is used by Ardor \cite{tasca2017ontology},
which is the approval of a randomly selected consensus participant
who must hold a stake in Ardor token units. 
\subsubsection{Write permission}

denotes who is allowed to write entries to the distributed ledger.
The value can either be \textit{restricted}, if participation
is restricted or \textit{public}, if it is not.

The Bitcoin consensus mechanism is public \cite{xu2016blockchain},
meaning that it allows everyone who has computing power to participate
\cite{sankar2017survey}. Conversely, the consensus mechanism of Ripple
is restricted \cite{xu2016blockchain}, meaning that only a few trusted
institutions can participate \cite{yeow2018decentralized}.

\subsubsection{Validate permission} 

signifies who is allowed to validate claims before they are written
to the distributed ledger.
The value can either be \textit{restricted}, if participation
is restricted or \textit{public,} if it is not.

In the case of Bitcoin, writers validate the correctness of claims before writing them to a block: hence, the validation
permission is public. In contrast, in the case of IOTA, a central
entity, the coordinator, validates transactions before they are collected
in an entry and written to the directed acyclic graph \cite{yeow2018decentralized}.

\subsubsection{Fee} 

denotes whether participants in the consensus (writers and validators)
are paid a fee for validating new entries and writing them to the distributed ledger.
The value can either be \textit{yes} or \textit{no}.

In contrast to Bitcoin, where writers/validators are rewarded with
fees \cite{sankar2017survey}, IOTA writers and validators receive
no fees \cite{yeow2018decentralized}. In the case of Ripple, consensus
participants are not rewarded with fees, although actors need to
pay a fee \cite{ripple_transaction}. This system layout is captured by the fee attribute in the Action component (Section \ref{sec:action}).

\subsection{Action}
\label{sec:action}

\begin{mydef}
	\textit{An action is one or more real-life activities, which can be digitally represented by a DLT system as a transaction.}  
\end{mydef}
In this sense, a transaction represents digitally a real-life action.
The attributes of action are \textit{actor permission}, \textit{read permission} and \textit{fee}.

\subsubsection{Actor permission}

denotes who can perform an action.
The value can either be \textit{restricted} if actors have
to fulfill special requirements before performing actions or \textit{public},
if anyone can perform actions.

Bitcoin allows everyone to create a private key to send and receive
token units \cite{tasca2017ontology}: hence, it has a public actor
permission. Ripple uses restricted access rights. In order to comply
with regulations (e.g. know-your-customer), actors need to register
\cite{tasca2017ontology}.

\subsubsection{Read permission} 

refers to which actors can read the contents of transactions from
the distributed ledger.
The value can either be \textit{restricted}, if preconditions
need to be fulfilled before permission is granted, 
or \textit{public}, if permission is not restricted.

Most DLT systems have public read access in the sense that everyone
can read the content of the actions, which have occurred, e.g. the
number of bitcoins transferred \cite{tasca2017ontology}. Systems utilizing privacy
coins often restrict read access to the actors involved in a transaction
(e.g. Zcash \cite{xu2017taxonomy}), usually by making an effort to
hide the number of token units transferred \cite{bonneau2015sok}.

\subsubsection{Fee} 

denotes whether an actor has to pay a fee for performing an action
that is unrelated to the consensus. The values are \textit{yes}
or \textit{no}.

Some DLT systems require actors to pay a fee, which is unrelated to
the consensus before they can store an action on the distributed
ledger. For instance, actors have to pay a fee in Augur, which is
not distributed to consensus participants \cite{peterson2018augur} but given to actors providing services in the system.
In the case of Bitcoin, no additional fee is required to perform an
action, except the fee paid to the consensus participants. Ripple
also requires actors to pay a fee for each action, which is not paid
to consensus participants but is subsequently destroyed \cite{ripple_transaction}.

\subsection{Token}
\label{sec:tax_token}
\begin{mydef}
	\textit{Token is a unit of value issued within a DLT system and which can be used as a medium of exchange or unit of account.}
\end{mydef}
The associated attributes are \textit{supply property}, \textit{burn
	property}, \textit{creation condition}, \textit{unconditional creation} and \textit{underlying}.

\subsubsection{Supply property} 

refers to the total quantity of token units made available. The value
can either be \textit{capped}, if the total supply is limited to a
finite number or \textit{uncapped} otherwise.

If demand increases for a token, a capped supply can cause the
perceived token value to appreciate and corresponds to a deflation in prices nominated in this token. Moreover,
it can result in an appreciated exchange rate with other tokens,
which in turn, increases the stability of a DLT system \cite{bonneau2015sok}.
Bitcoin has a capped supply of $21$ million units \cite{tasca2017ontology},
whereas Dodgecoin does not have an upper limit \cite{bonneau2015sok}.

\subsubsection{Burn property} 

denotes whether token supply is reduced by removing token units. The values are \textit{yes} or \textit{no}.

Some DLT systems destroy token units in a process referred to as `burn'.
If demand remains constant, this decrease in the money supply causes
token units to appreciate and hence, results in a better exchange rate with other
tokens. For example in the case of Ripple, paid fees are removed
from the total supply and are not returned \cite{ripple_transaction}.
In contrast, Bitcoin has no inherent mechanism to destroy token units.

\subsubsection{Transferability} 

refers to whether the ownership of a token unit can be changed. The
value can either be \textit{transferable}, if the token can
be transferred, or \textit{non-transferable} otherwise.

Bitcoin token units can be transferred between different actors. Akasha
plans to use non-transferable reputation tokens, so-called Mana and
Essence \cite{thoughts_new_nodate}.

\subsubsection{Creation condition} 

denotes whether the creation of new token units is linked to the incentivization
of the consensus mechanism and/ or an action. The value can
either be \textit{consensus}, if creation is linked to the consensus
mechanism, \textit{action}, if creation is linked to an action, \textit{both},
if creation is linked to the consensus mechanism as well as an action,
or \textit{none} otherwise.

In the case of Bitcoin, new tokens are created to incentivize the
consensus mechanism \cite{xu2017taxonomy}. Other systems create new
tokens to incentivize an action. For instance, Steemit creates new
steem to incentivize content creation on the platform (e.g. writing
blog articles) \cite{steemit_whitepaper}. Moreover, Ripple does not
use its token to incentivize the consensus mechanism or an action
\cite{ripple_supply}. Furthermore, hybrid versions are possible,
where new tokens are created to incentivize both the consensus mechanism
and an action. For instance, newly created token units in the DASH
system are awarded to both the consensus participants and the master
nodes, who perform actions such as mixing transactions to enable obfuscatable
address traceability \cite{dash_2018}.

\subsubsection{Unconditional creation}

refers to the number of new token units that can be created, which
do not serve to incentivize the consensus mechanism or an action.
The value can either be \textit{partial}, if some tokens
are created unconditionally, \textit{all}, if all tokens are created
unconditionally (e.g. 100 \% pre-mined tokens), or \textit{none} otherwise.

At the genesis of the Bitcoin system, no token units had previously
been mined and all tokens come into existence by incentivizing the
consensus \cite{bonneau2015sok}. On the
other hand, all Ripple tokens were created during the genesis of the
system.  
In the case of Augur, some tokens were created during the genesis
of the system~\cite{peterson2018augur}.

\subsubsection{Underlying}

denotes the source of a token value and what it consists of.

The value can either be \textit{token}, if the token grants access to another token; \textit{distributed ledger} if the
token grants access to the distributed ledger, e.g. if the token is
needed in order to use the storage or computing capacity of the distributed
ledger; \textit{consensus}, if the token grants access to the consensus
mechanism, e.g. in a proof-of-stake type system; \textit{action}, if the token grants
access to perform or receive actions, goods or services in the real world;
or \textit{none}, if the token has no underlying.

The first two values (distributed ledger and token) are considered
to be on-chain and the latter two are considered to be off-chain
underlyings of a token unit (as depicted in Figure \ref{fig:concept}).

The Ethereum token allows everyone to store data or smart contracts
on-chain \cite{xu2017taxonomy} and to access in this way the distributed
ledger of the network. Hence, the source of value of Ether token units is that
they grant access to the processing power of the distributed ledger.
In contrast to Ether, the Golem network token units allow holders
to access off-chain computations \cite{kakushadze2018blockchain}.
Thus, its underlying is action as the token provides access to a service in the real world (Action component). The Storj Token allows users to access
off-chain storage \cite{tasca2017ontology}, which again resides in the Action component. Siacoin allows for the storage of arbitrary data on
both its distributed ledger \cite{siahub_nodate} and its off-chain
network \cite{wu_democratic_2017}. Hence its underlyings reside in the DL and Action components.

\section{Experimental Methodology}
\label{sec:experimental_methodology}

\begin{figure}[!t]
	\centering
	\includegraphics[width=1.0\columnwidth]{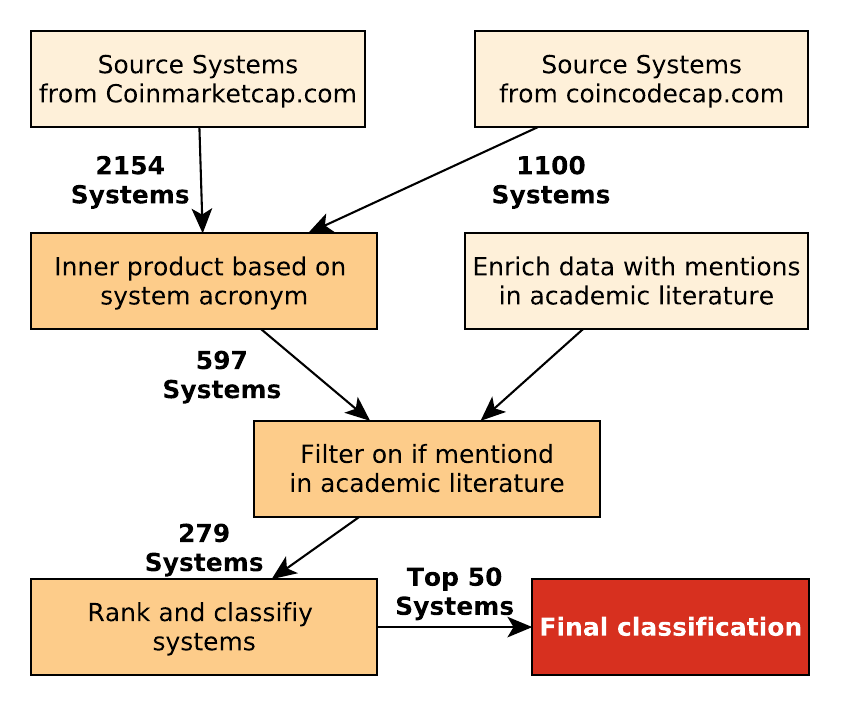}
	\caption{Identification and selection process of top $50$ systems for classification ranked according to Section \ref{sec:meth_classification}. The final classification is provided in the Supplementary Materials. 
	}
	\label{fig:refinement}
\end{figure}

Based on the introduced taxonomy, $50$ DLT systems are classified. The taxonomy and classification are evaluated by (i) the blockchain community via a survey and (ii) a quantitative analysis of real-world data. Furthermore, the quantitative analysis of the classification by the means of machine learning methods identifies key design choices in the observed DLT systems that structure modeling complexity at design phase. In the following, the methodologies of the classification (Section \ref{sec:meth_classification}), the blockchain community feedback (Section \ref{meth:survey}) and the machine learning analysis (Section \ref{meth:data_analysis}) are illustrated.
\subsection{Classification}
\label{sec:meth_classification}

The scope of the classification is to comprehensively capture the CED and DLT of viable, academically referenced and actively maintained DLT systems. Moreover, the classification aims at capturing the current state of a DLT system. In particular, features that are about to be released in the future are not considered. Finally, in the case that a system is 1\textsuperscript{st} layer (utilizing a native distributed ledger, e.g. a mainchain) and 2\textsuperscript{nd} layer (utilizing an external distributed ledger, e.g. sidechains), only the 1\textsuperscript{st} layer is classified. Likewise, if a system utilizes more than one token, only the main token is classified.

In order to guarantee reproducibility, objectivity, and comprehensiveness, a system selection process for the classification is designed.
Figure \ref{fig:refinement} depicts this process and visualizes the number of remaining systems per refinement step.
Two websites are used:

\begin{itemize}
	\item \textit{Coinmarketcap.com}: Central point of information inquiry in the cryptoeconomics field, listing DLT systems by their market capitalization. 
	The rationale is that the economic value of a system is a good proxy for its viability. Hence this source provides a ranked list of viable DLT systems.
	\item \textit{coincodecap.com}: This site lists Github indicators of DLT systems. In particular, it contains information about the number of code commitments, Github stars, and contributors to a DLT system. These indicators capture an active development of a system. 
\end{itemize}
The limitation of these data sources is that they only list systems that maintain a native cryptoeconomic token. Hence, Blockchain-as-a-Service systems\footnote{These are systems not utilizing a native distributed ledger, as defined in Table 2 of the Supplementary Material.}, such as Hyperledger Fabric are not considered. 
Moreover, depending on the development strategy of a system, commits might be merged externally and only pushed occasionally as major updates to Github. This may result in a lower rank of a DLT system, despite being actively maintained. This limitation is considered in the proposed ranking function (\ref{eq:ranking}).

Snapshots of the sources were taken on the $17^{th}$ April 2019 and are merged based on the systems acronym\footnote{A three-letter code identifying the token of a system.}. 
In order to account for academic relevance, the selection of the systems is enhanced with the number of mentions of DLT systems in Elseviers ScienceDirect database\footnote{Database of peer-reviewed literature, enabling full-text searches: https://www.elsevier.com/solutions/sciencedirect (last accessed: May 2019).} and then filtered based on the criterion of whether systems are actually mentioned in literature ($\#mentions > 0$). For the database search, the following search string is utilized on the API field \textit{qs}\footnote{Searches over all article excluding references: https://dev.elsevier.com/tecdoc\_sdsearch\_migration.html (last accessed: May 2019).}: "PROJECT NAME" AND (Blockchain OR Ledger).

The remaining systems are ranked based on the following ranking function

\begin{equation}
\label{eq:ranking}
\begin{aligned}
r(i) &= &&0.6 * \mathsf{m_{cap}(i)} + 0.3 * \mathsf{c_{commit}(i)} + \\ 
&&&0.1 * \mathsf{c_{contr}(i)}
\end{aligned}
\end{equation}
where $\mathsf{m_{cap}}$ is the rank based on the market capitalization of a system $i$, $\mathsf{c_{commit}}$ the commitment rank and $\mathsf{c_{contr}}$ the contributers rank. The weights are chosen to account for the limitation of the Github activity to be a proxy for active system maintenance, hence the lower weights.
The top $50$ systems are then classified, based on an extensive literature review performed by the first author and checked independently by the co-authors and the blockchain community. Sources for the classification are academic literature, DLT systems websites, and whitepapers. An overview of the final classified systems can be found in Table 1 of the Supplementary Material. Moreover, the actual classification of the systems is provided in Tables 3-6 of the Supplementary Material.

\subsection{Blockchain Community Feedback}
\label{meth:survey}

Participants
were invited based on their contributions to Github\footnote{Available at \url{https://github.com} (last accessed: July 2019).}
repositories of DLT systems and their official websites. Participants
received a personalized email invitation (Figure 1 in Supplementary Material) to participate in a scientific
survey to rate the classification of their DLT system
and the expressiveness (as defined in Section~\ref{sec:eval_taxonomy}) of the proposed taxonomy. A total of $326$ invitations were sent and $85$ practitioners
in the field responded (response rate $26.1\%$). 50 respondents completed
the survey (completion rate $58.8\%$). Only completed surveys are considered in the analysis. 
The participants were recruited during two phases each lasting two months: The first beginning on the 22\textsuperscript{nd} of March 2018 and the second on 24\textsuperscript{th} July 2019. The feedback of the first phase resulted in improvements of the taxonomy, as illustrated in Section B of the Supplementary Material, and the feedback of both phases resulted in improvements of the classification.

\subsubsection{Classification}
\label{sec:method_classification}
In the first part of the survey, the participants were shown the classification of the four components and 19 attributes of
the DLT system to which they contribute. Consult
Figure \ref{fig:tax_ontology} for an overview of the attributes and Tables 3-6 of the Supplementary Material
for the classification ratings.
The participants had the option to agree, disagree or state that they
were uncertain about the classification. They could always comment
on their decision, irrespective of their choice.

In order to calculate the consistency with which participants rated the classification of the same system, the consistency per attribute is calculated as follows: Assuming equidistance in the likert scale \cite{norman2010likert}, the participant responses are represented by a linear scale whereby $0$ denotes disagreement, $0.5$ denotes uncertainty,
and $1$ denotes agreement. Then, for each DLT system from which more than
one response was obtained as illustrated in Table \ref{tab:survey_dlt_project},
the consistency of responses is calculated for each system and attribute with the mean absolute error between the responses. 
Then, the average consistency for each attribute over all DLT systems
is obtained by calculating the weighted average value of the previously calculated mean absolute errors.

\subsubsection{Taxonomy}
\label{sec:feedback_taxonomy}
In the second part of the survey, the blockchain community is asked to evaluate the taxonomy.
Nickerson et al. propose five criteria to assess the \textit{usefulness} of a taxonomy \cite{nickerson2013method}. Namely, a taxonomy is 
\begin{itemize}
	\item \textit{concise}, if it uses a limited number of attributes,
	\item \textit{robust}, if it uses enough attributes to clearly \textit{differentiate} the objects of interest
	\item \textit{comprehensive}, if it can \textit{classify} all known objects within the domain under considerations,
	\item \textit{extensible}, if it allows for inclusion of additional attributes and attribute values when new types of objects appear,
	\item \textit{explanatory}, if it contains object attributes that do not model every possible detail of the objects but, rather, provide useful explanations of the nature of the objects under study or help to understand future objects.
\end{itemize}

The literature review (Section \ref{sec:backgorund}) reveals differences regarding how many attributes should be included
in a \textit{robust} taxonomy of DLT systems. Also, the scope of the classification is to \textit{comprehensively} classify the CED of all academically relevant systems. Thus, considering these two points, the taxonomy
is evaluated using the robustness and comprehensiveness criteria of
Nickerson et al. \cite{nickerson2013method}.
To this end, this paper introduces the concept of expressiveness: 

\begin{mydef}
	\textit{A taxonomy is expressive when it is robust and comprehensive.} 
\end{mydef}

\noindent where a robust and comprehensive taxonomy are given by Nickerson et. al \cite{nickerson2013method}. 

The perceived expressiveness of the developed taxonomy can be determined by asking the survey participants for each component and attribute:

\begin{quest} 
	\textit{"How \textit{expressive} is [component/attribute] to \textit{differentiate} between and \textit{classify} DLT systems".}
\end{quest}

This formulation neither exposes survey participants to the theory of expressiveness, comprehensiveness and robustness nor overloads them with a high number of questions.

The consistency calculation for the taxonomy feedback follows along the lines of the classification (Section \ref{sec:method_classification}): Despite utilizing a five-point Likert scale (from very non-expressive to very expressive) to create
values ranging from zero to one, the calculation of consistency remains
the same as the one for the classification.

\subsection{Machine Learning Analysis}\label{meth:data_analysis}

In order to mine the key design choices in the classified DLT systems, two unsupervised machine learning methods are applied to the classified systems:

\begin{enumerate}
	\item \textit{Mulitple Correspondence Analysis} (MCA) is a statistical method that is widely used in the social sciences. It can analyze data without a priori assumptions concerning the data, such as data falling into discrete clusters or variables being independent  \cite{greenacre1984correspondence} \cite{abdi2007multiple}. It is a generalization of the principal component analysis (PCA) for categorical data coded in the form of an indicator matrix or a Burt matrix \cite{greenacre2005multiple}, which aims at summarizing underlying structures in the fewest possible dimensions \cite{xiong2012dhcc}. In particular, it identifies new latent dimensions, which are a combination of the original dimensions and hence can explain information not directly observable \cite{greenacre2006multiple}. Moreover, similar to PCA, these dimensions are ordered by their importance to explain the amount of variance in the data \cite{abdi2007multiple}.

	\item \textit{Kmeans} \cite{jain2010data} for varying $k$ is applied on the classification to cluster the DLT systems based on their attribute values. The optimum number of clusters is derived by both, performing a bootstrap evaluation that determines the stability of the clusters \cite{hennig2007cluster} and by two well-known cluster evaluation metrics: Silhouette and Calinski-Harabasz \cite{rendon2011internal}. 
	
\end{enumerate}

\section{Experimental Evaluation}
\label{sec:experimental_evaluation}
The evaluation aims to identify key design choices that govern the modeling complexity of DLT systems at design phase. 
In order to base these insights on a strong footing, first, the taxonomy and classification are validated by feedback from the blockchain community (Section \ref{sec:community_evaluation}). Then two machine learning methods are applied on the classification to mine the design choices on a quantitative basis. (Section \ref{sec:machine_learning_analysis}).
\subsection{Blockchain Community Feedback}
\label{sec:community_evaluation}

The taxonomy (Section \ref{sec:taxonomy}) and classification (Table 3-6 of the     Supplementary Material) are evaluated using feedback from the blockchain community.

\subsubsection{Demographics}
\label{sec:eval_demo}

\begin{table}[!htb]
	\centering
	\caption{Survey participants per DLT system, their specific roles, and experience}
	
	\label{tab:survey_dlt_project}
	\begin{tabular}{ll}
		\toprule
		\textbf{DLT system} & \textbf{Total} \\ \midrule
		Aragon           & 2 \\
		Ark              & 1 \\
		Bitcoin          & 1 \\
		Bitcoin Cash     & 2 \\
		BitShares        & 2 \\
		Byteball         & 1 \\
		Cardano          & 3 \\
		Dash             & 6 \\
		Decred           & 1 \\
		DigiByte         & 1 \\
		Ethereum         & 1 \\
		Factom           & 1 \\
		Golem            & 1 \\
		IOTA             & 2 \\
		Komodo           & 1 \\
		MOAC-MotherChain & 1 \\
		Monacoin         & 1 \\
		Monero           & 4 \\
		NEM              & 1 \\
		NEO              & 1 \\
		Nexus            & 2 \\
		PIVX             & 1 \\
		ReddCoin         & 1 \\
		Siacoin          & 1 \\
		Skycoin          & 1 \\
		Steem            & 1 \\
		Stellar          & 1 \\
		Storj            & 1 \\
		Stratis          & 1 \\
		TRON             & 2 \\
		Verge            & 1 \\
		Waves            & 1 \\
		Zcash            & 2 \\
		\textbf{Total}      & \textbf{50}    \\ \bottomrule
	\end{tabular}
	\quad
	\begin{tabular}{ll}
		\toprule
		\multicolumn{1}{l}{\textbf{Role in Project}}       & \multicolumn{1}{l}{\textbf{Total}} \\ \midrule
		Project Lead                               & 7                                   \\
		Core/Team Developer                       & 21                                   \\
		Team Member                                & 8                                   \\
		Advisor                                    & 1                                   \\
		Community Developer                        & 4                                   \\
		Community Member                           & 2                                   \\
		Other                                      & 7                                   \\ 
		\multicolumn{1}{l}{\textbf{Total}} & \multicolumn{1}{l}{\textbf{50}}    \\ \bottomrule
		\toprule
		\multicolumn{1}{l}{\textbf{Experience}} & \multicolumn{1}{l}{\textbf{Total}} \\ \midrule
		$>$ 3 years                & 15                                  \\
		1-3 years                   & 29                                 \\
		$<$ 1 year                       & 6                                   \\ 
		\multicolumn{1}{l}{\textbf{Total}} & \multicolumn{1}{l}{\textbf{50}}    \\ 
		\bottomrule
	\end{tabular}

\end{table}

Table \ref{tab:survey_dlt_project} shows the demographics of
the survey participants. In particular, it shows participants specific roles for the systems and their experience. The 50 participants work in (core) technical (25 developers) and strategic (7 Project leads) positions.
Moreover, 15 participants have more than three years of experience,
29 participants have worked one to three years, and 6 participants
have worked for less than a year in the field of DLT systems. 
Moreover, Table \ref{tab:survey_dlt_project} illustrates that the participants are involved in 33 out of the 50 classified systems.

\subsubsection{Classification}
\label{sec:eval_class}

\begin{figure}[!htb]
	\centering
	
	\subfigure[Classification (N=50,46,26)]{\includegraphics[width=0.48\columnwidth]{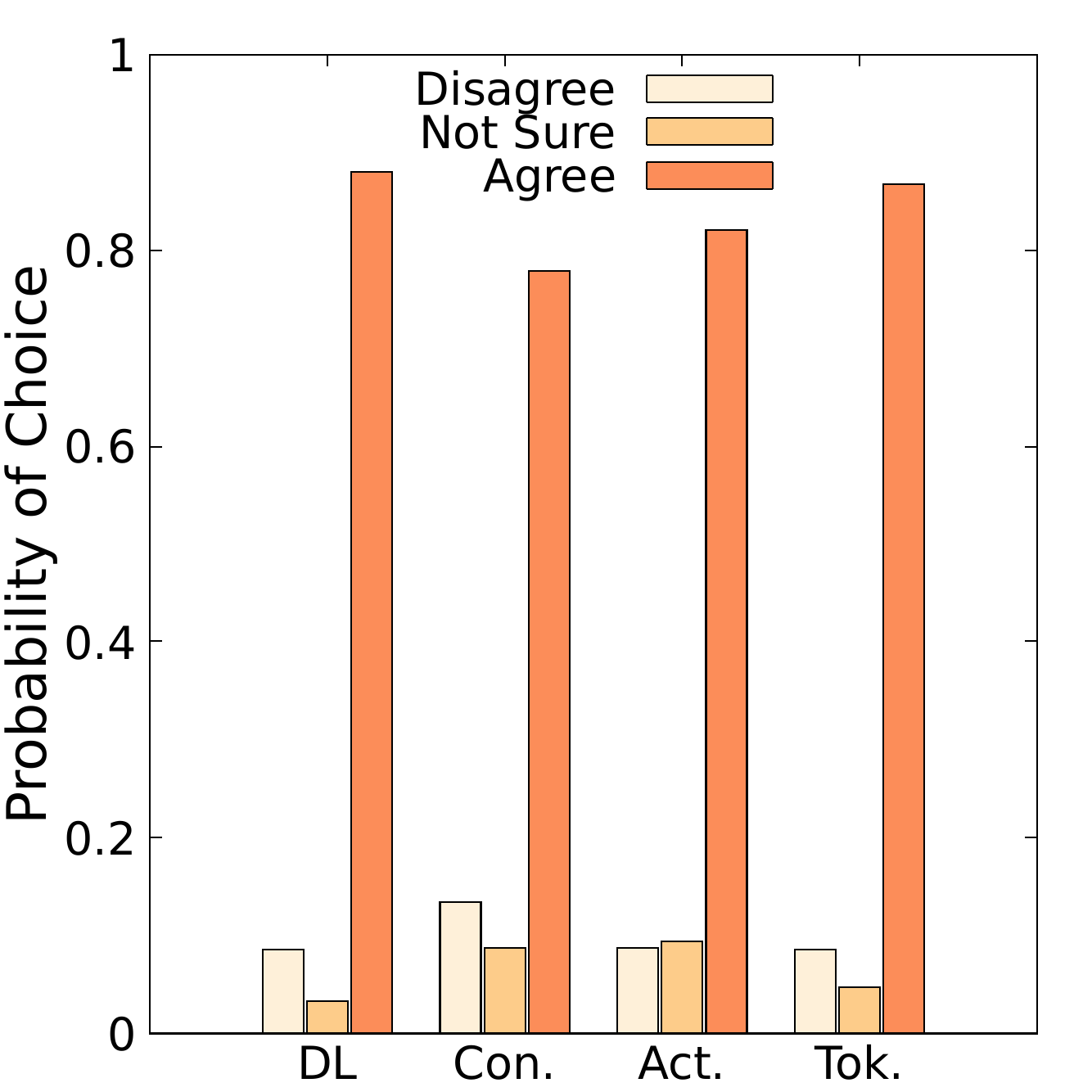}\label{fig:class_components}}
	\subfigure[Taxonomy (N=50)]{
		\includegraphics[width=0.48\columnwidth]{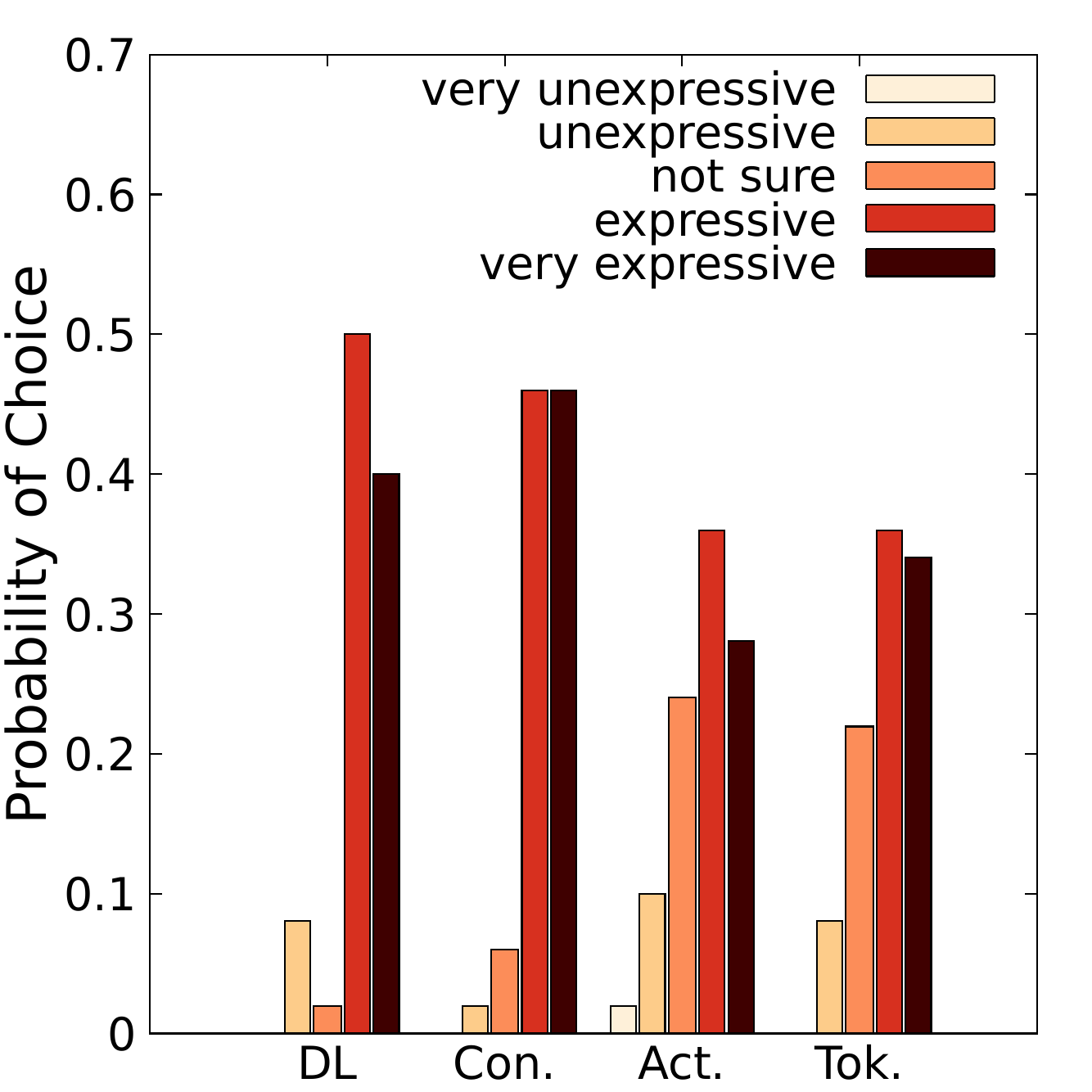}\label{fig:expressiveness_components}}
	\caption{Acceptance level of the classification and expressiveness of taxonomy components as perceived by survey participants.}
	\label{fig:expressiveness_and_agree}
\end{figure}

Figure \ref{fig:class_components} depicts the aggregate acceptance level for
each of the components. The Distributed Ledger component received
the highest acceptance level with $88.0\%$, followed by the Token component
($86.8\%$), Action component
($82.0\%$) and Consensus component ($77.8\%$).

\begin{figure*}[!htb]
	\centering
	\subfigure[Distributed Ledger (N=50,46,26)]{\includegraphics[width=0.24\textwidth]{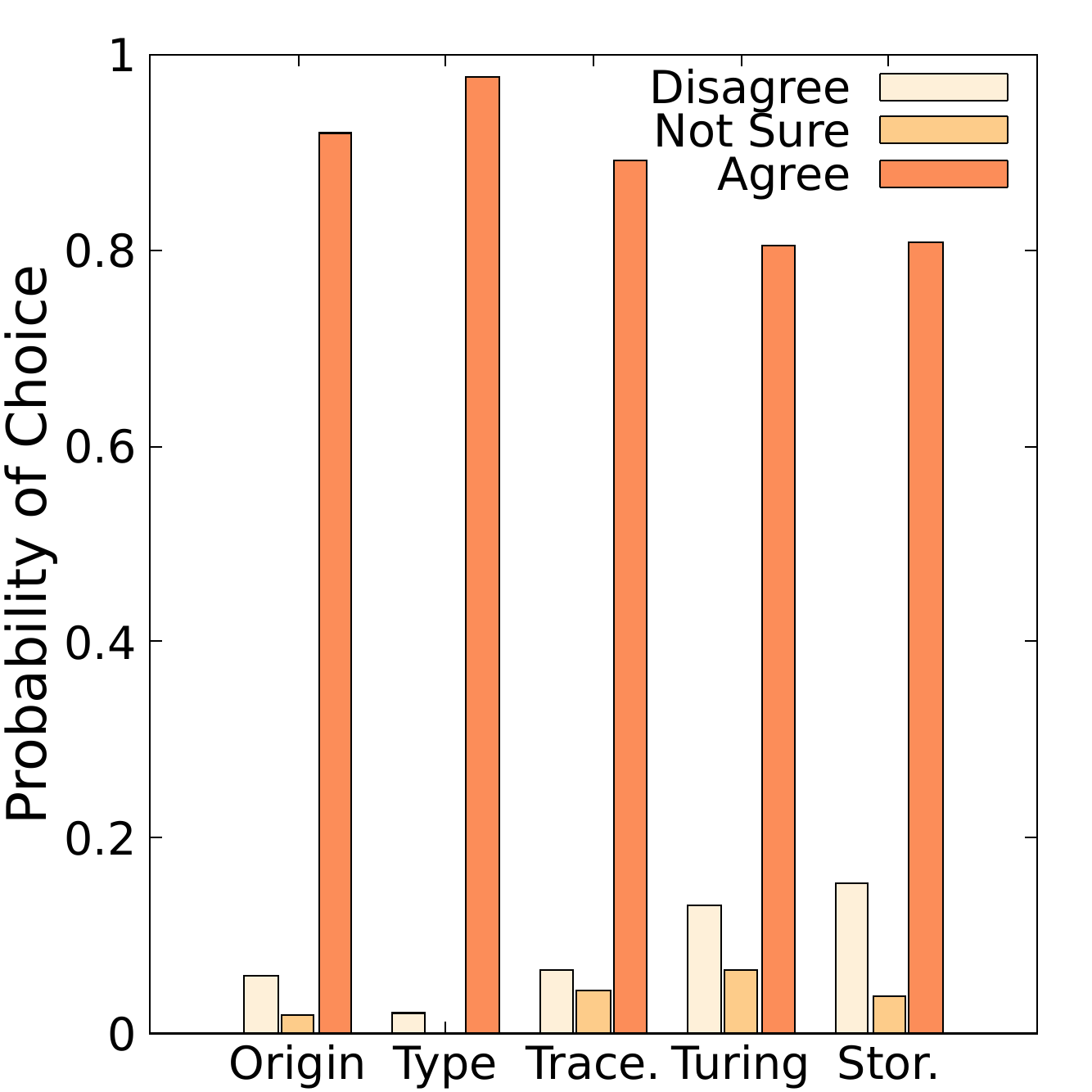}}
	\subfigure[Consensus (N=46)]{\includegraphics[width=0.24\textwidth]{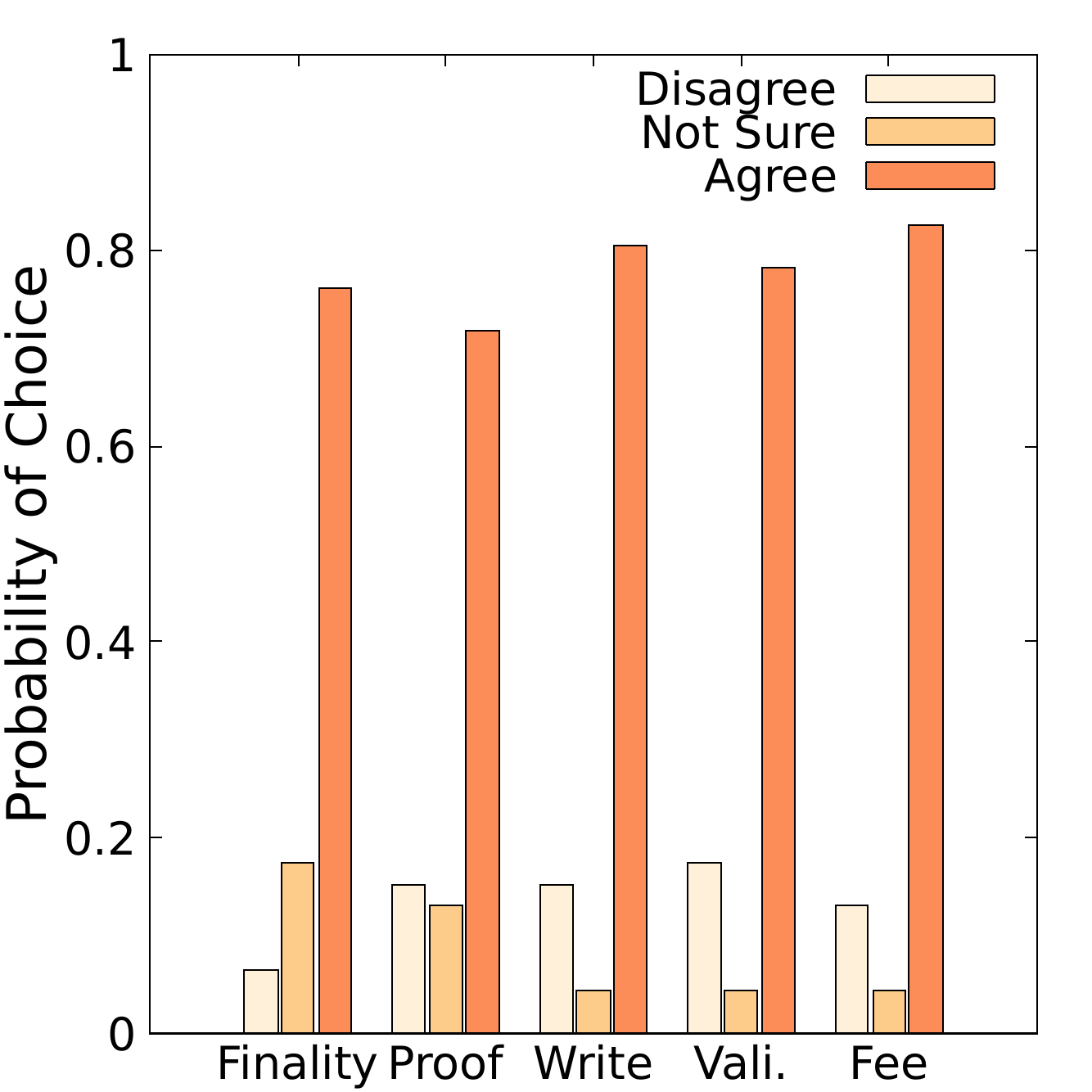}}
	\subfigure[Action (N=50)]{\includegraphics[width=0.24\textwidth]{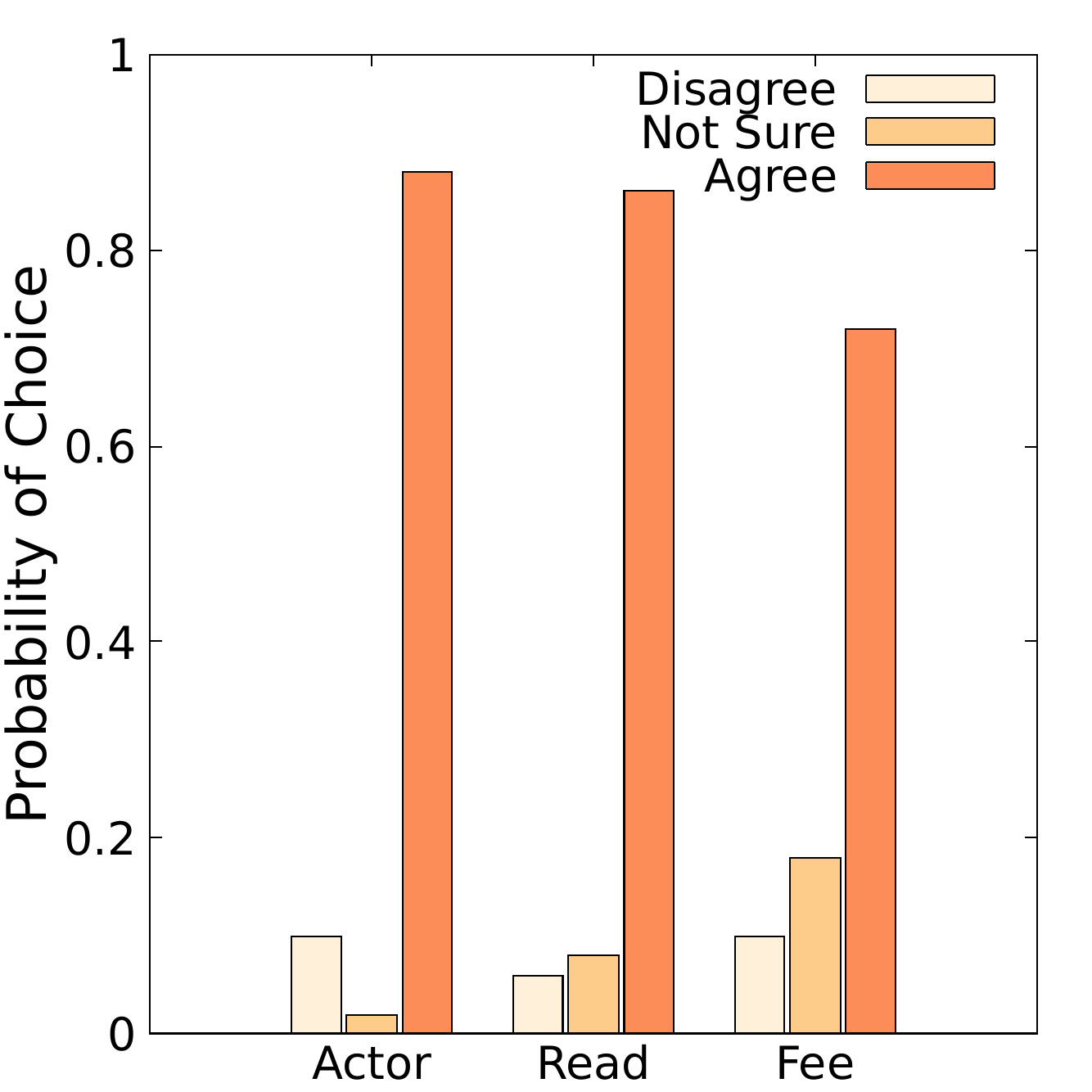}}
	\subfigure[Token (N=50,26)]{\includegraphics[width=0.24\textwidth]{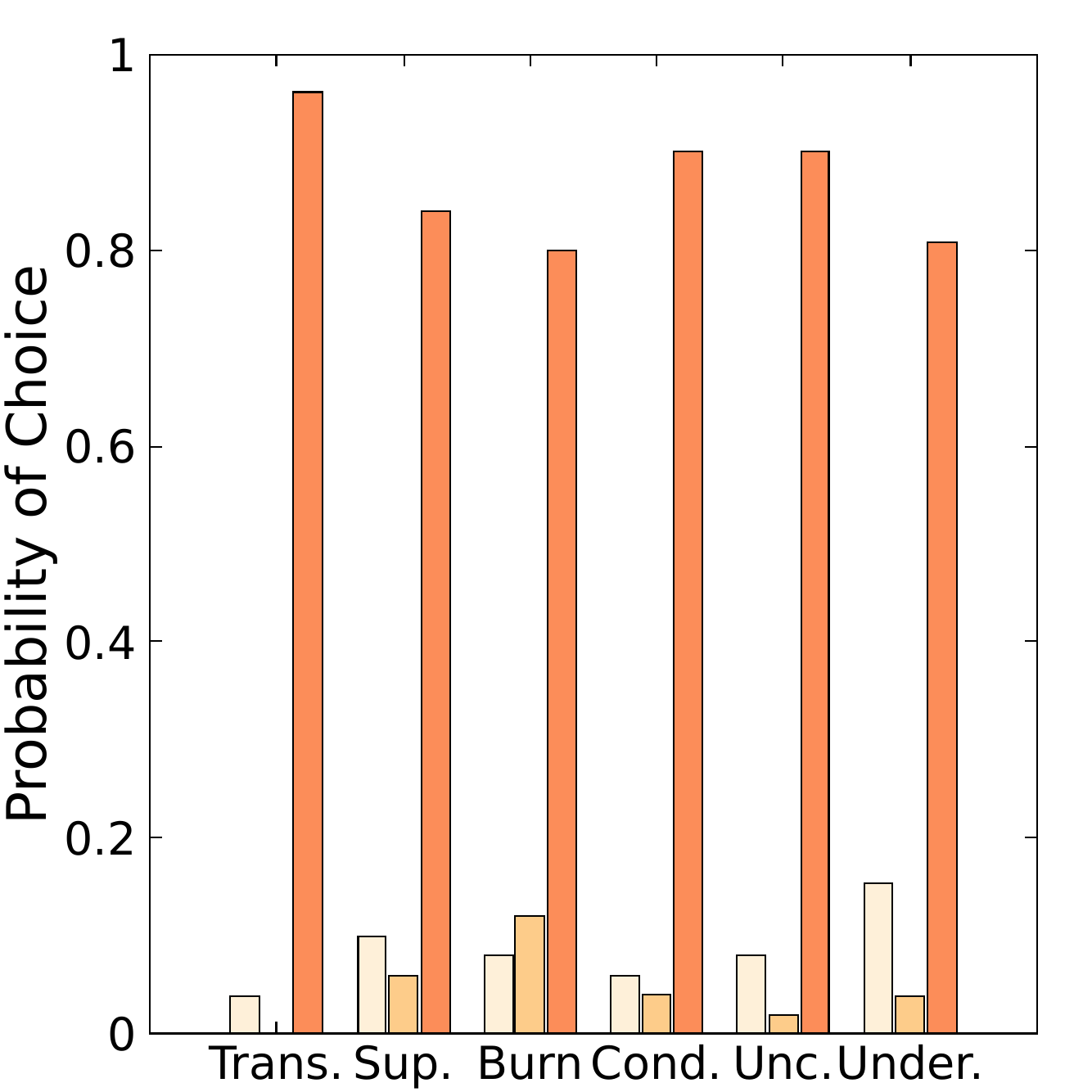}}
	\caption{Classification evaluation of the attributes, grouped component-wise.}
	\label{fig:class_attributes}    
	\subfigure[Distributed Ledger]{\includegraphics[width=0.24\textwidth]{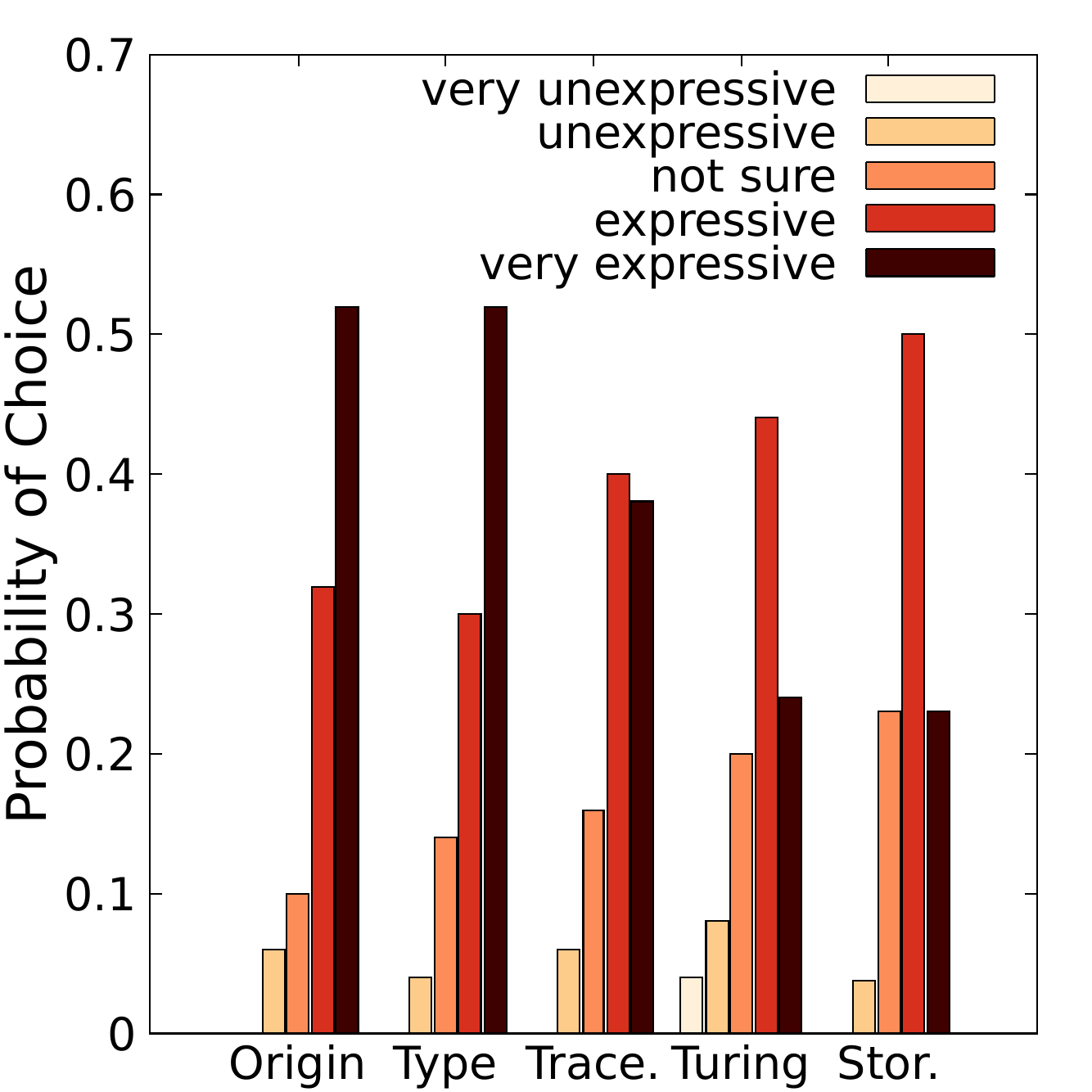}}
	\subfigure[Consensus]{\includegraphics[width=0.24\textwidth]{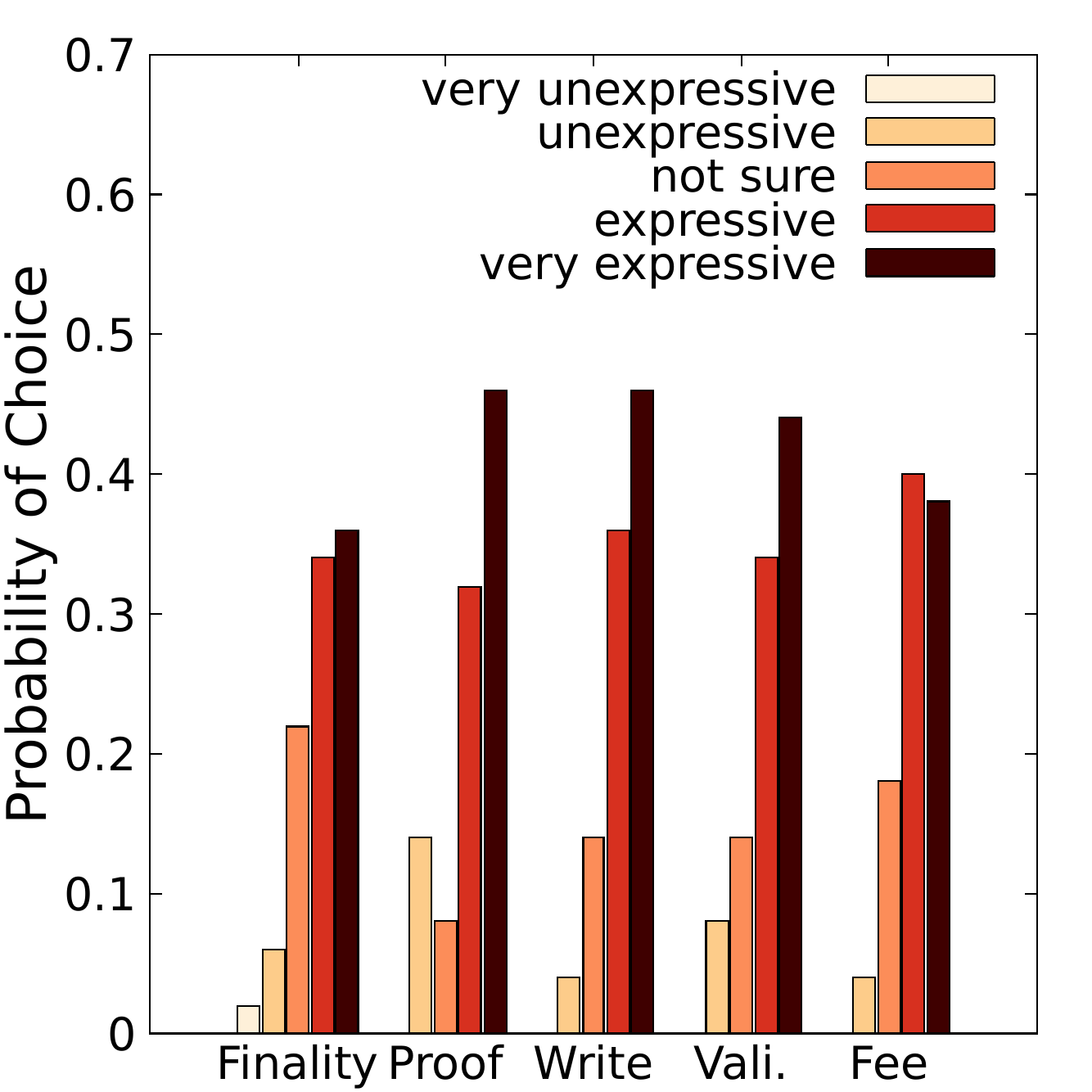}}
	\subfigure[Action]{\includegraphics[width=0.24\textwidth]{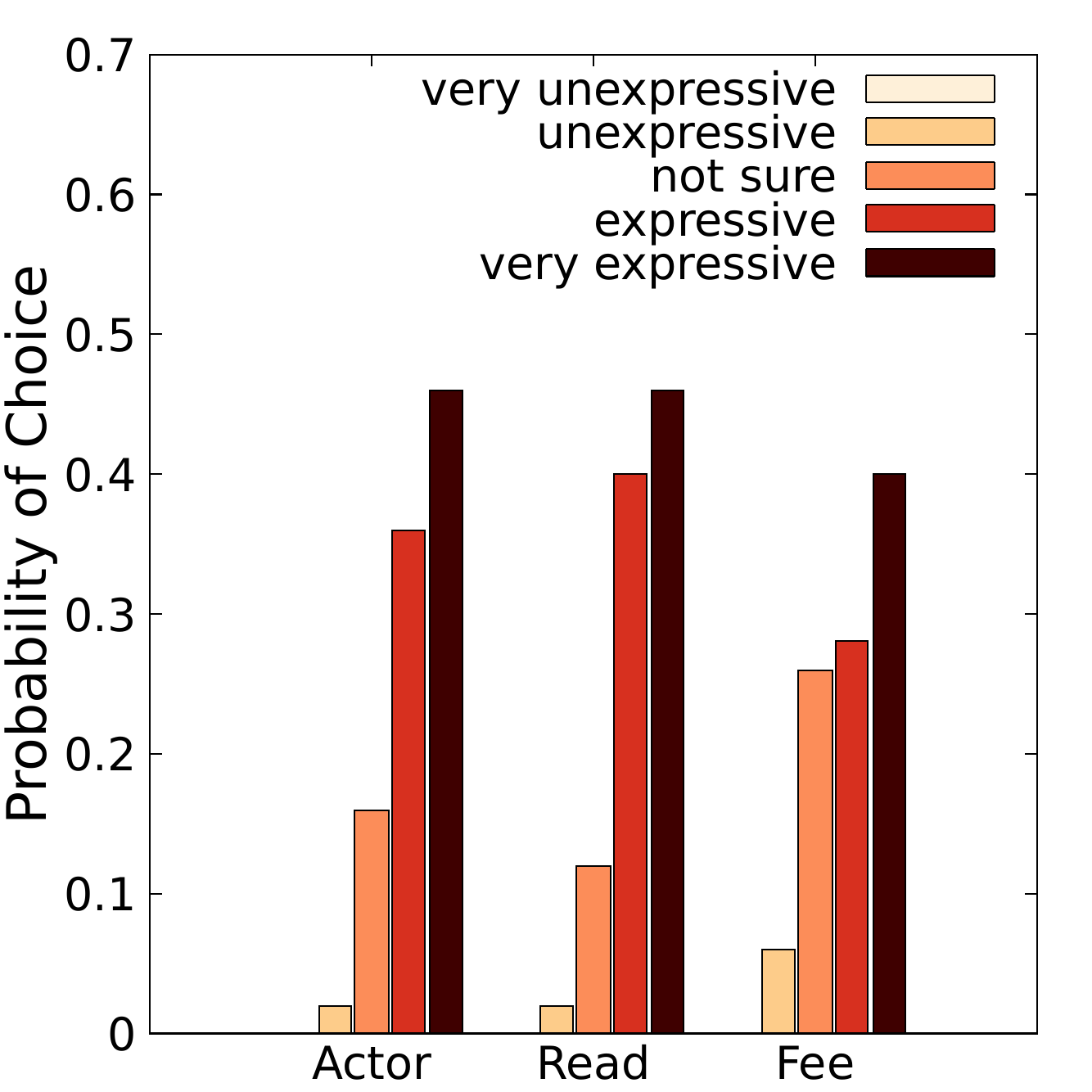}}
	\subfigure[Token]{\includegraphics[width=0.24\textwidth]{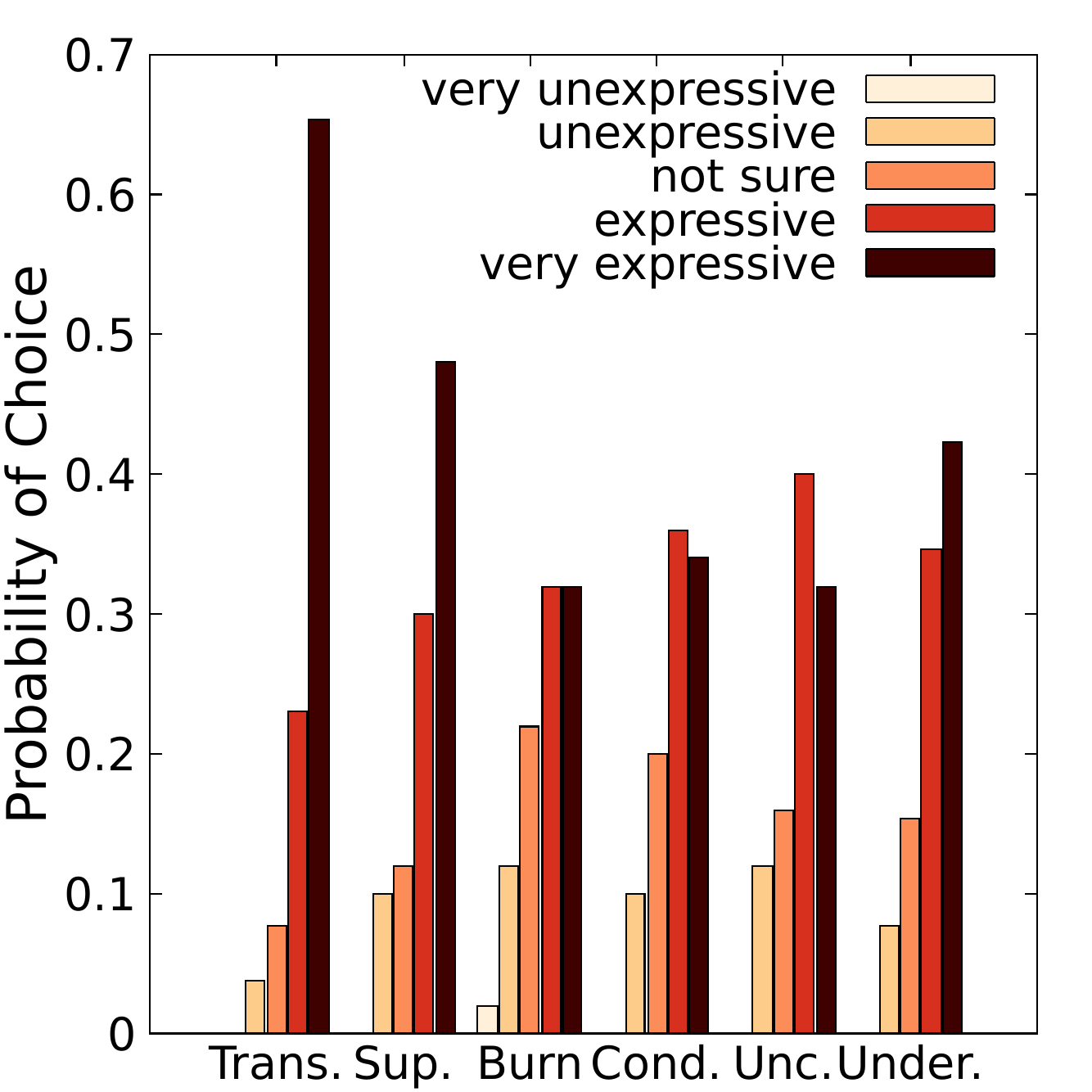}}
	\caption{Expressiveness evaluation of the attributes, grouped component-wise (N=50).}\label{fig:expressiveness_attributes}
\end{figure*}

Figure \ref{fig:class_attributes} illustrates the acceptance level for
each attribute of the four components. It is noteworthy that the average 
approval rating over all components is $83.7\%$. Five attributes are above $90\%$:
transferability ($96.2\%$), origin ($92.0\%$), DL type ($97.8\%$), creation condition ($90.0\%$) and unconditional creation
($90.0\%$). 

The figure shows that the highest disagreements relate to the validate
permission ($17.4\%$), underlying ($15.4\%)$ and storage ($15.4\%)$.
The highest degree of uncertainty is expressed regarding the action fee ($18.0\%$), consensus finality ($17.4\%$) and consensus proof ($13.0\%$) attributes.

\begin{figure}[!htb]
	\centering
	
	\subfigure[Agreement on classification of attributes (N=27)]{\includegraphics[width=0.48\columnwidth]{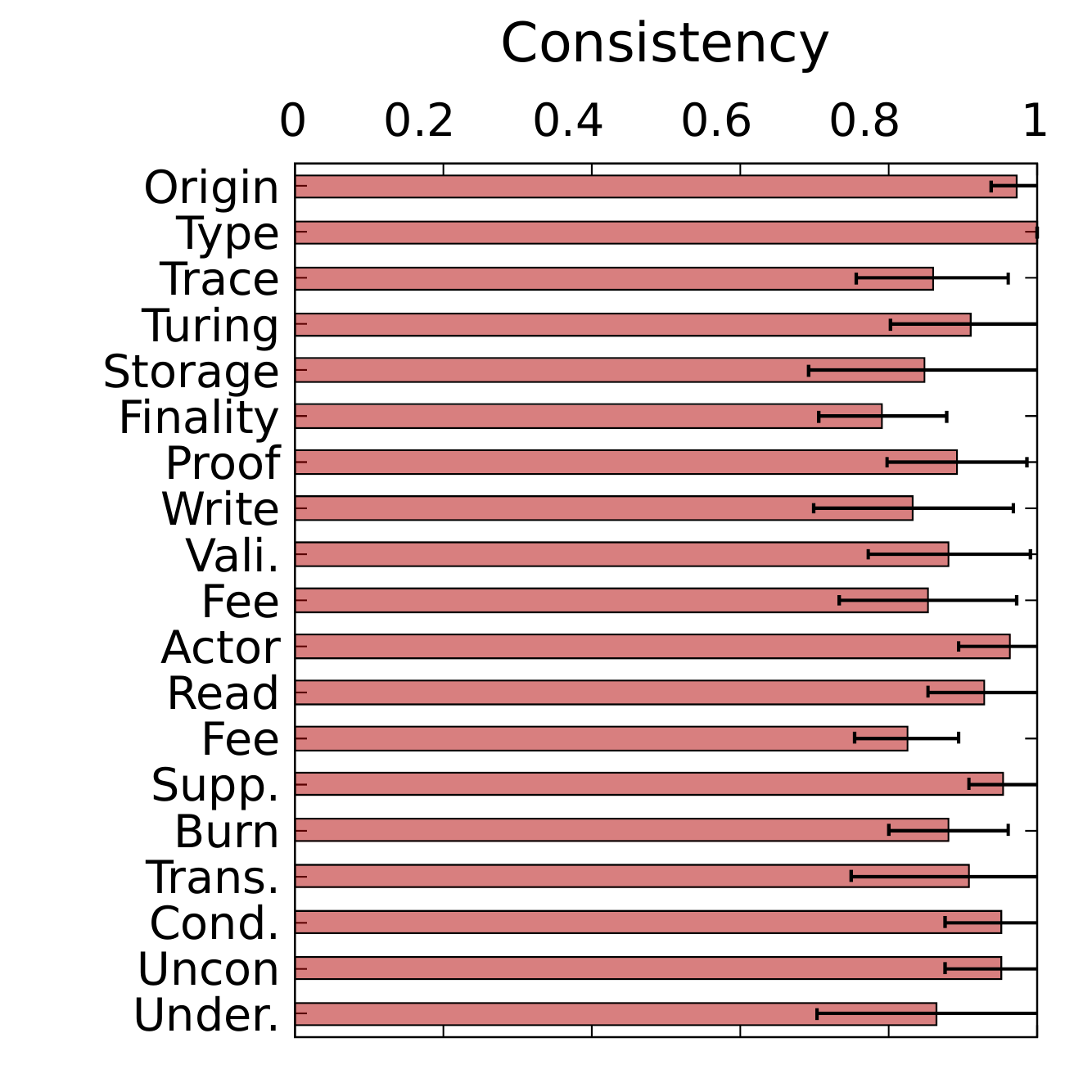}\label{fig:consistency_class}}
	\subfigure[Expressiveness of attributes (N=27)]{
		\includegraphics[width=0.48\columnwidth]{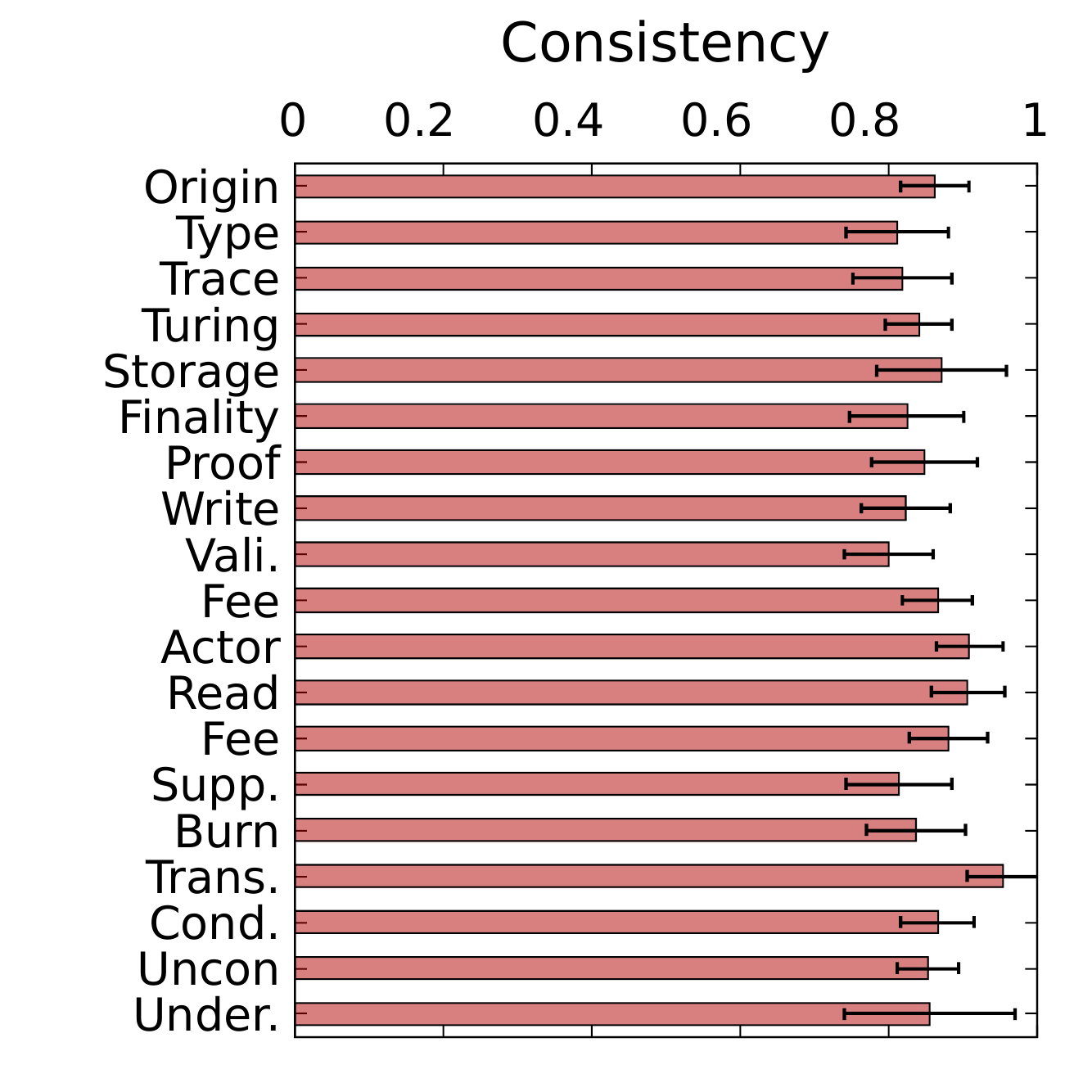}\label{fig:consistency_tax}}
	\caption{Weighted average of consistency calculation per attribute, using DLT systems consistency values of which more than one response is obtained.}
	\label{fig:consistency_class_and_tax}
\end{figure}

In order to investigate the consistency of the responses, the weighted consistency averages
for each attribute are depicted in Figure \ref{fig:consistency_class_and_tax}.
The overall consistency is on average $89.9\%$. The lowest consistency
measured relates to the consensus type ($79.2\%$) and action fee ($82.4\%$), correlating
with the higher degree of disagreement observed earlier. The highest
consistencies are observed for the DL type ($100.0\%$), origin ($97.3\%$), actor permission ($96.4\%$), supply property ($95.8\%$), creation condition ($95.6\%$) and unconditional creation ($95.6\%$) attributes.

In a nutshell, the acceptance level of $83.7 \%$ over all components and the average consistency of $89.9\%$ indicates the acceptance of the classification by the community.

\subsubsection{Taxonomy}
\label{sec:eval_taxonomy}

Figure \ref{fig:expressiveness_components} depicts the expressiveness
of the four components as perceived by the survey participants. The
Consensus component is seen as the most expressive ($92.0\%$), followed
by Distributed Ledger ($90.0\%$), Token ($70.0\%$) and Action component
($64.0\%$). The highest uncertainty relates to the Action ($24.0\%$)
and Token ($22.0\%$) components. 
The Action component consists of the lowest number of attributes, which
may decrease its perceived expressiveness. In particular, the reduced
number of attributes seems to hinder differentiation between DLT systems. Moreover, the literature review reveals, that Consensus is included in all taxonomies (Section \ref{sec:backgorund}). Thus this component might have been the most familiar to the participants resulting in higher expressiveness.

15 participants commented on the expressiveness of the components.
They stated that a component depicting the governance of a system should be illustrated by the taxonomy ($26.6\%$)\footnote{In brackets are depicted the percentage for which this responds type
	accounts for the overall received comments. Please note, that the percentages
	do not add up to $100\%$ as a survey participants comments can account for more than one responds type.}, including the funding of a DLT system.
Three participants ($20\%$) mention that the Action component is not expressive enough to illustrate specific features of a
system, such as the distribution of actors. Similar statements were made about the Token component ($20.0\%$). In particular, it has been stated, that inter-token dynamics should be covered and that further attributes are required to illustrate the creation conditions and 1\textsuperscript{st} and 2\textsuperscript{nd} layer tokens ($20.0\%$). Moreover, the quality of code implementation, type of programming language, strategy of code development and scalability of the system has been mentioned ($26.6\%$) as expressive attributes missing in the taxonomy. One participant stated, that the underlying attribute should be more sharply defined\footnote{Since the participant's feedback the definition of the underlying has been revised. Please refer to Section B of the Supplementary Material}, and another used the opportunity to further elaborate on the system functioning.
Finally, some participants made statements endorsing the construction of the taxonomy ($13.3\%$).

Figure \ref{fig:expressiveness_attributes} depicts the perceived
expressiveness of the 19 attributes. The five most expressive attributes
are deemed to be transferability ($88.5\%$), read permission ($86.0\%$), origin ($84.0\%$), actor permission ($82.0\%$), write permission ($82.0\%$) and DL type ($82.0\%$). Action fee ($26.0\%$), storage ($23.1\%$), consensus type ($22.0\%$) and burn property ($22 \%$) raise the highest degree of uncertainty.
The least expressive attributes are deemed to be the consensus proof ($14.0\%$), burn property ($14.0\%$) and Turing completeness/ unconditional creation (each $12.0 \%$) attributes. 
Despite the
Action component being the least expressive component, two of its attributes are amongst the top five most expressive attributes. This supports the consideration to extend the action component by adding further attributes. A similar observation is made for the Token component: transferability is the most expressive attribute, but the perceived expressiveness of its component is lower than for the DL and Consensus components, which suggests extending the attributes of the Token component.

The assessment of the feedback regarding the attributes provided by the survey participants during the first recruitment phase lead to an inclusion of further attributes into the taxonomy. The nature and reasoning of these adjustments can be found in Section B of the Supplementary Material. This inclusion of new attributes indicates that the taxonomy is extensible \cite{nickerson2013method}.

Figure \ref{fig:consistency_tax} depicts the consistency with which
the participants evaluated the expressiveness of the taxonomy attributes.
The average consistency over all
attributes is $85.5\%$, meaning that survey respondents from the
same DLT systems rated the expressiveness of the taxonomy similarly
to each other. In particular, they diverge from each other just $14.5\%$ on average, that is less than one choice difference
on the aforementioned Likert scale.

In a nutshell, the average expressiveness rating of $79 \%$ over all components and the average consistency of $85.5\%$ indicates that the taxonomy is expressive.

\subsection{Machine Learning Analysis}
\label{sec:machine_learning_analysis}

\begin{figure*}[!htb]
	\centering
	\subfigure[\textit{2nd} Layer (1. Dim.) ]{\includegraphics[width=0.24\textwidth]{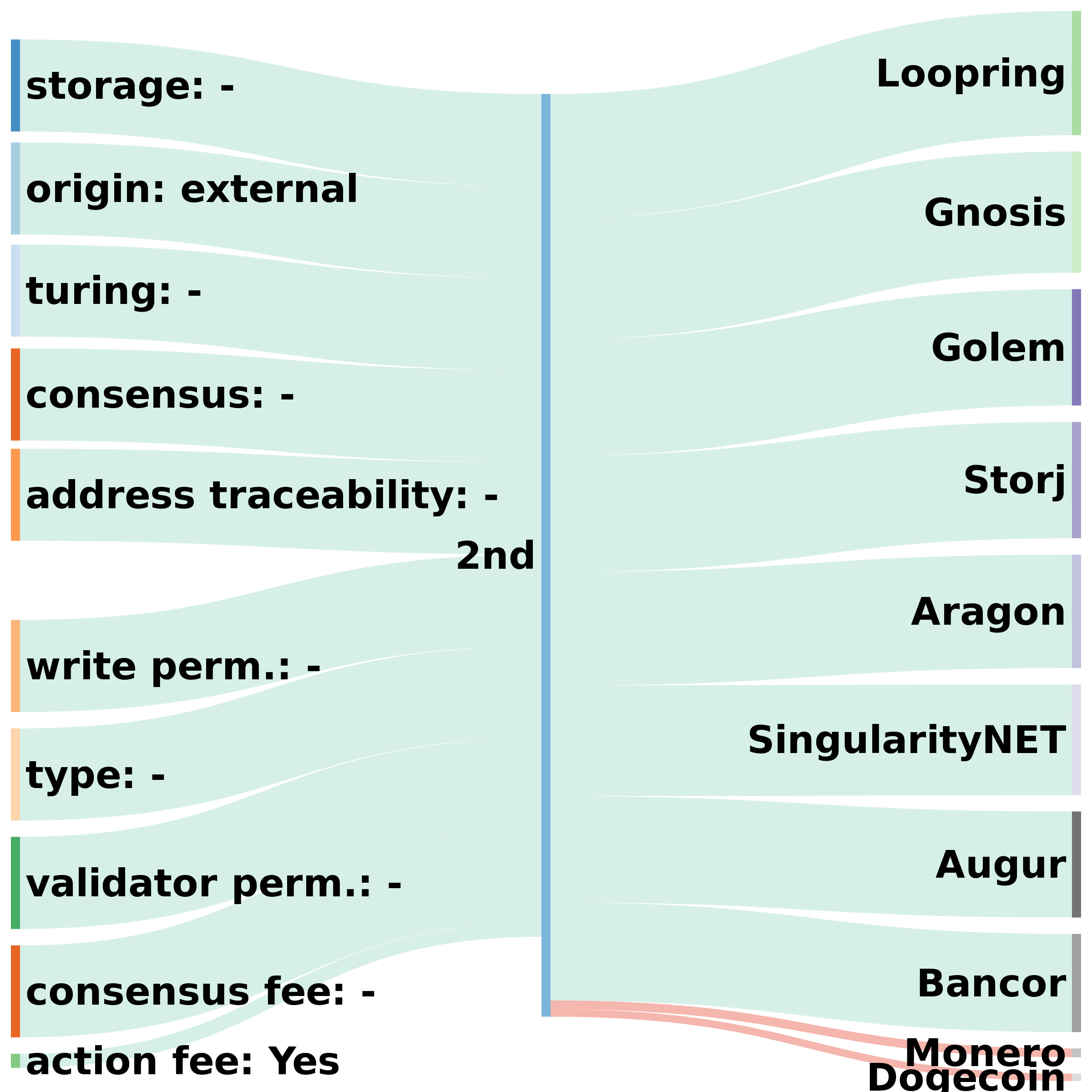}}
	\subfigure[\textit{Permissioned} Participation (2. Dim)]{\includegraphics[width=0.24\textwidth]{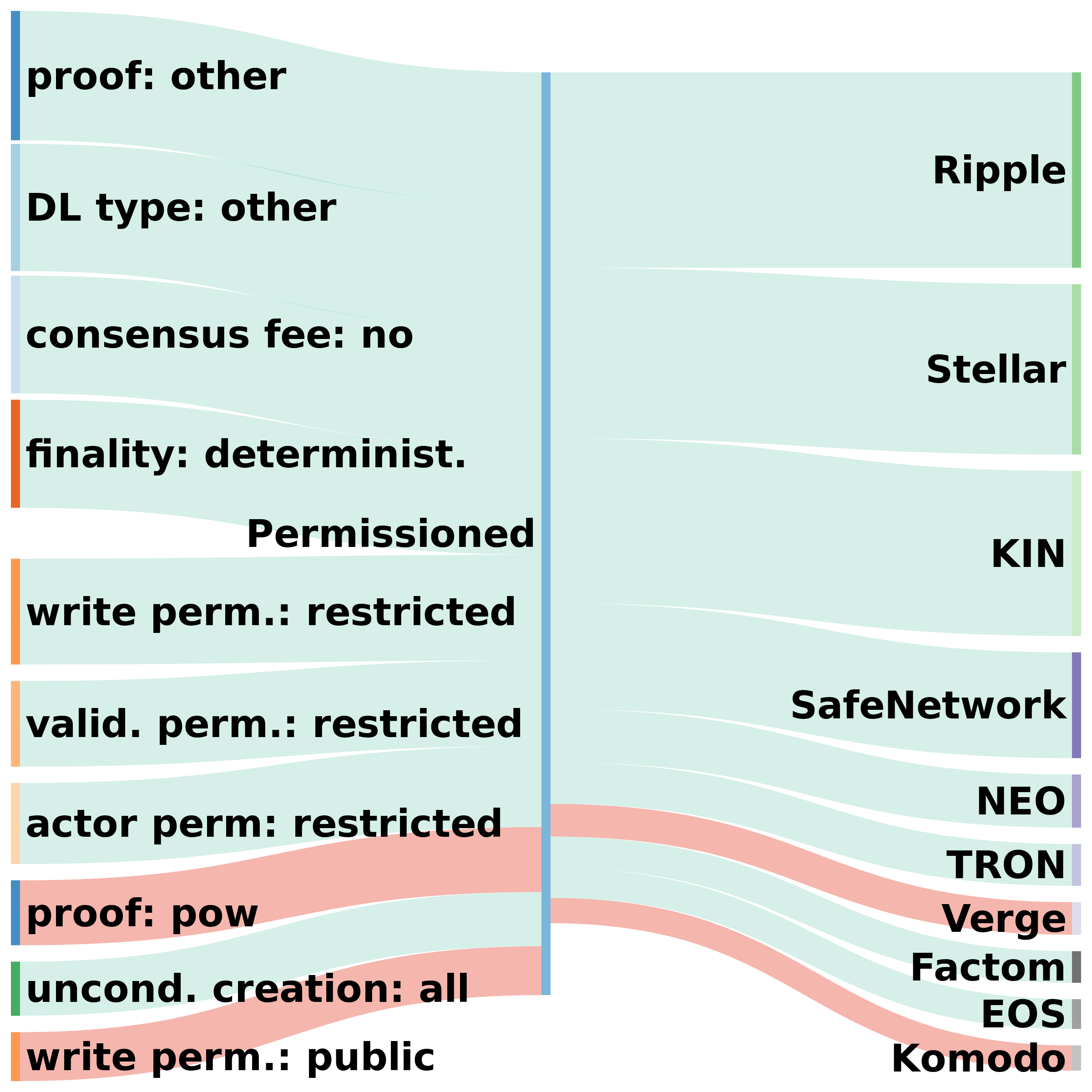}}\label{fig:sankey_participation}
	\subfigure[\textit{High} Stakig Capability (3. Dim)]{\includegraphics[width=0.24\textwidth]{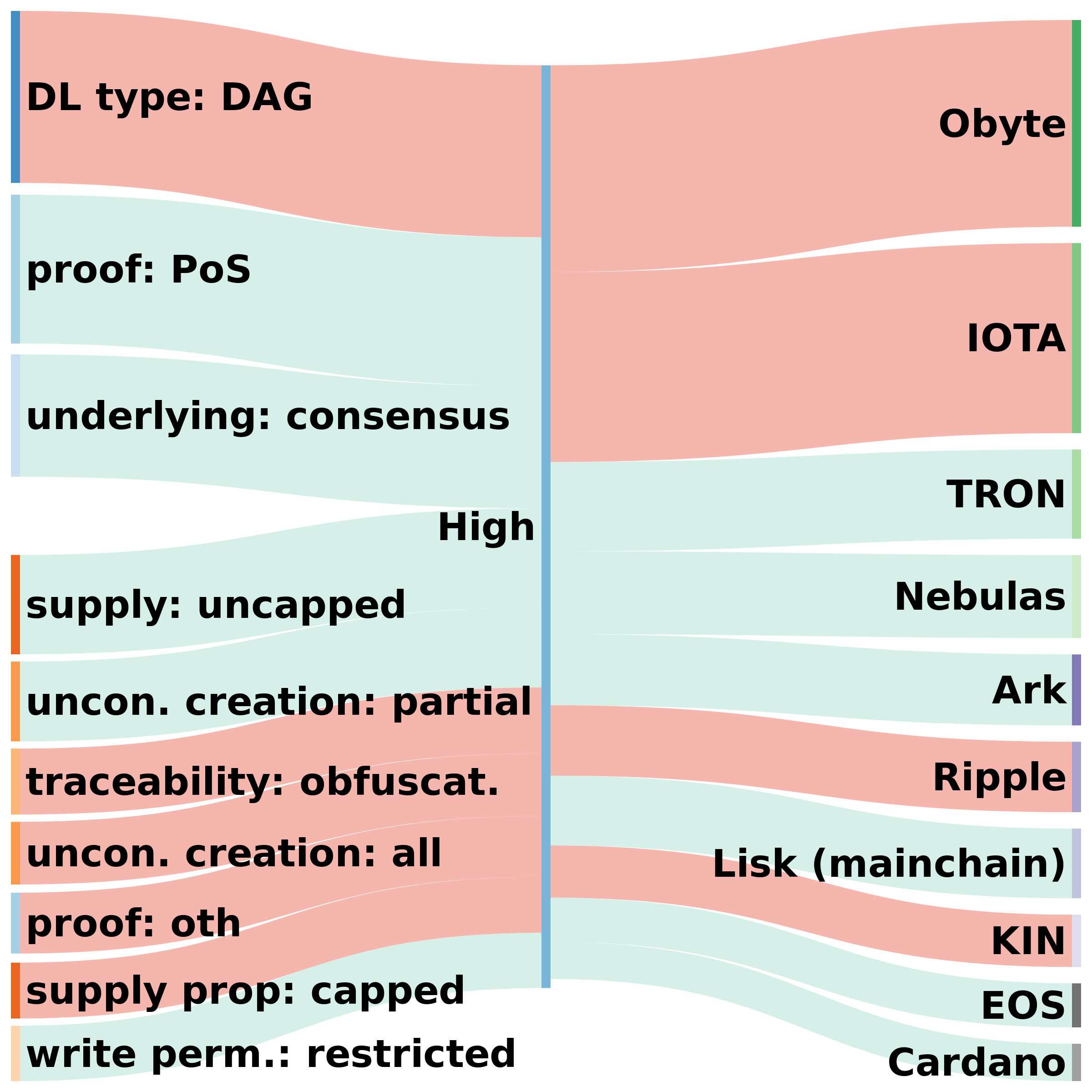}}
	\subfigure[\textit{Complex} Cryptoeconomic Complexity (4. Dim)]{\includegraphics[width=0.24\textwidth]{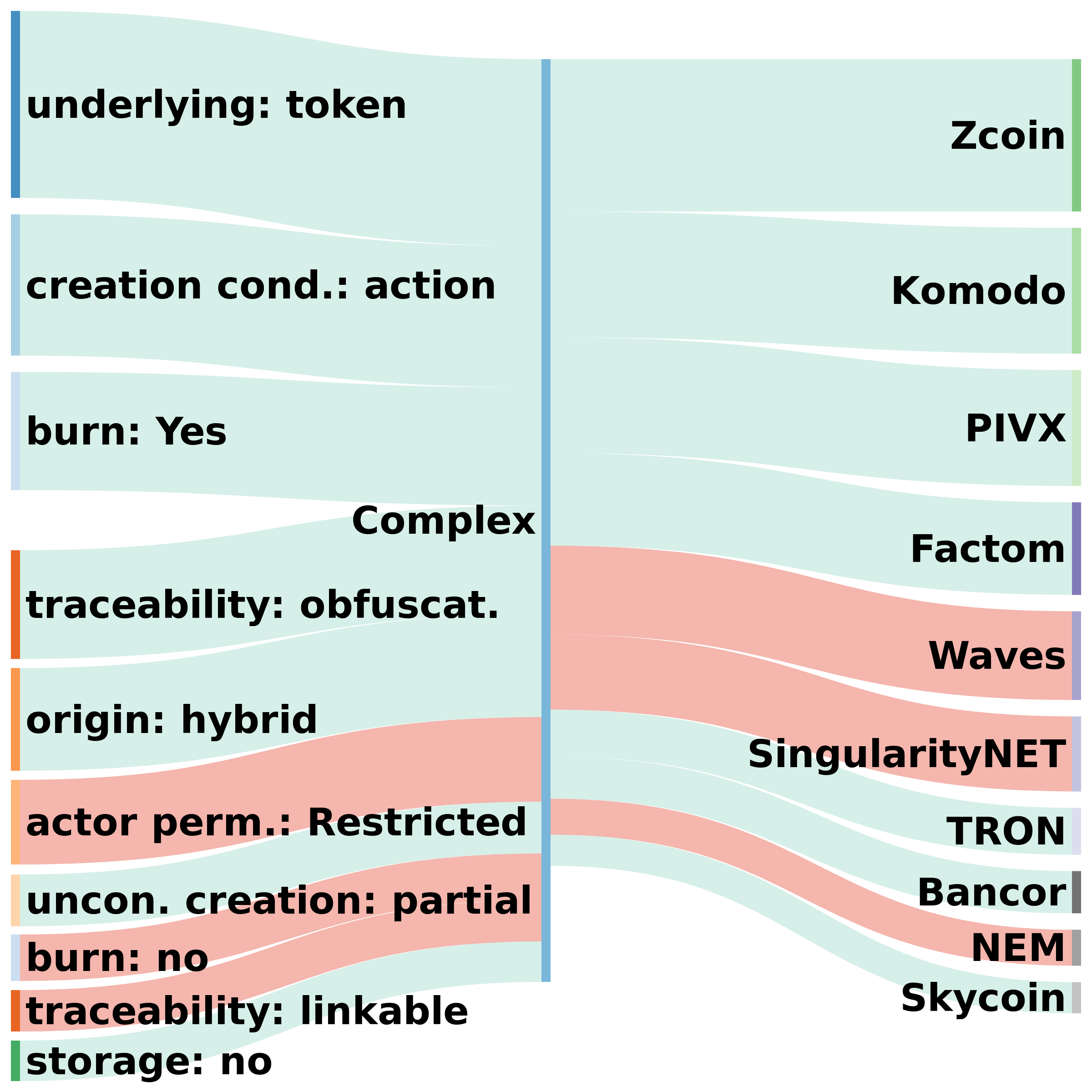}}
	\caption{Absolute Contribution (flow's thickness) of attribute values (left) and systems (right) to value of new dimension (middle/ italics underneath the figures). The color code depicts if attribute value/ system contributes negatively (red/ dark) or positively (green/ light).}
	\label{fig:sankey}

\end{figure*}

\begin{table}[]
	
	\caption{Eigenvalues and corresponding explained variances of MCA dimensions after applying Benzecri and Greenacre correction}
	\label{tab:mca}
	\centering
	\begin{tabular}{lllll}
		\toprule
		\multirow{2}{*}{\textbf{Dim}} & \multirow{2}{*}{\textbf{Description}} & \multirow{2}{*}{\textbf{Eigenvalue}} & \multicolumn{2}{l}{\textbf{Corrected Variances}} \\ 
		&           &         & \textit{Benzceri}  & \textit{Greenacre}      \\                     \midrule
		\textit{1}            &    Layering          & 0.311                                   & 0.764                                 & 0.679                                  \\
		\textit{2}     & Participation                     & 0.060                                   & 0.148                                 & 0.132                                  \\
		\textit{3}      & Staking capability                    & 0.013                                   & 0.032                                 & 0.029                                  \\
		\textit{4}       & Cryptoec. complexity                  & 0.007                                   & 0.018                                  & 0.016                                  \\
		\textit{5}        &                  & 0.006                                   & 0.014                                 & 0.012                                  \\
		\textit{6}         &                 & 0.003                                   & 0.008                                  & 0.007                                  \\
		\textit{7}          &                & 0.002                                  & 0.006                                 & 0.005                                  \\
		\textit{8}           &               & 0.002                                   & 0.004                                 & 0.004                                  \\
		\textit{9}            &              & 0.001                                   & 0.003                                 & 0.002                                  \\
		\textit{10}            &             & 0.001                                   & 0.002                                 & 0.001                                  \\
		\textit{11}             &            & 0.001                                     & 0.001                                 & 0.001                                  \\
		\textit{12}              &           & 0                                       & 0.001                                 & 0.000                                  \\
		\textit{13}               &          & 0                                       & 0                                     & 0                  \\ \bottomrule
	\end{tabular}
	
\end{table}

The multiple correspondence analysis is utilized to identify underlying design choices in the classified systems. In particular, the method identifies new latent dimensions, which are a combination of the original attributes of the taxonomy. In Table \ref{tab:mca} these twelve latent dimensions and their contribution to the explained variance in the data after applying Benzceri (optimistic) and Greenacre (pessimistic) corrections are depicted in decreasing order of importance. 
The first four dimensions account for $96.2\%$ of total variation (for the Benzecri correction) and thus are considered significant to explain the variance in the data.

Figure \ref{fig:sankey} depicts how these four dimensions are determined by both, the original attribute values of the taxonomy and the classified 50 systems. The four significant dimensions in the new vector space are in descending order of explained variance:

\begin{itemize}
	\item \textit{Dimension 1}: Illustrates if a system is layered. In particular, if the system uses a native distributed ledger or an external one and thus corresponds to the origin attribute of the taxonomy. 
	\item \textit{Dimension 2}: Illustrates the participation level in a system. In particular, the degree of openness is represented ranging from \textit{permissioned} (e.g. restricted Actor permission) to \textit{permissionless} systems. 
	\item \textit{Dimension 3}: Illustrates the capability to stake, e.g., if the system utilizes a PoS typical layout such as a token providing access to participate in the consensus.
	\item \textit{Dimension 4}: Illustrates the level of cryptoeconomic complexity. The values range from complex (e.g. token interactions) to simple (e.g. tokens not burnable).
\end{itemize} 
The second, third and fourth dimensions are not trivially determined by studying the classified systems visually, as the determining attribute values span over several components. Moreover, the differentiation between permissioned and permissionless systems \cite{wust2018you} \cite{vukolic2017rethinking} and the degree of staking capability \cite{bentov2016cryptocurrencies} \cite{kiayias2017ouroboros} reflect ongoing discussion of the community on the effective design of DLT systems. The actor permission attribute contributes significantly to the construction of the permissionless dimensions, as depicted in Figure \ref{fig:sankey_participation}, and hence this dimension extends the permissionless concept from the consensus to the Action component. Neither Bitcoin nor Ethereum contributes significantly to the construction of the new dimensions, despite being studied the most in academic literature ($588$, respectively $296$ citations in Elsevier's ScienceDirect database). 
This might be due to other systems adopting the design of these well-known DLT systems and hence their design does not contribute significantly to the variance in the data.
Additionally, observing the systems, which contribute the most to the 4\textsuperscript{th} dimension (level of cryptoeconomic complexity), one notices that these are systems, which address a specific domain, respectively address a particular challenge and hence require an elaborated CED: PVIX and Zcoin are privacy chains; Komodo and Bancor use token interactions and a creation condition requiring actions; Factom and Komodo utilize a hybrid system layout consisting of a natively maintained distributed ledger and an external one. 

\begin{figure}[!tb]
	\centering
	
	\subfigure[Dimension 1 and 2]{\includegraphics[width=0.48\columnwidth]{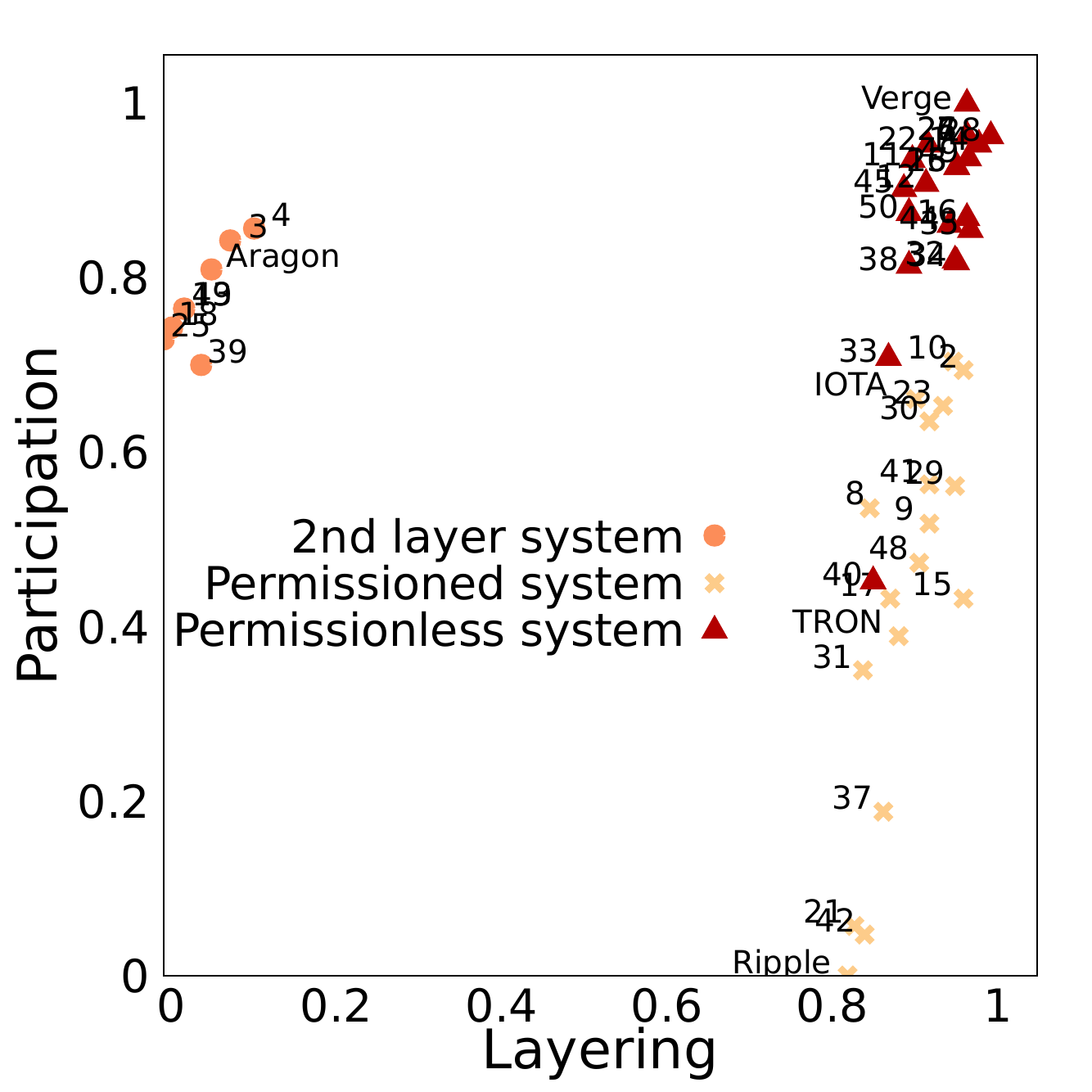}\label{fig:dim_1_2}}
	\subfigure[Dimensions 2 and 3]{
		\includegraphics[width=0.48\columnwidth]{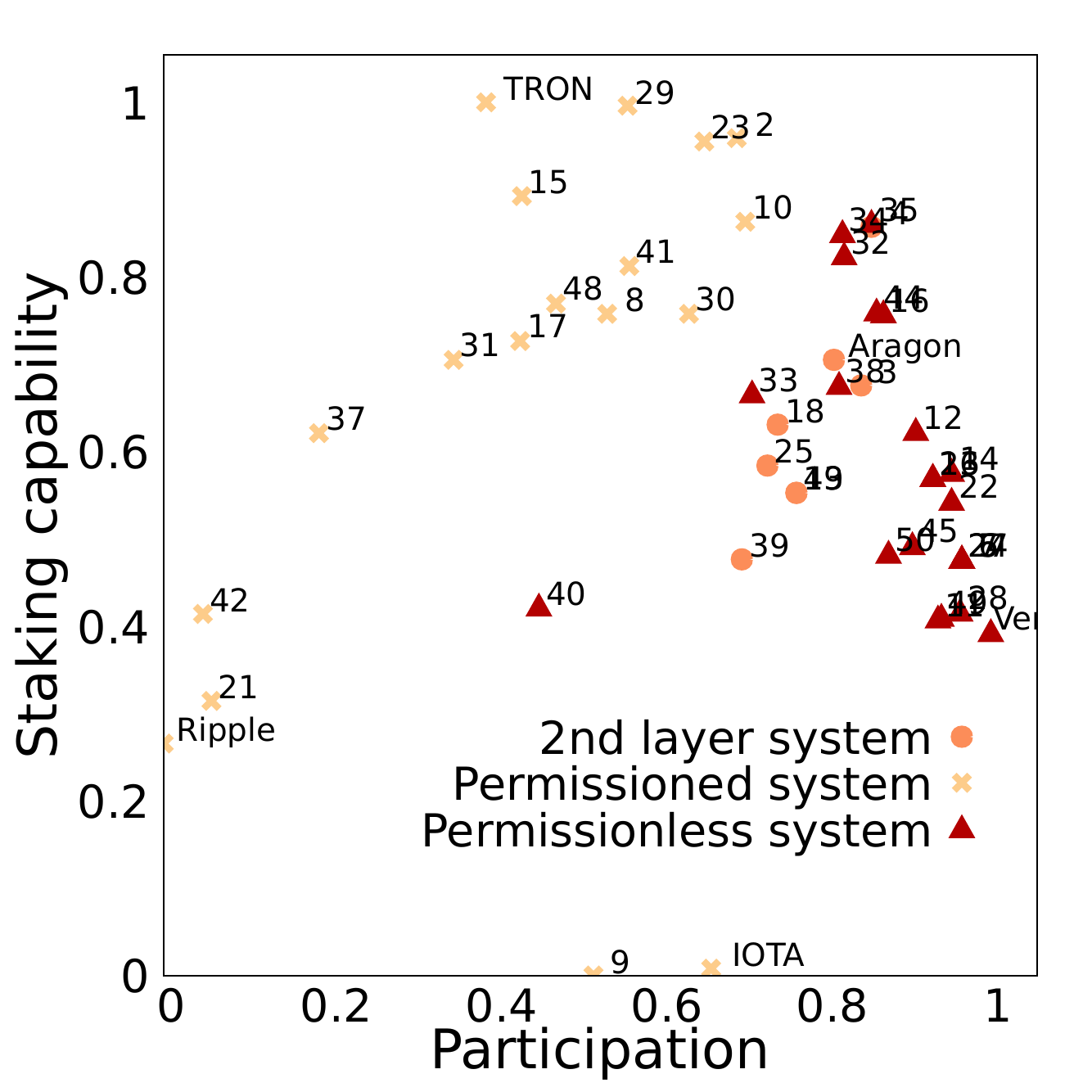}\label{fig:dim_3_4}}
	\caption{DLT systems in the latent dimensions, as identified by MCA. The labels are determined by the k-means clustering algorithm. The translation of the identifiers to DLT systems can be found in Table 1 of the Supplementary Material. Moreover, Figure 3 in the Supplementary Material illustrates other combinations of dimensions.}
	\label{fig:systems_in_new_dimensions}
\end{figure}

Figure \ref{fig:systems_in_new_dimensions} depicts the 50 DLT systems in the new dimensions. A strong clustering of systems can be observed for the first two dimensions (Figure \ref{fig:dim_1_2}), and a weaker for the 2\textsuperscript{nd} and 3\textsuperscript{rd} dimensions (Figure \ref{fig:dim_3_4}), which is explained due to to the lower explained variance in the data by the latter dimension.

\begin{table}[]
	\caption{Bootstrap statistics of identified clusters when applying kmeans with varying $k$ on classification: $k=3$ results in the most stable     clusters}
	\label{tab:bootstrap}
	\centering
	\begin{tabular}{llllllll}
		\toprule
		\multirow{2}{*}{\textbf{k}} & \multirow{2}{*}{\textbf{Boot}} & \multicolumn{6}{c}{\textbf{Cluster}}                                       \\ 
		&                                & \textit{1} & \textit{2} & \textit{3} & \textit{4} & \textit{5} & \textit{6} \\ \midrule
		\multirow{2}{*}{\textbf{2}} & \textit{mean}                  & 0.91       & 0.95       & -          & -          & -          & -          \\
		& \textit{brd}                   & 12         & 1          & -          & -          & -          & -          \\
		\multirow{2}{*}{\textbf{3}} & \textit{mean}                  & 0.96       & 0.97       & 1          & -          & -          & -          \\
		& \textit{brd}                   & 0          & 0          & 0          & -          & -          & -          \\
		\multirow{2}{*}{\textbf{4}} & \textit{mean}                  & 0.75       & 0.91       & 0.99       & 75          & -          & -          \\
		& \textit{brd}                   & 19         & 1          & 1         & 21          & -          & -          \\
		\multirow{2}{*}{\textbf{5}} & \textit{mean}                  & 0.71       & 0.64       & 0.43       & 0.62       & 1          & -          \\
		& \textit{brd}                   & 25         & 32         & 80         & 25         & 0          & -          \\
		\multirow{2}{*}{\textbf{6}} & \textit{mean}                  & 0.82       & 1          & 0.70       & 0.65      & 0.5        & 0.64       \\
		& \textit{brd}                   & 19         & 0          & 23         & 33         & 68         & 44  \\ \midrule      
	\end{tabular}
	
\end{table}

Table \ref{tab:bootstrap} outlines the cluster stability and the number of dissolved clusters when applying k-means for various $k$ on the classified attribute values of the $50$ DLT systems. Comparing the bootmean\footnote{"Highly stable" clusters should yield values of 0.85 and above: https://rdrr.io/cran/fpc/man/clusterboot.html (Accessed: 2020-01-08).} (cluster-wise average Jaccard similarity) and bootbrd (cluster-wise number of times a cluster is dissolved) identifies three clusters as the most stable separation of the classification. This is further validated by the Silhouette and Calinski-Harabasz score, which identify two or three clusters to be optimal, as depicted in Figure 4 of the Supplementary Material.

In Figure \ref{fig:systems_in_new_dimensions}, the DLT systems are labeled based on these clusters. One notices, considering the distribution of labels in Figure \ref{fig:dim_1_2}, that the three clusters can be identified as 2\textsuperscript{nd} layer systems, permissioned systems, and permissionless systems. 
Likewise, utilizing the same labeling in Figure \ref{fig:dim_3_4}, it is noticed that these three clusters form distinct groups: 2\textsuperscript{nd} layer systems being in the center, followed by permissionless and permissioned systems.

\begin{figure}[!t]
	\centering
	\includegraphics[width=0.5\columnwidth]{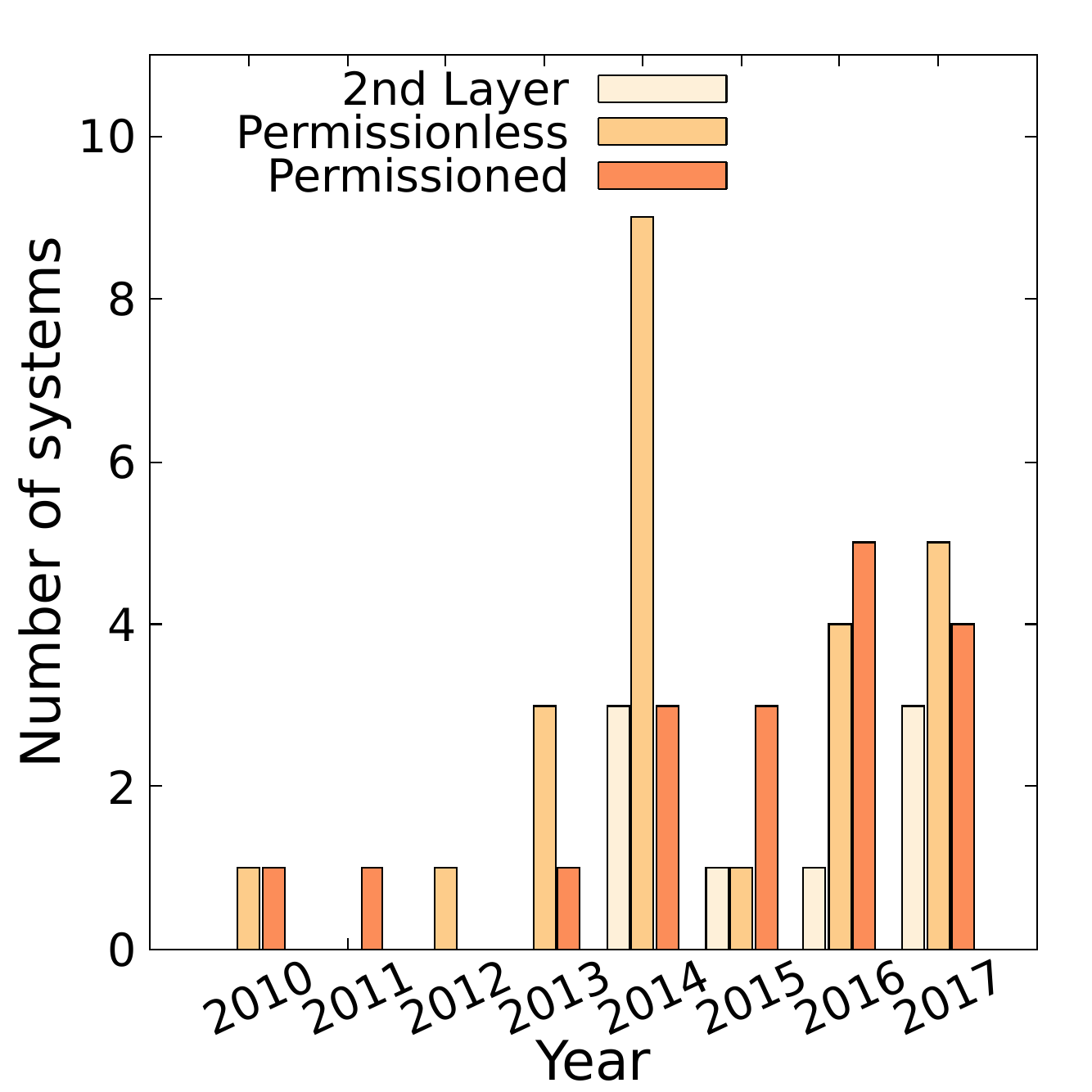}
	\caption{Number of Github repository creations of classified DLT systems for the clusters identified by k-means. 
	}
	\label{fig:number_of_systems_per_year_clusters}
\end{figure}
Figure \ref{fig:number_of_systems_per_year_clusters} depicts the number of new systems per year and cluster. The number of newly introduced systems peaked in 2014, when in total 15 of the 50 systems were introduced. This high number is mainly due to the introduction of permissionless systems. In recent years, the probability of introducing a permissioned or permissionless system is equal, while introducing a 2\textsuperscript{nd} layer system has been lower.

The analysis concludes, that two key design choices in DLT systems are identified method-independently: \textit{layering} and \textit{participation level}.
Moreover, \textit{staking capability} and \textit{cryptoeconomic complexity} are identified by MCA.
The key design choices are not apparent in the taxonomy but are still captured by a combination of attribute values, which is an indication of the rich information the taxonomy can encode and explain. Hence, those findings support the explanatory capacity of the taxonomy as defined in earlier taxonomy theory \cite{nickerson2013method}. 
Moreover, the combination of attribute values into key design choices identified by the analysis limits the system configuration options and as a result reduces modeling complexity of DLT systems at design phase.

\section{Summary of findings: A Design Guideline for Distributed Ledgers}
\label{sec:summary_findings_design_guideline}

\begin{figure*}[!htb]
	\centering
	\includegraphics[width=1.0\textwidth]{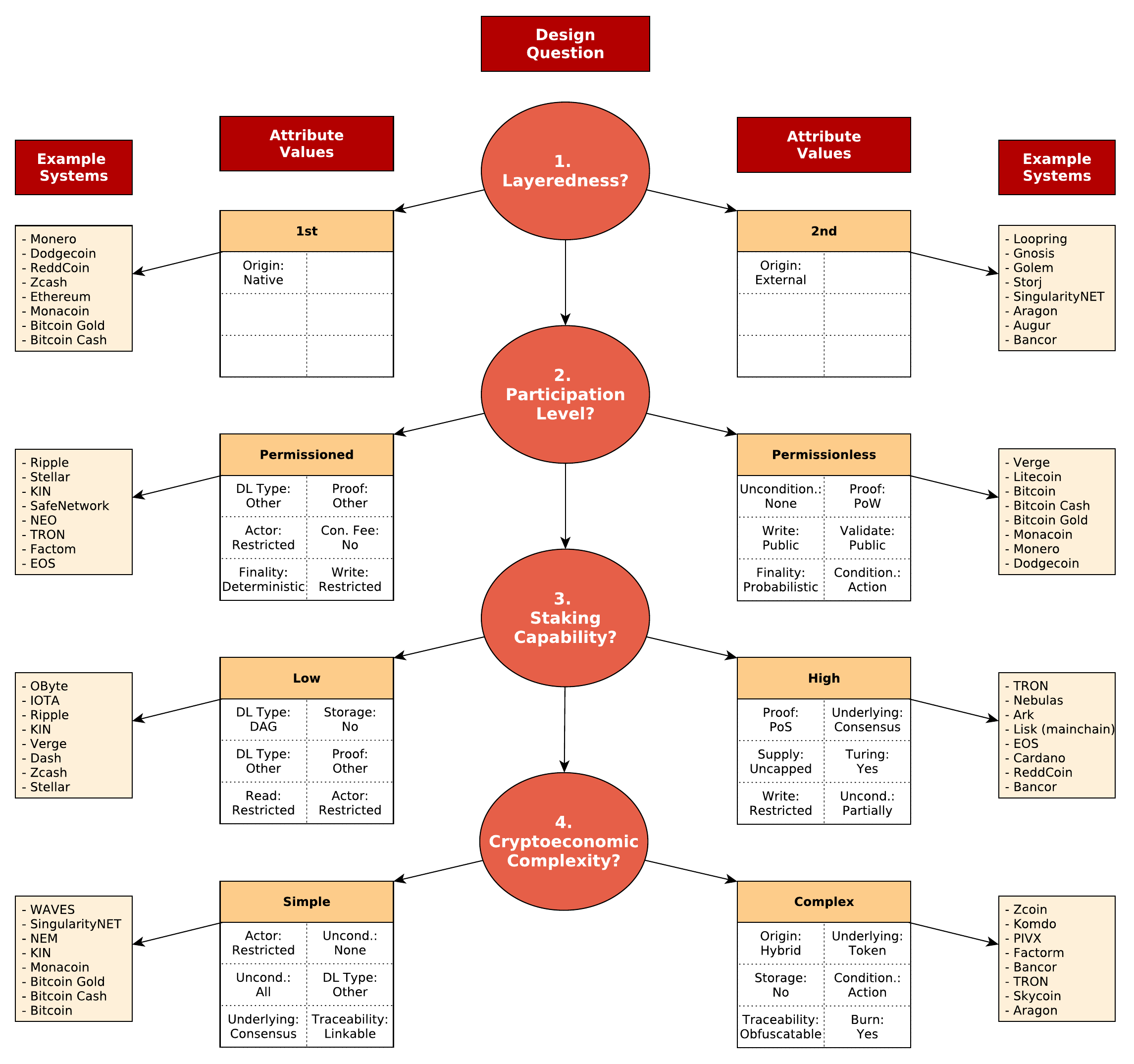}
	\caption{A design guideline of the key design choices in DLT systems, suggesting an order with which a designer may determine system configuration. The questions, attribute values, example systems, and order are a result of analysis conducted using real-world data and machine learning methods (Section \ref{sec:machine_learning_analysis}). For each design decision, based on the MCA analysis, attribute values and the corresponding example systems are stated, which match best the respective design decision.  }\label{fig:design_guideline}        
\end{figure*}

The key findings of the performed experiments are summarized as follows:

\begin{itemize}
	\item The proposed taxonomy is \textit{useful}, as defined in earlier taxonomy literature \cite{nickerson2013method}. In particular, the blockchain community validates the taxonomy as robust and comprehensive (on average $79\%$ expressiveness, Section \ref{sec:community_evaluation}). Moreover, the taxonomy is extensible (Section \ref{sec:community_evaluation}) and explanatory (Section \ref{sec:machine_learning_analysis}), as found by analyzing the blockchain community feedback and applying machine learning methods on the classification.
	\item The classification of $50$ viable and actively maintained DLT systems is accepted by the blockchain community (on average $83.7 \%$ acceptance over all components, Section \ref{sec:eval_class}).
	\item The quantitative analysis of the classification identifies four key design choices that structure the modeling complexity of DLT systems at design phase (Section \ref{sec:machine_learning_analysis}). Each of these choices combines several attribute values and thus reduces the configuration complexity of DLT systems.

\end{itemize}

Based on these findings, a design guideline is derived, which is depicted in Figure \ref{fig:design_guideline}. The key design choices are determined quantitatively by applying machine learning algorithms on empirical data. The order is determined by the level of explained variance, as calculated by MCA. Each question corresponds to a binary design decision. For each decision, the six attribute values, which contribute the most to this design choice are illustrated. 
Moreover, for each choice, the systems, which match best the attribute value configuration are depicted.
The significance of this approach lies in the fact, that the design guideline is derived quantitatively by reasoning based on validated empirical data: the viable and actively maintained DTL systems classified according to the taxonomy. 

The findings demonstrate that the contributions of this paper support system designers to research and design DLT systems: 
The conceptual architecture and taxonomy map the space of possible design configurations and thus assist researchers to position a system in the DLT landscape.
Finally, the quantitatively derived guideline determines which design choices are key for a DLT system. Therefore such a guide can provide a more tailored understanding of the DLT architectural elements and limit the modeling complexity of DLT systems at design phase. The identified key design choices are the ones with the impact of having been derived from existing viable and actively maintained DLT systems.



\section{Conclusion and Future Work} 
\label{sec:conclusion}

This paper concludes that the evolving complexity of distributed ledgers can be better understood via a proposed taxonomy of DLT systems of high \emph{usefulness}~\cite{nickerson2013method}. This is feasible by validating the classification of DLT systems into the taxonomy using wisdom of the crowd and machine learning methods fed with real-world data. Ultimately, data from the classification encode information with which a novel design guideline is derived that identifies key design choices that govern the complexity of distributed ledgers. This guideline can explain and provide new insights for researcher, practitioners and entrepreneurs about which possible design choices have the highest impact, where there is space for innovation and which systems have competitive features or shared functionality.


The results point to various avenues for future research. 
Firstly, the findings of this paper suggest that the taxonomy can be further extended with additional Action and Token attributes.
Also, a component modeling the governance of the systems may become critical in deciding if a system has a decentralized organization (e.g. no trusted party).

Secondly, although the taxonomy represents the current state of viable and actively maintained DLT systems, the proposed methods to evaluate its usefulness are general.
Hence, future research can quantify with the introduced methodology the extent to which the suggested extensions affect the usefulness of the proposed taxonomy.

Thirdly, the initial cluster analysis demonstrates that key design choices can be derived quantitatively by analyzing empirical data of viable and actively maintained DLT systems. This suggests to extend the classification in future work (e.g. with Blockchain-as-a-Service systems) or to apply different statistical methods to the data in order to validate and further identify key design choices.

\section*{Acknowledgment}

This work is supported by the Swiss National Science Foundation (grant no. 170226) for the European FLAG ERA project 'FuturICT 2.0 - Large scale experiments and simulations for the second generation of FuturICT' (https://futurict2.eu/) and the European Community’s H2020 Program under the scheme 'ICT-10-2015 RIA', grant agreement \#688364 'ASSET: Instant Gratification for Collective Awareness \& Sustainable Consumerism' (http://www.asset-consumerism.eu).

In addition, the authors would like to thank Michael Noack, Qusai Jouda and Max Roessner for their support in the questionnaire development and the proofreading of the classification.
Moreover, the authors would like to thank Coinmonks and CoinCodeCap.com for access to their data and code.

%
%
%









\bibliographystyle{unsrt}

\bibliography{refs} 

\end{document}


	%
	\title{Supplementary Material to \\ Decrypting Distributed Ledger Design - \\Taxonomy, Classification and Blockchain Community Evaluation}

	\author{\IEEEauthorblockN{Mark C. Ballandies}
		\IEEEauthorblockA{Computational Social Science \\
			ETH Zurich\\
			Zurich, Switzerland\\ 
			bmark@ethz.ch}
		\and
	\IEEEauthorblockN{Marcus M. Dapp}
		\IEEEauthorblockA{Computational Social Science \\
			ETH Zurich\\
			Zurich, Switzerland\\ 
			mdapp@ethz.ch}
		\and
		\IEEEauthorblockN{Evangelos Pournaras}
		\IEEEauthorblockA{School of Computing\\
			University of Leeds\\
			Leeds LS2 9JT UK\\ 
			e.pournaras@leeds.ac.uk}}

	
	%


	\maketitle

	\IEEEpeerreviewmaketitle

\subsection{Survey Participants Invitation Email}

%
%

\label{sec:invitation_mail}
\begin{figure}[!htb]
	\centering
	
	\subfigure{\includegraphics[width=0.99\columnwidth]{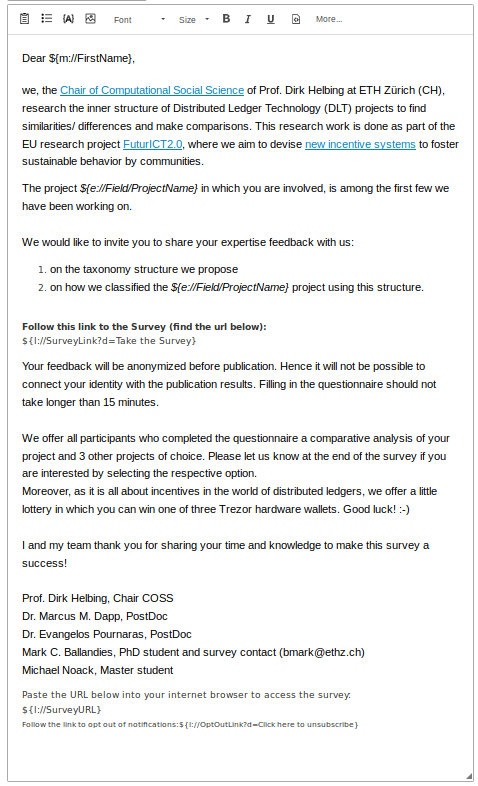}}
	
	\caption{Survey participants invitation email.}
	\label{fig:invitation_mail}
\end{figure}

Figure \ref{fig:invitation_mail} depicts the email inviting participants to the survey.

\subsection{Incorporating Blockchain Community Feedback}\label{sec:appendix_adjustments}

\begin{figure}[!tb]
	\centering
	
	\subfigure[Classification (N=24)]{\includegraphics[width=0.48\columnwidth]{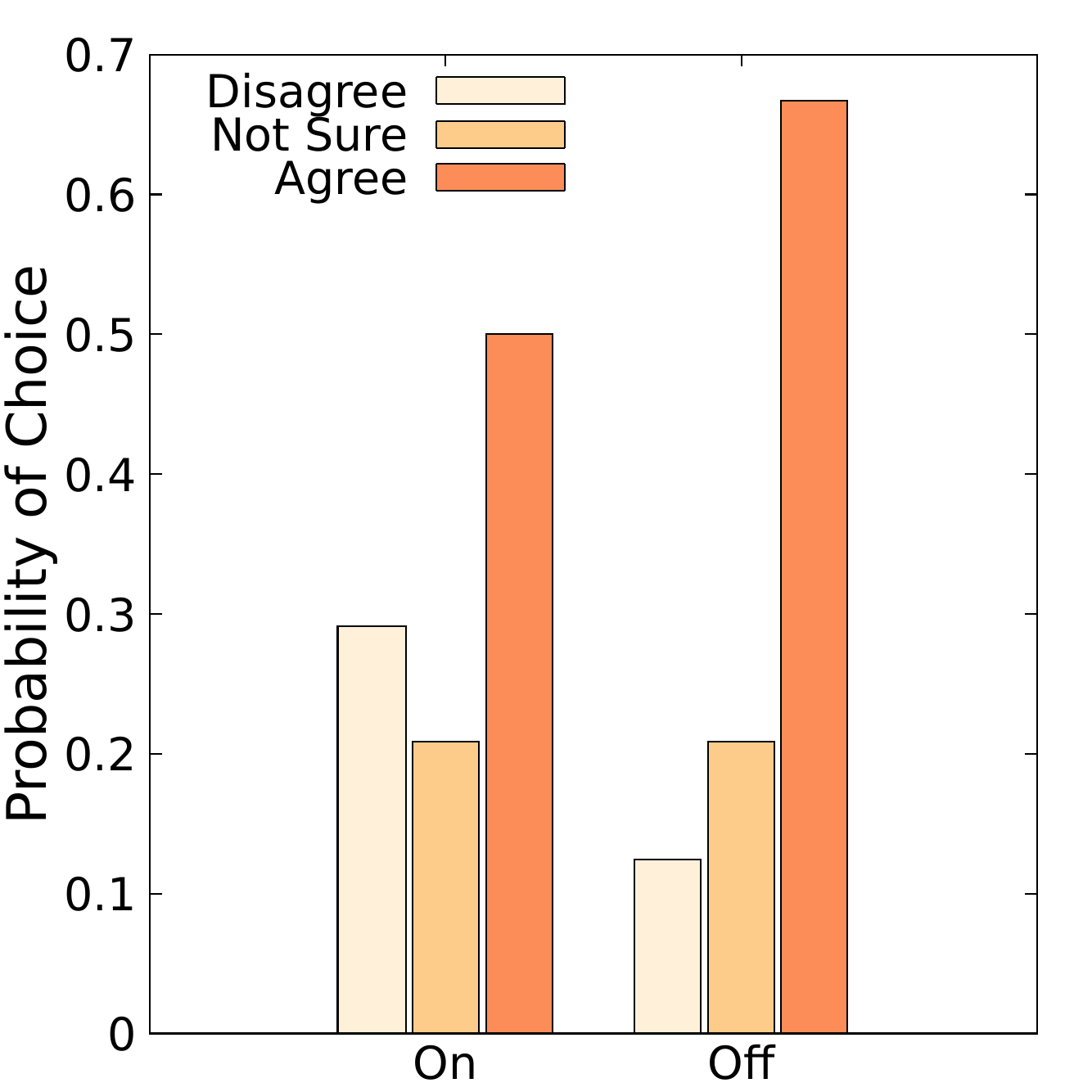}\label{fig:consistency_class_under}}
	\subfigure[Expressiveness (N=24)]{
		\includegraphics[width=0.48\columnwidth]{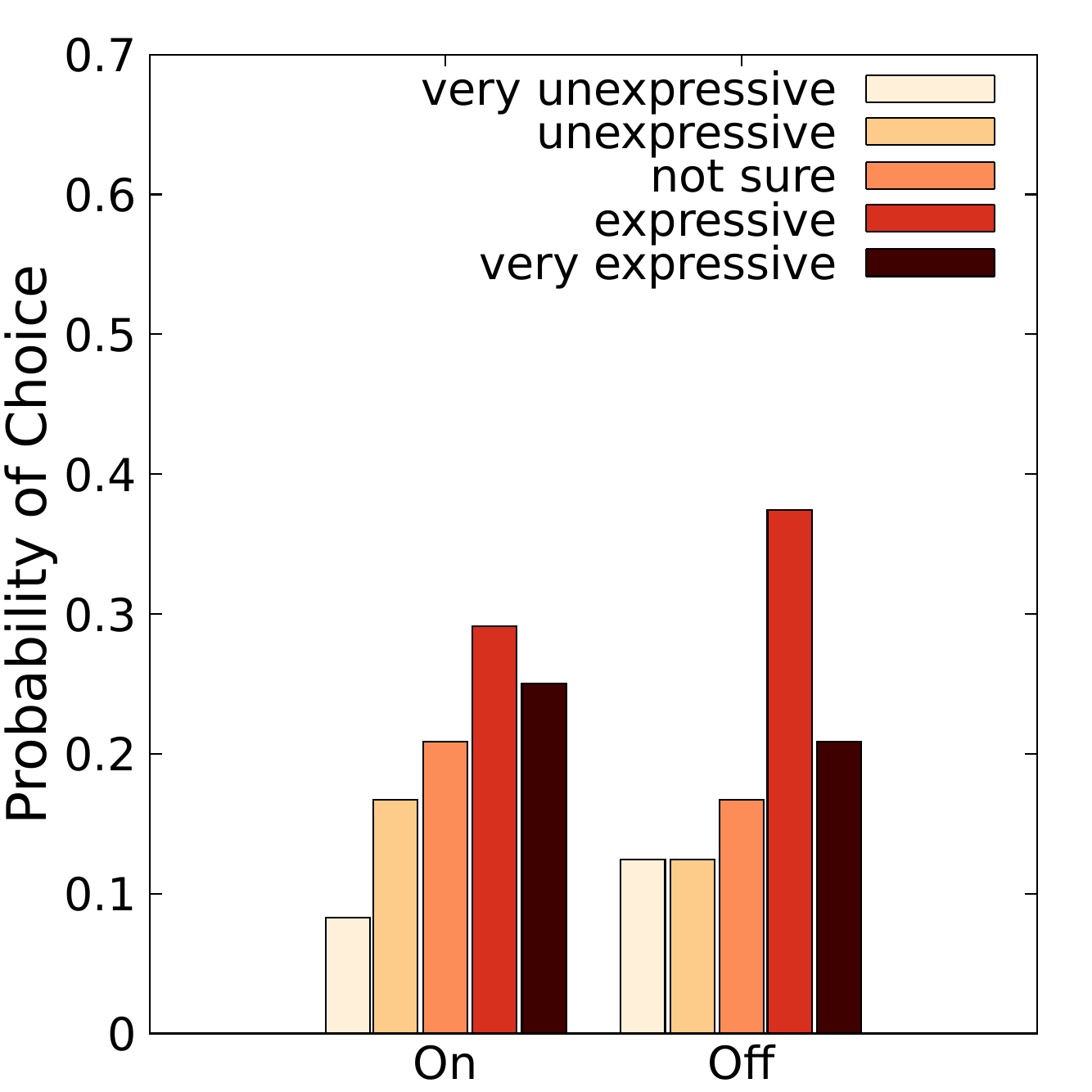}\label{fig:consistency_tax_under}}
	\caption{Acceptance level of the classification and expressiveness of the On-chain and Off-chain underlying as perceived by survey participants.}
	\label{fig:on_off_underlying_under}
\end{figure}


In the initial taxonomy and classification in the first phase of the survey, the underlying attribute was replaced with the on-chain underlying and off-chain underlying attributes to illustrate in finer detail the source of value of a token. In particular, in this version the underlying specifies the concrete object of value (e.g. storage or computation) and where it resides (on-chain or off-chain). Nevertheless, these attributes obtained the lowest acceptance level of all attributes (Figure \ref{fig:on_off_underlying_under}). Detailed feedback regarding the attributes is summarized in the following and illustrated in the greater detail in Section \ref{sec:detailed_feedback}. Namely, the main areas of disagreement or uncertainty are: 
\begin{enumerate}
	\item Not distinguishing between on-chain and off-chain underlying
	\item Not distinguishing the underlying from services which can be assessed in the DLT system
	\item Mixing up the underlying with the possibility to use a token as a currency
	\item Considering future or past features of the DLT systems
	\item Rejecting on-chain storage of hashes as storage
\end{enumerate}
Based on this feedback, three changes have been made to the taxonomy.

First, in order to reduce the ambiguity regarding the difference between on-chain underlying, off-chain underlying and other services in the system (and hence addressing point one and two of the previous
summary), the token underlying has been more clearly mapped to the
components of the conceptual framework (Figure 1 in Paper).
To achieve this, the off-chain and on-chain underlyings have been merged into one attribute and their values have been abstracted
to include values from the framework where the underlying resides, as described in Section IV-D-6 of the paper.
For instance, in the case of the Ethereum token, instead of expressing
that the token grants access to the on-chain underlying \textit{computation},
the token is now said to provide access to the \textit{distributed
	ledger}, which in turn implies granting access to computation, because Ethereum has a Turing complete distributed ledger.

Second, to emphasize the option to store arbitrary
data on distributed ledgers (e.g hashes), which, for instance, enables
Bitcoin to function as the overarching infrastructure for other systems,
the storage attribute has been added to the Distributed Ledger component.
This addresses point number five.

Third, in order to address point number three, the transferability
attribute has been added to the Token component, which emphasizes
the possibility to use the token as a currency. 

Finally, as the scope of the classification is to capture the
current state of DLT systems and not future possible
extensions or past configurations, point number four is not addressed.

The new transferability attribute is considered as the most expressive amongst all attributes by the community (Figure 6 in Paper). Moreover, the restructuring of the underlying attribute and the inclusion of the storage attribute improves the underlying assessment by the community significantly. These findings justify the modifications to the taxonomy.
Also, the inclusion of new attributes into the taxonomy indicates that the taxonomy is extensible as defined in earlier taxonomy theory \cite{nickerson2013method}.

\subsection{Detailed blockchain community feedback on previous underlying attributes}
\label{sec:detailed_feedback}
After the first two-month phase in which participants obtained the survey, the feedback has been analyzed which lead to changes in the taxonomy, as summarized in Section B. 
The detailed feedback is illustrated in the following.

When examining the comments provided by the participants who disagree
with the classification of the on-chain underlying attribute\footnote{In brackets are depicted the percentage for which this responds type
	accounts for the overall disagreements. Please note, that the percentages
	do not add up to $100\%$ as a survey participant could state several
	reasons for disagreement}, it is noteworthy that these respondents do not regard the on-chain
storage of hashes as storage ($22.2\%$). In some cases, respondents make contradictory comments about the on-chain value of their system
token ($11.1\%$). Some participants mix up the on-chain underlying
of the token with the overall services that the DLT system provides
($22.2\%$), which do not necessarily require to be accessed via the
token. Some participants disagree with the classification because it does not consider plans to implement on-chain underlyings in the
future ($33.3\%$). 
Finally, some mix up the on-chain
underlying with the off-chain underlying ($11.1\%$).

Among respondents who express uncertainty, some do not distinguish
between their current implementation of on-chain underlyings and possible
or planned future implementations ($50.0\%$), which are not in the scope of the classification. Other respondents mix up the possibility
to use a token as a currency with its underlying ($25.0\%$). Some
state that the question is not formulated clearly enough ($25.0\%$).

In the case of the off-chain underlying, a similar picture can be
observed. In the following, the responses expressing disagreement and uncertainty
are combined. Some respondents disagree with the classification because
it does not consider past and future off-chain underlyings ($30.0\%$).
Some understood the off-chain underlying to be an exclusive right
conferred by the token ($10.0\%$). Moreover, as in the case of the
on-chain underlying, some participants link the possibility to use
the token as a currency to its underlying ($20.0\%)$. Some participants
mix up on-chain and off-chain value ($10.0\%$), others do not
understand the question ($10.0\%$) or do not respond ($20.0\%$).

\begin{figure*}[!htb]
	\centering
	\subfigure[Dimension 1 and 3 ]{\includegraphics[width=0.24\textwidth]{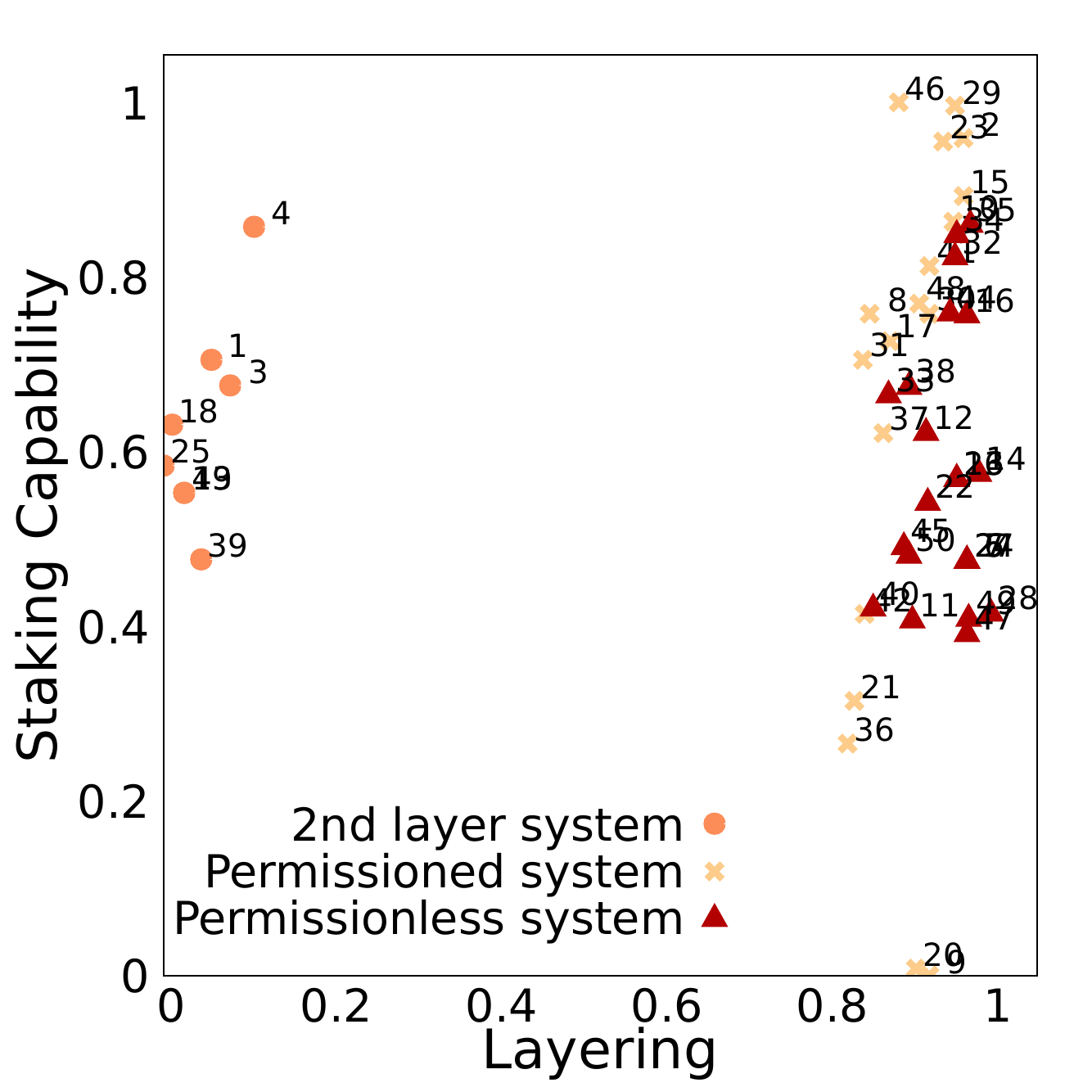}}
	\subfigure[Dimension 1 and 4]{\includegraphics[width=0.24\textwidth]{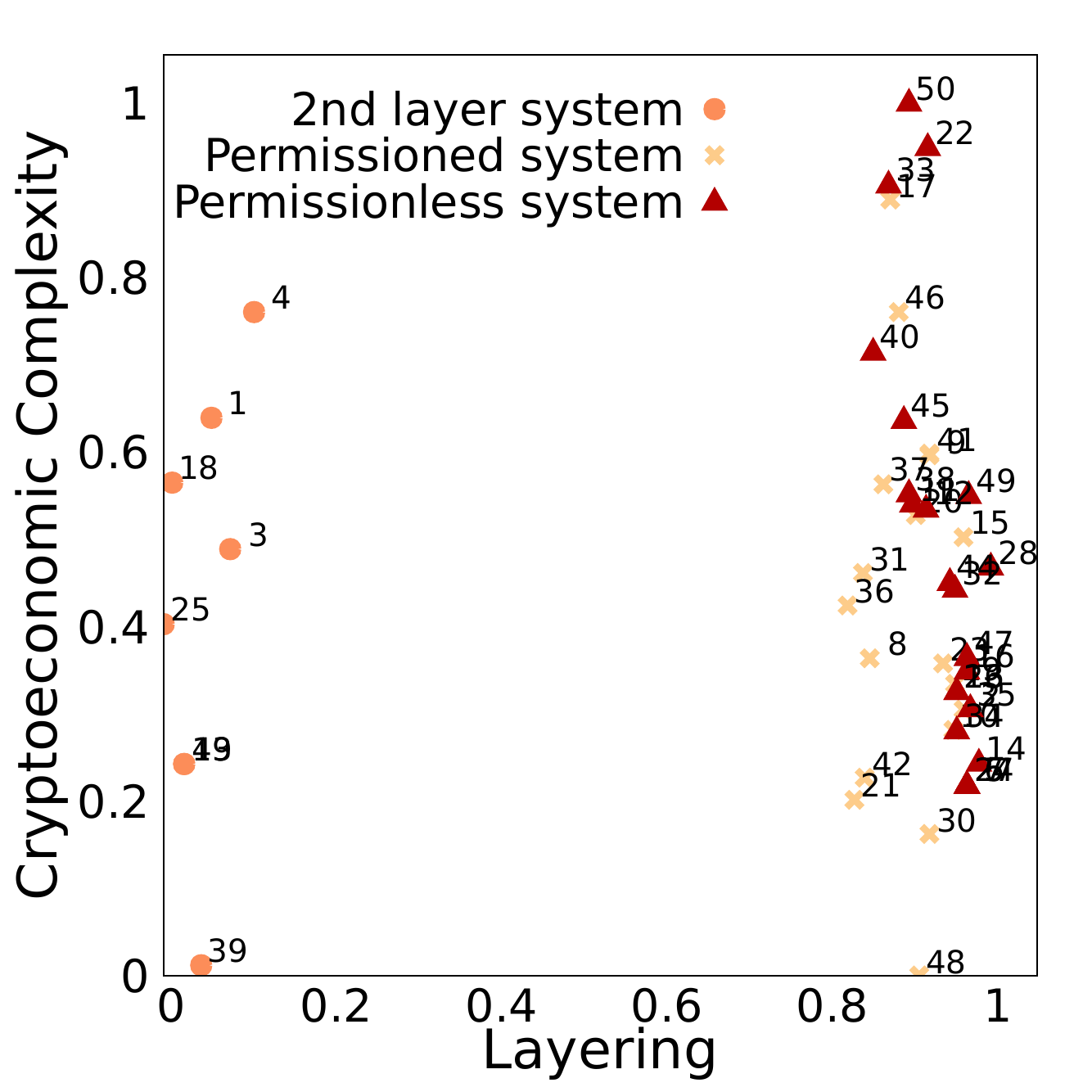}}\label{fig:factor_14_sup}
	\subfigure[Dimensions 2 and 4]{\includegraphics[width=0.24\textwidth]{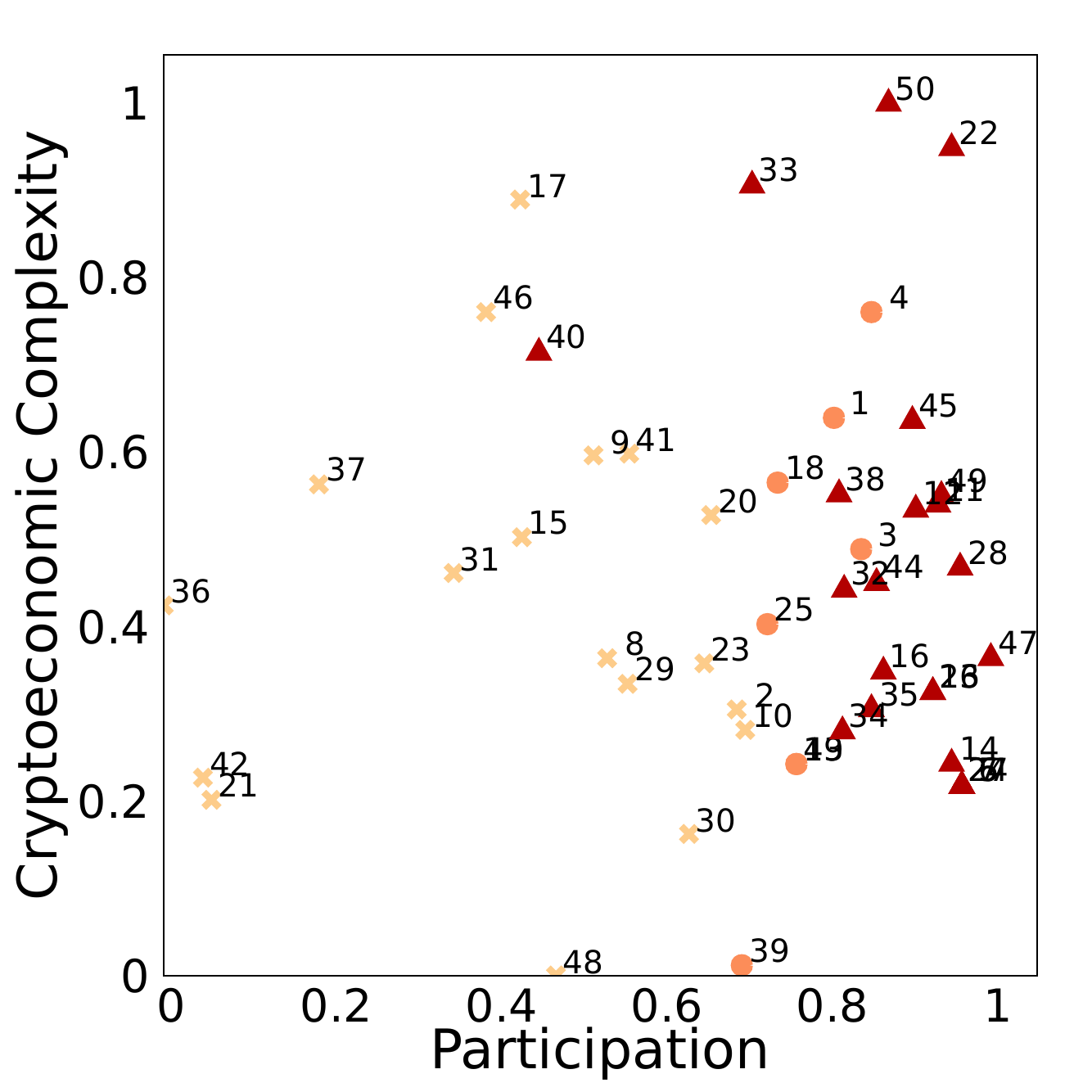}}
	\subfigure[Dimension 3 and 4]{\includegraphics[width=0.24\textwidth]{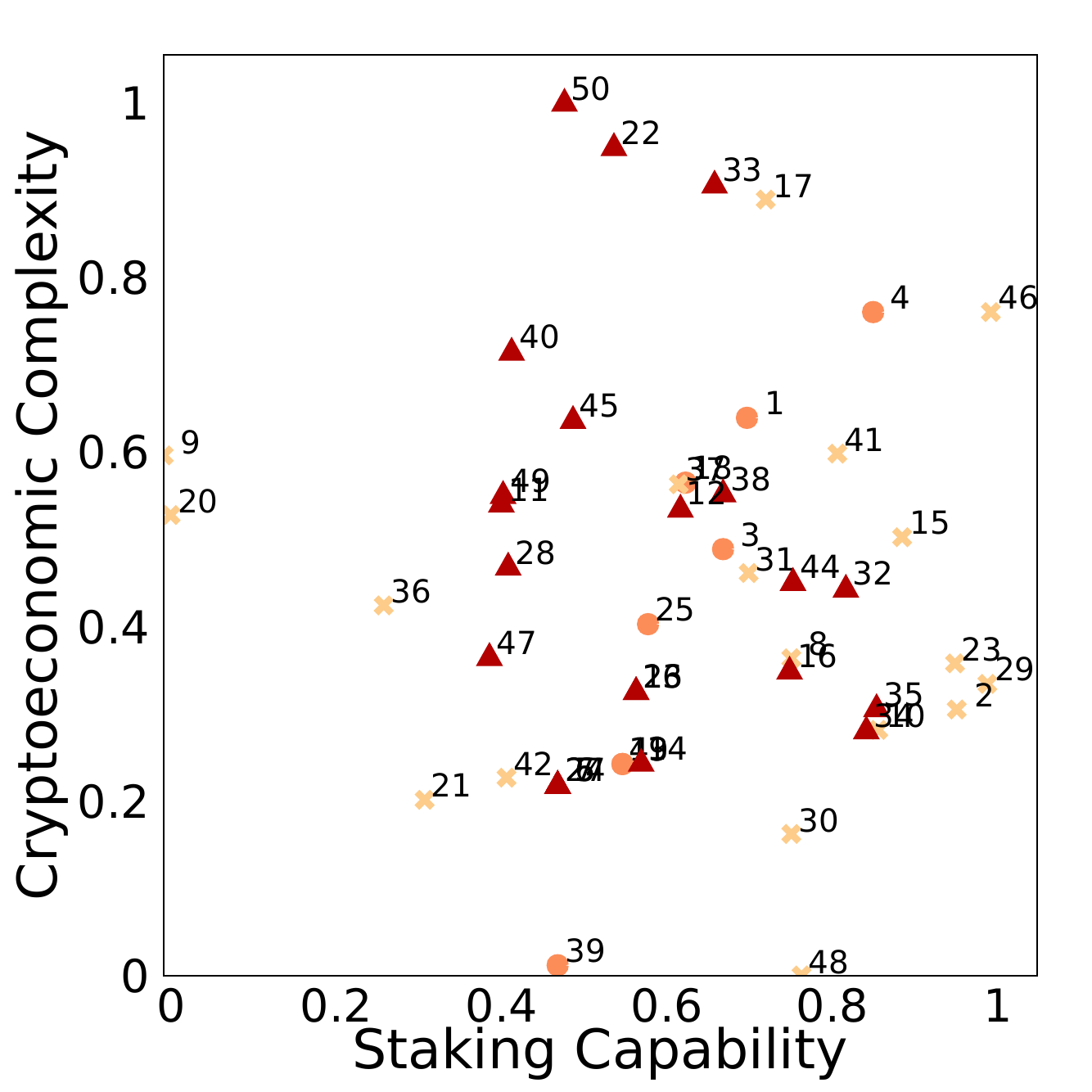}}
	\caption{DLT systems in the latent dimensions, as identified by MCA. The labels are determined by the k-means clustering algorithm. The translation of the identifiers to DLT systems can be found in the Appendix of the paper.}
	\label{fig:systems_in_new_dimensions_sup}
\end{figure*}

\subsection{Machine Learning Analysis}

Table \ref{tab:mca_systems} depicts the classified DLT systems in the new latent dimensions as identified by MCA (Section VI-B in Paper). The cluster associations are identified by k-means (Section VI-B in Paper).

Figure \ref{fig:systems_in_new_dimensions_sup} illustrate the classified DLT systems in various combinations of the latent Dimensions, as identified by MCA (Section VI-B in Paper).

\begin{figure}[!htb]
	\centering
	
	\subfigure[Silhouette]{\includegraphics[width=0.48\columnwidth]{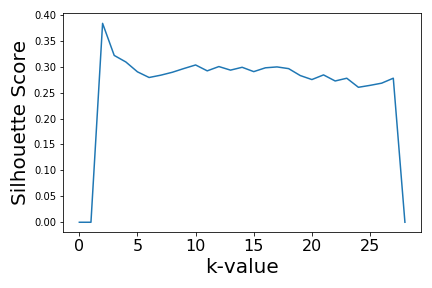}\label{fig:silhoutte}}
	\subfigure[Calinski-Harabasz]{
		\includegraphics[width=0.48\columnwidth]{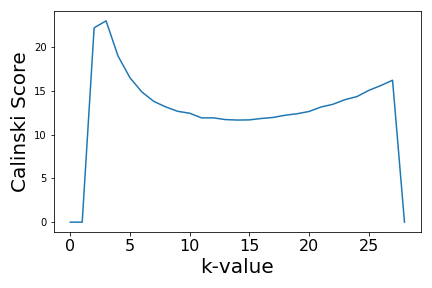}\label{fig:calinski}}
	\caption{Silhouette and Calinski-Harabasz score when applying kmeans for varying $k$ on the original classification.}
	\label{fig:k_means_validity}
\end{figure}

Figure \ref{fig:k_means_validity} illustrates the Silhouette and Calinski-Harabasz scores when applying k-means for varying k on the classified systems (Section VI-B in Paper).

\begin{table}[]
	\caption{DLT systems in the latent dimensions as identified by MCA. The clusters are identified by k-means.}
	\label{tab:mca_systems}
	\resizebox{0.5\textwidth}{!}{
		\begin{tabular}{lllllll} \toprule
			\textbf{id} & \textbf{DLT system} & \textbf{Cluster} & \textbf{Layer} & \textbf{Participation} & \textbf{Stakeability} & \textbf{Complexity} \\ \midrule
			1           & Aragon              & 1                & 0.06           & 0.81                   & 0.71                  & 0.64                \\
			2           & Ark                 & 0                & 0.97           & 0.69                   & 0.96                  & 0.31                \\
			3           & Augur               & 1                & 0.08           & 0.84                   & 0.68                  & 0.49                \\
			4           & Bancor              & 1                & 0.11           & 0.86                   & 0.86                  & 0.76                \\
			5           & Bitcoin             & 2                & 0.97           & 0.96                   & 0.48                  & 0.22                \\
			6           & Bitcoin Cash        & 2                & 0.97           & 0.96                   & 0.48                  & 0.22                \\
			7           & Bitcoin Gold        & 2                & 0.97           & 0.96                   & 0.48                  & 0.22                \\
			8           & BitShares           & 0                & 0.85           & 0.54                   & 0.76                  & 0.36                \\
			9           & Byteball            & 0                & 0.93           & 0.52                   & 0                     & 0.6                 \\
			10          & Cardano             & 0                & 0.95           & 0.7                    & 0.86                  & 0.28                \\
			11          & Dash                & 2                & 0.91           & 0.94                   & 0.41                  & 0.54                \\
			12          & Decred              & 2                & 0.92           & 0.91                   & 0.62                  & 0.54                \\
			13          & DigiByte            & 2                & 0.96           & 0.93                   & 0.57                  & 0.33                \\
			14          & Dogecoin            & 2                & 0.99           & 0.95                   & 0.58                  & 0.24                \\
			15          & EOS                 & 0                & 0.97           & 0.43                   & 0.89                  & 0.5                 \\
			16          & Ethereum            & 2                & 0.97           & 0.87                   & 0.76                  & 0.35                \\
			17          & Factom              & 0                & 0.88           & 0.43                   & 0.73                  & 0.89                \\
			18          & Gnosis              & 1                & 0.01           & 0.74                   & 0.63                  & 0.57                \\
			19          & Golem               & 1                & 0.03           & 0.76                   & 0.55                  & 0.24                \\
			20          & IOTA                & 0                & 0.91           & 0.66                   & 0.01                  & 0.53                \\
			21          & KIN                 & 0                & 0.83           & 0.06                   & 0.31                  & 0.2                 \\
			22          & Komodo              & 2                & 0.92           & 0.95                   & 0.54                  & 0.95                \\
			23          & Lisk-mainchain      & 0                & 0.94           & 0.65                   & 0.96                  & 0.36                \\
			24          & Litecoin            & 2                & 0.97           & 0.96                   & 0.48                  & 0.22                \\
			25          & Loopring            & 1                & 0              & 0.73                   & 0.59                  & 0.4                 \\
			26          & MOAC-MotherChain    & 2                & 0.96           & 0.93                   & 0.57                  & 0.33                \\
			27          & Monacoin            & 2                & 0.97           & 0.96                   & 0.48                  & 0.22                \\
			28          & Monero              & 2                & 1              & 0.96                   & 0.42                  & 0.47                \\
			29          & Nebulas             & 0                & 0.96           & 0.56                   & 1                     & 0.34                \\
			30          & NEM                 & 0                & 0.93           & 0.63                   & 0.76                  & 0.16                \\
			31          & NEO                 & 0                & 0.84           & 0.35                   & 0.71                  & 0.46                \\
			32          & Nexus               & 2                & 0.96           & 0.82                   & 0.82                  & 0.44                \\
			33          & PIVX                & 2                & 0.88           & 0.71                   & 0.67                  & 0.91                \\
			34          & Qtum                & 2                & 0.96           & 0.82                   & 0.85                  & 0.28                \\
			35          & ReddCoin            & 2                & 0.97           & 0.86                   & 0.86                  & 0.31                \\
			36          & Ripple              & 0                & 0.83           & 0                      & 0.27                  & 0.42                \\
			37          & SafeNetwork         & 0                & 0.87           & 0.19                   & 0.62                  & 0.56                \\
			38          & Siacoin             & 2                & 0.9            & 0.82                   & 0.68                  & 0.55                \\
			39          & SingularityNET      & 1                & 0.04           & 0.7                    & 0.48                  & 0.01                \\
			40          & Skycoin             & 2                & 0.86           & 0.45                   & 0.42                  & 0.72                \\
			41          & Steem               & 0                & 0.93           & 0.56                   & 0.81                  & 0.6                 \\
			42          & Stellar             & 0                & 0.85           & 0.05                   & 0.41                  & 0.23                \\
			43          & Storj               & 1                & 0.03           & 0.76                   & 0.55                  & 0.24                \\
			44          & Stratis             & 2                & 0.95           & 0.86                   & 0.76                  & 0.45                \\
			45          & Syscoin             & 2                & 0.89           & 0.9                    & 0.49                  & 0.64                \\
			46          & TRON                & 0                & 0.89           & 0.39                   & 1                     & 0.76                \\
			47          & Verge               & 2                & 0.97           & 1                      & 0.39                  & 0.37                \\
			48          & Waves               & 0                & 0.91           & 0.47                   & 0.77                  & 0                   \\
			49          & Zcash               & 2                & 0.97           & 0.94                   & 0.41                  & 0.55                \\
			50          & Zcoin               & 2                & 0.9            & 0.88                   & 0.48                  & 1          \\ \bottomrule        
		\end{tabular}
	}
\end{table}

\subsection{Cryptoeconomic Reasoning using Boolean Algebra}
\label{sec:terminology_def}

\begin{table}[!tb]
	\caption{Formal definitions of cryptoeconomic terms by reasoning using Boolean Algebra and the proposed taxonomy.}
	\resizebox{0.5\textwidth}{!}{
		\label{tab:term_def}
		\centering
		\begin{tabular}{ll}
			\toprule
			\textbf{System Term}  & \textbf{Formal Definition}                                              \\ \midrule
			Blockchain                 & \textit{blockchain} type\\
			1\textsuperscript{st} layer                & \textit{native} ownership                   \\
			2\textsuperscript{nd} layer                & \textit{external} ownership                 \\
			Permissioned             & \textit{restricted} write permission OR     \\
			& \textit{restricted} validator permission    \\
			Permissionless           & \textit{public} write permission AND        \\
			& \textit{public} validator permission        \\
			Public                   & \textit{$\langle$permissionless DLT system$\rangle$} AND      \\
			& \textit{public} actor permission            \\
			Private                  & \textit{$\langle$permissioned DLT system$\rangle$} AND        \\
			& \textit{restricted} actor permission      \\
			Privacy                  & \textit{obfuscatable} traceability AND            \\
			& \textit{$\langle$public DLT System$\rangle$} \\
			Infrastructure                  & \textit{yes} turing completeness OR            \\
			& \textit{yes} storage \\
			Blockchain-as-a-Service         & $\nexists$ Action component attribute AND         \\  
			& $\nexists$ Token component attribute            \\ 
			\midrule
			\textbf{Cryptoeconomic Term}    \\ \midrule 
			Utility token                  & \textit{distributed ledger} underlying OR                  \\
			& \textit{action} underlying           \\
			Asset token                  & \textit{token} underlying OR                \\
			& \textit{action} (physical good) underlying         \\
			Payment token                 & \textit{yes} transferability            \\ \bottomrule            
		\end{tabular}
	}
\end{table}

The taxonomy introduced in Section IV of the paper allows a number of widely used terms in the field of DLT systems to be more systematically defined by combining the values of specific attributes with
operators from Boolean algebra. As demonstrated in Table \ref{tab:term_def}, which features an illustrative subset of these terms,
this method enables the delineation of terms such as permissioned/ permissionless
blockchains, as well as asset/utility tokens.
In particular, the latter pair has been identified by the Swiss Financial
Market Supervisory Authority (FINMA) as important for determining whether a token should be classified as a security token. This is of interest
to market participants because it has regulatory implications \cite{finma_finma_nodate}.

\subsection{Classification}
Tables \ref{tab:classification_dl}-\ref{tab:classification_token} illustrate the classification of the 50 DLT systems in the four components.

\begin{table*}[!htb]
	\centering
	
	\caption{Systems Classification according to the Distributed-Ledger component}
	
	\label{tab:classification_dl}
	\begin{tabular}{lllllll} \toprule
		\textbf{ID} & \textbf{DLT system} & \textbf{Origin}                & \textbf{Type} & \textbf{\begin{tabular}[c]{@{}l@{}}Address \\ Traceability\end{tabular}} & \textbf{\begin{tabular}[c]{@{}l@{}}Turing\\ Completeness\end{tabular}} & \textbf{Storage} \\ \midrule
		1           & Aragon              & External (Ethereum)            & -             & -                                                                        & -                                                                      & -                \\
		2           & Ark                 & Native                         & Blockchain    & Linkable                                                                 & No                                                                     & Yes              \\
		3           & Augur               &          External (Ethereum)                      & -             & -                                                                        & -                                                                      & -                \\
		4           & Bancor              & External (Ethereum and EOS)    & -             & -                                                                        & -                                                                      & -                \\
		5           & Bitcoin             & Native                         & Blockchain    & Linkable                                                                 & No                                                                     & Yes              \\
		6           & Bitcoin Cash        & Native                         & Blockchain    & Linkable                                                                 & No                                                                     & Yes              \\
		7           & Bitcoin Gold        & Native                         & Blockchain    & Linkable                                                                 & No                                                                     & Yes              \\
		8           & BitShares           & Native                         & Blockchain    & Linkable                                                                 & No                                                                     & Yes              \\
		9           & Byteball            & Native                         & DAG           & Obfuscatable                                                             & No                                                                     & Yes              \\
		10          & Cardano             & Native                         & Blockchain    & Linkable                                                                 & No                                                                     & Yes              \\
		11          & Dash                & Native                         & Blockchain    & Obfuscatable                                                             & No                                                                     & Yes              \\
		12          & Decred              & Native                         & Blockchain    & Linkable                                                                 & No                                                                     & No               \\
		13          & DigiByte            & Native                         & Blockchain    & Linkable                                                                 & No                                                                     & Yes              \\
		14          & Dogecoin            & Native                         & Blockchain    & Linkable                                                                 & No                                                                     & Yes              \\
		15          & EOS                 & Native                         & Blockchain    & Linkable                                                                 & Yes                                                                    & Yes              \\
		16          & Ethereum            & Native                         & Blockchain    & Linkable                                                                 & Yes                                                                    & Yes              \\
		17          & Factom              & Hybrid (Bitcoin)               & Blockchain    & Linkable                                                                 & No                                                                     & Yes              \\
		18          & Gnosis              & External (Ethereum)            & -             & -                                                                        & -                                                                      & -                \\
		19          & Golem               & External (Ethereum)            & -             & -                                                                        & -                                                                      & -                \\
		20          & IOTA                & Native                         & DAG           & Linkable                                                                 & No                                                                     & No               \\
		21          & KIN                 & Native                         & Other         & Linkable                                                                 & No                                                                     & Yes              \\
		22          & Komodo              & Hybrid (Bitcoin)               & Blockchain    & Obfuscatable                                                             & No                                                                     & Yes              \\
		23          & Lisk (Mainchain)    & Native                         & Blockchain    & Linkable                                                                 & No                                                                     & Yes              \\
		24          & Litecoin            & Native                         & Blockchain    & Linkable                                                                 & No                                                                     & Yes              \\
		25          & Loopring            & \begin{tabular}[c]{@{}l@{}}External\\ (Ethereum, Qtum, NEO)\end{tabular}  & -             & -                                                                        & -                                                                      & -                \\
		26          & MOAC (MotherChain)  & Native                         & Blockchain    & Linkable                                                                 & No                                                                     & Yes              \\
		27          & Monacoin            & Native                         & Blockchain    & Linkable                                                                 & No                                                                     & Yes              \\
		28          & Monero              & Native                         & Blockchain    & Obfuscatable                                                             & No                                                                     & Yes              \\
		29          & Nebulas             & Native                         & Blockchain    & Linkable                                                                 & Yes                                                                    & Yes              \\
		30          & NEM                 & Native                         & Blockchain    & Linkable                                                                 & No                                                                     & Yes              \\
		31          & NEO                 & Native                         & Blockchain    & Linkable                                                                 & Yes                                                                    & Yes              \\
		32          & Nexus               & Native                         & Blockchain    & Linkable                                                                 & No                                                                     & Yes              \\
		33          & PIVX                & Native                         & Blockchain    & Obfuscatable                                                             & No                                                                     & Yes              \\
		34          & Qtum                & Native                         & Blockchain    & Linkable                                                                 & Yes                                                                    & Yes              \\
		35          & ReddCoin            & Native                         & Blockchain    & Linkable                                                                 & No                                                                     & Yes              \\
		36          & Ripple              & Native                         & Other         & Linkable                                                                 & No                                                                     & Yes              \\
		37          & SafeNetwork         & Native                         & Other         & Linkable                                                                 & Yes                                                                    & Yes              \\
		38          & Siacoin             & Native                         & Blockchain    & Linkable                                                                 & No                                                                     & Yes              \\
		39          & SingularityNET      & External (Ethereum)            & -             & -                                                                        & -                                                                      & -                \\
		40          & Skycoin             & Native                         & Blockchain    & Obfuscatable                                                             & Yes                                                                    & Yes              \\
		41          & Steem               & Native                         & Blockchain    & Linkable                                                                 & No                                                                     & Yes              \\
		42          & Stellar             & Native                         & Other         & Linkable                                                                 & No                                                                     & Yes              \\
		43          & Storj               & External (Ethereum)            & -             & -                                                                        & -                                                                      & -                \\
		44          & Stratis             & Native                         & Blockchain    & Obfuscatable                                                             & No                                                                     & Yes              \\
		45          & Syscoin             & Native                         & Blockchain    & Obfuscatable                                                             & No                                                                     & Yes              \\
		46          & TRON                & Native                         & Blockchain    & Linkable                                                                 & Yes                                                                    & Yes              \\
		47          & Verge               & Native                         & Blockchain    & Obfuscatable                                                             & No                                                                     & Yes              \\
		48          & Waves               & Native                         & Blockchain    & Linkable                                                                 & Yes                                                                    & Yes              \\
		49          & Zcash               & Native                         & Blockchain    & Obfuscatable                                                             & No                                                                     & Yes              \\
		50          & Zcoin               & Native                         & Blockchain    & Obfuscatable                                                             & No                                                                     & No     \\ \bottomrule          
	\end{tabular}
	
\end{table*}

\begin{table*}[!htb]
	\centering
	
	\caption{Systems Classification according to the Consensus component}
	
	\label{tab:class_consensus}
	\begin{tabular}{lllllll} \toprule
		\textbf{ID} & \textbf{DLT system} & \textbf{Finality} & \textbf{Proof} & \textbf{\begin{tabular}[c]{@{}l@{}}Write\\ Permission\end{tabular}} & \textbf{\begin{tabular}[c]{@{}l@{}}Validate\\ Permission\end{tabular}} & \textbf{Fee} \\ \midrule \\
		1           & Aragon              & -             & -              & -                                                                   & -                                                                      & -            \\
		2           & Ark                 & Probabilistic & PoS            & Restricted                                                          & Restricted                                                             & Yes          \\
		3           & Augur               & -             & -              & -                                                                   & -                                                                      & -            \\
		4           & Bancor              & -             & -              & -                                                                   & -                                                                      & -            \\
		5           & Bitcoin             & Probabilistic & PoW            & Public                                                              & Public                                                                 & Yes          \\
		6           & Bitcoin Cash        & Probabilistic & PoW            & Public                                                              & Public                                                                 & Yes          \\
		7           & Bitcoin Gold        & Probabilistic & PoW            & Public                                                              & Public                                                                 & Yes          \\
		8           & BitShares           & Deterministic & PoS            & Restricted                                                          & Public                                                                 & Yes          \\
		9           & Byteball            & Deterministic & Other          & Public                                                              & Restricted                                                             & Yes          \\
		10          & Cardano             & Probabilistic & PoS            & Restricted                                                          & Restricted                                                             & Yes          \\
		11          & Dash                & Probabilistic & PoW            & Public                                                              & Public                                                                 & Yes          \\
		12          & Decred              & Probabilistic & Hybrid         & Public                                                              & Public                                                                 & Yes          \\
		13          & DigiByte            & Probabilistic & PoW            & Public                                                              & Public                                                                 & Yes          \\
		14          & Dogecoin            & Probabilistic & PoW            & Public                                                              & Public                                                                 & Yes          \\
		15          & EOS                 & Deterministic & PoS            & Restricted                                                          & Restricted                                                             & No           \\
		16          & Ethereum            & Probabilistic & PoW            & Public                                                              & Public                                                                 & Yes          \\
		17          & Factom              & Probabilistic & Other          & Restricted                                                          & Restricted                                                             & No           \\
		18          & Gnosis              & -             & -              & -                                                                   & -                                                                      & -            \\
		19          & Golem               & -             & -              & -                                                                   & -                                                                      & -            \\
		20          & IOTA                & Probabilistic & PoW            & Public                                                              & Restricted                                                             & No           \\
		21          & KIN                 & Deterministic & Other          & Restricted                                                          & Restricted                                                             & No           \\
		22          & Komodo              & Probabilistic & PoW            & Public                                                              & Public                                                                 & Yes          \\
		23          & Lisk (Mainchain)    & Probabilistic & PoS            & Restricted                                                          & Restricted                                                             & Yes          \\
		24          & Litecoin            & Probabilistic & PoW            & Public                                                              & Public                                                                 & Yes          \\
		25          & Loopring            & -             & -              & -                                                                   & -                                                                      & -            \\
		26          & MOAC (MotherChain)  & Probabilistic & PoW            & Public                                                              & Public                                                                 & Yes          \\
		27          & Monacoin            & Probabilistic & PoW            & Public                                                              & Public                                                                 & Yes          \\
		28          & Monero              & Probabilistic & PoW            & Public                                                              & Public                                                                 & Yes          \\
		29          & Nebulas             & Deterministic & Pos            & Restricted                                                          & Restricted                                                             & Yes          \\
		30          & NEM                 & Probabilistic & PoS            & Restricted                                                          & Restricted                                                             & Yes          \\
		31          & NEO                 & Deterministic & Other          & Restricted                                                          & Restricted                                                             & Yes          \\
		32          & Nexus               & Deterministic & Hybrid         & Public                                                              & Public                                                                 & Yes          \\
		33          & PIVX                & Deterministic & PoS            & Public                                                              & Public                                                                 & Yes          \\
		34          & Qtum                & Probabilistic & PoS            & Public                                                              & Public                                                                 & Yes          \\
		35          & ReddCoin            & Probabilistic & PoS            & Public                                                              & Public                                                                 & Yes          \\
		36          & Ripple              & Deterministic & Other          & Restricted                                                          & Restricted                                                             & No           \\
		37          & SafeNetwork         & Deterministic & Other          & Restricted                                                          & Restricted                                                             & No           \\
		38          & Siacoin             & Probabilistic & PoW            & Public                                                              & Public                                                                 & Yes          \\
		39          & SingularityNET      & -             & -              & -                                                                   & -                                                                      & -            \\
		40          & Skycoin             & Deterministic & Other          & Public                                                              & Public                                                                 & No           \\
		41          & Steem               & Deterministic & PoS            & Restricted                                                          & Restricted                                                             & No           \\
		42          & Stellar             & Deterministic & Other          & Restricted                                                          & Restricted                                                             & No           \\
		43          & Storj               & -             & -              & -                                                                   & -                                                                      & -            \\
		44          & Stratis             & Probabilistic & PoS            & Public                                                              & Public                                                                 & Yes          \\
		45          & Syscoin             & Probabilistic & PoW            & Public                                                              & Public                                                                 & Yes          \\
		46          & TRON                & Deterministic & PoS            & Restricted                                                          & Restricted                                                             & No           \\
		47          & Verge               & Probabilistic & PoW            & Public                                                              & Public                                                                 & Yes          \\
		48          & Waves               & Deterministic & PoS            & Restricted                                                          & Public                                                                 & Yes          \\
		49          & Zcash               & Probabilistic & PoW            & Public                                                              & Public                                                                 & Yes          \\
		50          & Zcoin               & Probabilistic & PoW            & Public                                                              & Restricted                                                             & Yes         \\ \bottomrule
	\end{tabular}
	
\end{table*}

\begin{table*}[!htb]
	\centering
	
	\caption{Systems Classification according to the Action component}
	
	\label{tab:classification_action}
	\begin{tabular}{lllll} \toprule
		\textbf{ID} & \textbf{DLT system} & \textbf{\begin{tabular}[c]{@{}l@{}}Actor\\ Permission\end{tabular}} & \textbf{\begin{tabular}[c]{@{}l@{}}Read\\ Permission\end{tabular}} & \textbf{Fee} \\ \midrule\\
		1           & Aragon              & Public                                                              & Public                                                             & Yes          \\
		2           & Ark                 & Public                                                              & Public                                                             & No           \\
		3           & Augur               & Public                                                              & Public                                                             & Yes          \\
		4           & Bancor              & Public                                                              & Public                                                             & No           \\
		5           & Bitcoin             & Public                                                              & Public                                                             & No           \\
		6           & Bitcoin Cash        & Public                                                              & Public                                                             & No           \\
		7           & Bitcoin Gold        & Public                                                              & Public                                                             & No           \\
		8           & BitShares           & Public                                                              & Public                                                             & Yes          \\
		9           & Byteball            & Public                                                              & Restricted                                                         & No           \\
		10          & Cardano             & Public                                                              & Public                                                             & No           \\
		11          & Dash                & Public                                                              & Public                                                             & Yes          \\
		12          & Decred              & Public                                                              & Public                                                             & No           \\
		13          & DigiByte            & Public                                                              & Public                                                             & No           \\
		14          & Dogecoin            & Public                                                              & Public                                                             & No           \\
		15          & EOS                 & Public                                                              & Restricted                                                         & No           \\
		16          & Ethereum            & Public                                                              & Public                                                             & No           \\
		17          & Factom              & Public                                                              & Public                                                             & Yes          \\
		18          & Gnosis              & Public                                                              & Public                                                             & Yes          \\
		19          & Golem               & Public                                                              & Public                                                             & Yes          \\
		20          & IOTA                & Public                                                              & Public                                                             & No           \\
		21          & KIN                 & Restricted                                                          & Public                                                             & Yes          \\
		22          & Komodo              & Public                                                              & Public                                                             & No           \\
		23          & Lisk (Mainchain)    & Public                                                              & Public                                                             & Yes          \\
		24          & Litecoin            & Public                                                              & Public                                                             & No           \\
		25          & Loopring            & Public                                                              & Public                                                             & Yes          \\
		26          & MOAC (MotherChain)  & Public                                                              & Public                                                             & No           \\
		27          & Monacoin            & Public                                                              & Public                                                             & No           \\
		28          & Monero              & Public                                                              & Restricted                                                         & No           \\
		29          & Nebulas             & Public                                                              & Public                                                             & No           \\
		30          & NEM                 & Public                                                              & Public                                                             & No           \\
		31          & NEO                 & Public                                                              & Public                                                             & Yes          \\
		32          & Nexus               & Public                                                              & Public                                                             & No           \\
		33          & PIVX                & Public                                                              & Restricted                                                         & Yes          \\
		34          & Qtum                & Public                                                              & Public                                                             & No           \\
		35          & ReddCoin            & Public                                                              & Public                                                             & No           \\
		36          & Ripple              & Restricted                                                          & Restricted                                                         & Yes          \\
		37          & SafeNetwork         & Public                                                              & Public                                                             & No           \\
		38          & Siacoin             & Public                                                              & Public                                                             & Yes          \\
		39          & SingularityNET      & Restricted                                                          & Public                                                             & No           \\
		40          & Skycoin             & Public                                                              & Public                                                             & Yes          \\
		41          & Steem               & Public                                                              & Public                                                             & No           \\
		42          & Stellar             & Restricted                                                          & Public                                                             & Yes          \\
		43          & Storj               & Public                                                              & Public                                                             & Yes          \\
		44          & Stratis             & Public                                                              & Public                                                             & No           \\
		45          & Syscoin             & Public                                                              & Public                                                             & Yes          \\
		46          & TRON                & Public                                                              & Public                                                             & Yes          \\
		47          & Verge               & Public                                                              & Public                                                             & No           \\
		48          & Waves               & Restricted                                                          & Public                                                             & No           \\
		49          & Zcash               & Public                                                              & Restricted                                                         & No           \\
		50          & Zcoin               & Public                                                              & Public                                                             & No          \\ \bottomrule
	\end{tabular}
	
\end{table*}

\begin{table*}[!htb]
	\centering
	
	\caption{systems Classification according to the Token component}
	
	\label{tab:classification_token}
	\begin{tabular}{llllllll} \toprule
		\textbf{ID} & \textbf{DLT system} & \textbf{\begin{tabular}[c]{@{}l@{}}Supply\\ Property\end{tabular}} & \textbf{\begin{tabular}[c]{@{}l@{}}Burn\\ Property\end{tabular}} & \textbf{Transferability} & \textbf{\begin{tabular}[c]{@{}l@{}}Creation\\ Condition\end{tabular}} & \textbf{\begin{tabular}[c]{@{}l@{}}Unconditional\\ Creation\end{tabular}} & \textbf{Underlying}  \\ \midrule        \\
		1           & Aragon              & Capped                                                             & Yes                                                              & transferable             & Action                                                                  & Partially                                                                 & Action                       \\
		2           & Ark                 & Uncapped                                                           & No                                                               & transferable             & Consensus                                                               & Partially                                                                 & DL, Consensus                \\
		3           & Augur               & Capped                                                             & No                                                               & transferable             & Action                                                                  & Partially                                                                 & Action                       \\
		4           & Bancor              & Uncapped                                                           & Yes                                                              & transferable             & Action                                                                  & Partially                                                                 & Token                        \\
		5           & Bitcoin             & Capped                                                             & No                                                               & transferable             & Consensus                                                               & None                                                                      & DL                           \\
		6           & Bitcoin Cash        & Capped                                                             & No                                                               & transferable             & Consensus                                                               & None                                                                      & DL                           \\
		7           & Bitcoin Gold        & Capped                                                             & No                                                               & transferable             & Consensus                                                               & None                                                                      & DL                           \\
		8           & BitShares           & Capped                                                             & Yes                                                              & transferable             & None                                                                    & All                                                                       & DL, Consensus, Action        \\
		9           & Byteball            & Capped                                                             & No                                                               & transferable             & None                                                                    & All                                                                       & DL                           \\
		10          & Cardano             & Capped                                                             & No                                                               & transferable             & Consensus                                                               & Partially                                                                 & DL, Consensus                \\
		11          & Dash                & Capped                                                             & No                                                               & transferable             & Both                                                                    & None                                                                      & DL, Action                   \\
		12          & Decred              & Capped                                                             & No                                                               & transferable             & Both                                                                    & Partially                                                                 & Consensus, Action            \\
		13          & DigiByte            & Capped                                                             & No                                                               & transferable             & Consensus                                                               & Partially                                                                 & DL                           \\
		14          & Dogecoin            & Uncapped                                                           & No                                                               & transferable             & Consensus                                                               & None                                                                      & DL                           \\
		15          & EOS                 & Uncapped                                                           & No                                                               & transferable             & Consensus                                                               & Partially                                                                 & DL                           \\
		16          & Ethereum            & Uncapped                                                           & No                                                               & transferable             & Consensus                                                               & Partially                                                                 & DL                           \\
		17          & Factom              & Uncapped                                                           & Yes                                                              & transferable             & Consensus                                                               & Partially                                                                 & DL, Action                   \\
		18          & Gnosis              & Capped                                                             & Yes                                                              & transferable             & None                                                                    & All                                                                       & Token                        \\
		19          & Golem               & Capped                                                             & No                                                               & transferable             & None                                                                    & All                                                                       & Action                       \\
		20          & IOTA                & Capped                                                             & No                                                               & transferable             & None                                                                    & All                                                                       & None                         \\
		21          & KIN                 & Capped                                                             & No                                                               & transferable             & None                                                                    & All                                                                       & DL, Action                   \\
		22          & Komodo              & Capped                                                             & No                                                               & transferable             & Both                                                                    & Partially                                                                 & DL, Token                    \\
		23          & Lisk (Mainchain)    & Uncapped                                                           & No                                                               & transferable             & Consensus                                                               & Partially                                                                 & DL, Consensus                \\
		24          & Litecoin            & Capped                                                             & No                                                               & transferable             & Consensus                                                               & None                                                                      & DL                           \\
		25          & Loopring            & Capped                                                             & Yes                                                              & transferable             & None                                                                    & All                                                                       & Action                       \\
		26          & MOAC (MotherChain)  & Capped                                                             & No                                                               & transferable             & Consensus                                                               & Partially                                                                 & DL                           \\
		27          & Monacoin            & Capped                                                             & No                                                               & transferable             & Consensus                                                               & None                                                                      & DL                           \\
		28          & Monero              & Uncapped                                                           & No                                                               & transferable             & Consensus                                                               & None                                                                      & DL                           \\
		29          & Nebulas             & Uncapped                                                           & No                                                               & transferable             & Consensus                                                               & Partially                                                                 & DL, Consensus                \\
		30          & NEM                 & Capped                                                             & No                                                               & transferable             & None                                                                    & All                                                                       & DL, Consensus                \\
		31          & NEO                 & Capped                                                             & No                                                               & transferable             & None                                                                    & All                                                                       & Consensus, Action, Token     \\
		32          & Nexus               & Uncapped                                                           & No                                                               & transferable             & Both                                                                    & Partially                                                                 & DL, Consensus                \\
		33          & PIVX                & Capped                                                             & Yes                                                              & transferable             & Both                                                                    & Partially                                                                 & DL, Consensus, Action, Token \\
		34          & Qtum                & Capped                                                             & No                                                               & transferable             & Consensus                                                               & Partially                                                                 & DL, Consensus                \\
		35          & ReddCoin            & Uncapped                                                           & No                                                               & transferable             & Consensus                                                               & Partially                                                                 & DL, Consensus                \\
		36          & Ripple              & Capped                                                             & Yes                                                              & transferable             & None                                                                    & All                                                                       & DL, Action                   \\
		37          & SafeNetwork         & Capped                                                             & Yes                                                              & transferable             & Consensus                                                               & Partially                                                                 & Action                       \\
		38          & Siacoin             & Uncapped                                                           & Yes                                                              & transferable             & Consensus                                                               & Partially                                                                 & DL, Action                   \\
		39          & SingularityNET      & Capped                                                             & No                                                               & transferable             & None                                                                    & All                                                                       & Action                       \\
		40          & Skycoin             & Capped                                                             & No                                                               & transferable             & None                                                                    & All                                                                       & Token                        \\
		41          & Steem               & Uncapped                                                           & No                                                               & transferable             & Both                                                                    & None                                                                      & Token                        \\
		42          & Stellar             & Uncapped                                                           & No                                                               & transferable             & None                                                                    & All                                                                       & DL, Action                   \\
		43          & Storj               & Capped                                                             & No                                                               & transferable             & None                                                                    & All                                                                       & Action                       \\
		44          & Stratis             & Uncapped                                                           & No                                                               & transferable             & Consensus                                                               & Partially                                                                 & DL, Consensus, A             \\
		45          & Syscoin             & Capped                                                             & No                                                               & transferable             & Both                                                                    & Partially                                                                 & DL, Action                   \\
		46          & TRON                & Uncapped                                                           & Yes                                                              & transferable             & Consensus                                                               & Partially                                                                 & DL, Token                    \\
		47          & Verge               & Capped                                                             & No                                                               & transferable             & Consensus                                                               & None                                                                      & DL                           \\
		48          & Waves               & Capped                                                             & No                                                               & transferable             & None                                                                    & All                                                                       & DL, Consensus                \\
		49          & Zcash               & Capped                                                             & No                                                               & transferable             & Consensus                                                               & Partially                                                                 & DL                           \\
		50          & Zcoin               & Capped                                                             & Yes                                                              & transferable             & Both                                                                    & Partially                                                                 & DL, Token \\ \bottomrule                  
	\end{tabular} 
	
\end{table*}

\cleartooddpage	

\subsection{Questionnaire}
In the following the questionnaire as perceived by the survey participants of the second recruitment phase is displayed.
\cleartooddpage

\includepdf[pages=1-, scale=0.8]{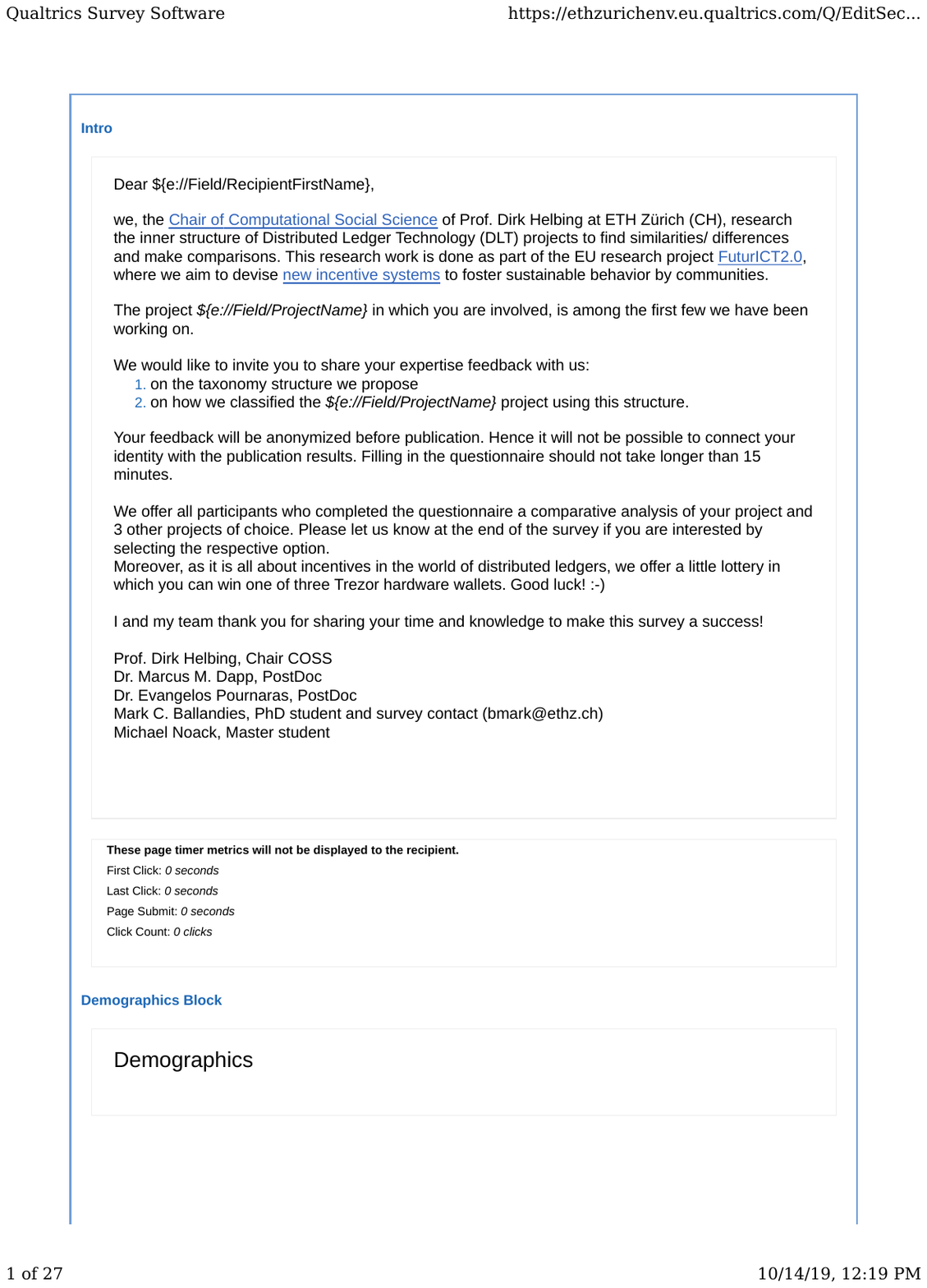}



\bibliographystyle{unsrt}
\bibliography{refs} 